\documentclass[fleqn,usenatbib]{mnras}

\usepackage{newtxtext,newtxmath}

\usepackage[T1]{fontenc}

\DeclareRobustCommand{\VAN}[3]{#2}
\let\VANthebibliography\thebibliography
\def\thebibliography{\DeclareRobustCommand{\VAN}[3]{##3}\VANthebibliography}

\usepackage{multirow}                               
\usepackage{longtable}                              
\usepackage{comment}
\usepackage{subcaption}                             
\usepackage{filecontents,catchfile}                 
\usepackage[export]{adjustbox}                      
\usepackage[dvipsnames]{xcolor}                     

\usepackage{graphicx}	
\usepackage{amsmath}	

\usepackage{ulem}                                               

\title[Interacting clusters]{
Search for interacting galaxy clusters from SDSS DR-17 employing optimized friend-of-friend algorithm and multi-messenger tracers}

\author[Oak et al.]{
 Tejas Oak,$^{1,2}$ and Surajit Paul,$^{3,2,1}$\thanks{E-mail: surajit.paul@manipal.edu}\\
 $^{1}$Department of Physics, Savitribai Phule Pune University, Ganeshkhind, Pune 411007, India\\
 $^{2}$Raman Research Institute, Sadashivnagar, Bengaluru 560080, India\\
 $^{3}$Manipal Centre for Natural Sciences, Centre of Excellence, Manipal Academy of Higher Education, Manipal 576104, India 
}

\date{Accepted 2024 January 16. Received 2024 January 16; in original form 2023 May 25}

\pubyear{2024}

\begin{document}

\label{firstpage}
\pagerange{\pageref{firstpage}--\pageref{lastpage}}
\maketitle

\begin{abstract}

In the theoretical framework of hierarchical structure formation, galaxy clusters evolve through continuous accretion and
mergers of substructures. Cosmological simulations have revealed the best picture of the Universe as a 3-D filamentary
network of dark-matter distribution called the cosmic web. Galaxy clusters are found to form at the nodes of this network
and are the regions of high merging activity. Such mergers being highly energetic, contain a wealth of information about the
dynamical evolution of structures in the Universe. Observational validation of this scenario needs a colossal effort to identify numerous events from all-sky surveys. Therefore, such efforts are sparse in literature and tend to focus on individual systems. In this work, we present an improved search algorithm for identifying interacting galaxy clusters and have successfully produced a comprehensive list of systems from SDSS DR-17. By proposing a set of physically motivated criteria, we classified these interacting clusters into two broad classes, ’merging’ and ’pre-merging/postmerging’ systems. Interestingly, as predicted by simulations, we found that most cases show cluster interaction along the prominent cosmic filaments of galaxy distribution (i.e., the proxy for DM filaments), with the most violent ones at their nodes. Moreover, we traced the imprint of interactions through multi-band signatures, such as diffuse cluster emissions in radio or X-rays. Although we could not find direct evidence of diffuse emission from connecting filaments and ridges; our catalogue of interacting clusters will ease locating such faintest emissions as data from sensitive telescopes like eROSITA or SKA, becomes accessible.

\end{abstract}




\section{Introduction}
The most widely accepted cosmological model, $\Lambda$-CDM features a hierarchical merging mechanism for structure formation at large scales \citep{whiteandressStructform,CDMStructform}. Here, the most fundamental unit of cosmological structure is a dark matter halo \citep{Navarro_1996_halos}. It can withstand cosmic expansion by virtue of gravitationally bound matter. Dark matter halo can also possess a virialized substructure,
that gravitationally interacts with the ordinary matter enabling the formation of stars and galaxies \citep{DMHaloGalForm}. Galaxies undergo merger to form galaxy groups; several galaxy groups under the action of gravity may form clusters of galaxies \citep{Bykov_2015_gals_and_gclust}. At large scales, the observed cosmological structure forms a web-like pattern and galaxies are seen to be distributed along the filamentary connections of this web which are typically of a few to a few tens of Mpc in length \citep{LSS_SDSS}. As most of the matter is present along the filaments in the cosmic web \citep{Lovisari_2021}, it is expected that most of the clusters also form and interact on these filaments. The nodes where the filaments merge are also the locations of the interaction of galaxy clusters and occasionally that even may lead to the formation of super-clusters. 

A cluster merger event is highly energetic, with the release of up to $10^{65}$ ergs of binding energy during mergers. Such an event creates very high-pressure gradients at the cluster core and heats the intra-cluster medium (ICM). As a result, the ICM starts expanding and can often give rise to shocks \citep{Paul_2011_ICM_Shocks} and chaotic regions of magneto-hydrodynamic turbulence, which are driven into the ICM. A strong thermal bremsstrahlung X-ray radiation can also be observed from these nodes of hot plasma where galaxy clusters are formed (\citet{formanxrayICM}). Galaxy clusters also exhibit a non-thermal component of the emission. Due to the cluster-size magnetic fields, possibly amplified by the merger turbulence \citep{Miniati_2016, Miniati2015Nature}, the particles accelerated to relativistic velocities emit synchrotron radiation which can be observed in radio wavelengths \citep{Feretti_2012,van_Weeren_2019}. Thus, the study of interacting galaxy clusters is insightful to understanding multiple phenomena such as the energy evolution in the Universe, the nature of dark matter, dynamical processes of plasma and large-scale structure as a whole.

Studying structures in the Universe at large scales is limited by the availability of large survey data. While X-ray observations are so far the best method to probe the intracluster medium, there is no deep all-sky survey in X-rays available in the public domain and would have to wait till eROSITA data becomes public \citep{eRosita_main}. Currently, the most detailed all-sky survey data of galaxies is available in optical wavelengths. However, optical data does not provide direct information on the diffuse gravitationally bound baryonic matter of large-scale structures, such as the medium in groups or clusters of galaxies. Rather, the identification of clusters in optical studies requires the implementation of techniques and has been best done so far by running clustering algorithms on the galaxy survey data \citep[e.g.,][]{huchra1982groups,Berlind_fof}. In this context, the most commonly used clustering algorithm is the conventional friends of friends (FoF) and has been able to detect galaxy clusters fairly reliably \citep{Tempel_2014}. However, the implementation of the FoF algorithm to search galaxy clusters is not always straightforward. As galaxy clusters can have a wide range of mass and physical dimensions, the parameters of the clustering algorithm need to be tuned such that most of the real clusters are robustly detected over the entire sky. The usual FoF method uses a framework to form a connected group in which the friends and the friends of friends are searched within a predefined radius known as the linking length. 

Normally, the linking length is chosen to be a weighted mean distance of the member elements in a search volume \citep{huchra1982groups}. There are a few heuristic arguments to determine the best value of the weight factor based on the spherical collapse model - normally, this weight factor or the linking parameter is taken to be 0.2 \citep{White_B_parameter}. However, different works carried out have undertaken substantially different values which better suited their respective science goals \citep{tagoInitialLL}. 

While, the FoF methods are often used to find structures from large galaxy redshift surveys \citep[][e.g.]{Botzler_fof,Murphy_fof} and cosmological dark matter simulations, efforts to locate interacting systems are very rare. There have been some previous efforts for a large volume algorithmic detection of interacting clusters, most recently by \citep{TempleMerger}. However, given the highly complex nature of the interaction, the authors concluded that visually inspecting each system remains the best way over algorithmic detection. The authors also pointed out that there were known merging systems that were not classified by their algorithm. Additionally, these efforts remained limited only to merging systems and do not further discuss interacting substructures, or pre/post-merging scenarios for instance.
Though there are a few known events of interacting galaxy clusters, they are mainly found in studies on a case-by-case basis \citep{merger_example_1,merger_example_2}. These systems were usually spotted by examining merging signatures from deep observations of multi-band diffuse emissions. A thorough detection of such systems will only be possible when the sensitive multi-band all-sky survey images become available or approximate locations of probable sources can be identified by running some algorithm on large sensitive survey data. However, for direct detection, carrying out all-sky surveys at high sensitivity (required to capture diffuse emissions) is a daunting task and may become possible only in the far future. On the other hand, there has been no all-sky algorithmic search for interacting clusters that cross-validate multiband signatures, creating a void in understanding the dynamical evolution of structures in the Universe. 

In this work, we attempt to search such interacting clusters and their multi-waveband features. However, they are known to be highly complex, where the role of dark matter and gravitationally trapped gas cannot be neglected. For example, it is known that the DM being collisionless, will undergo passage much earlier than the baryonic matter \citep{Dawson_DM_Shift}. In this sense, galaxy redshift surveys contain only a partial picture of interacting clusters as galaxies are a minor part of the total baryonic gas present in the Universe. On the other hand, a huge advantage of redshift surveys is that they are one of the richest and most complete datasets to work with, and departing from this resource would not be desirable. Our efforts were therefore focused on the following question - can the linking parameters in the clustering algorithm be informed by a supporting dataset which is closer to real, physically interacting clusters? We try to answer this question by formulating a novel optimization problem by implementing an optimized, physically informed FOF algorithm, developing reasonable interaction criteria to detect and classify interacting clusters into two broad phases, and finally, providing a multi-wavelength overview of a few of the systems which stand as validation to the attempt to detect diffuse intergalactic material, and provides a more complete picture of the interaction scenario.

The structure of the paper is as follows. Section~\ref{methedology} covers the details of the algorithm we developed for this work and a brief analysis of its performance. The data used in this work has also been detailed here. We present the interesting candidates from our search and their multi-band signatures in section \ref{results}.  Finally, we summarise and conclude the paper in Section~\ref{discussion}. 
Throughout this paper we assume , Hubble constant $H_0$ = 70.00 km s$^{-1}$ Mpc $^{-1}$, matter density $\Omega_m$ = 0.3,dark energy density $\Omega_\Lambda$ = 0.7.  

\section{Searching interacting galaxy clusters}
\label{methedology}
\begin{figure}
	\includegraphics[scale=0.5]{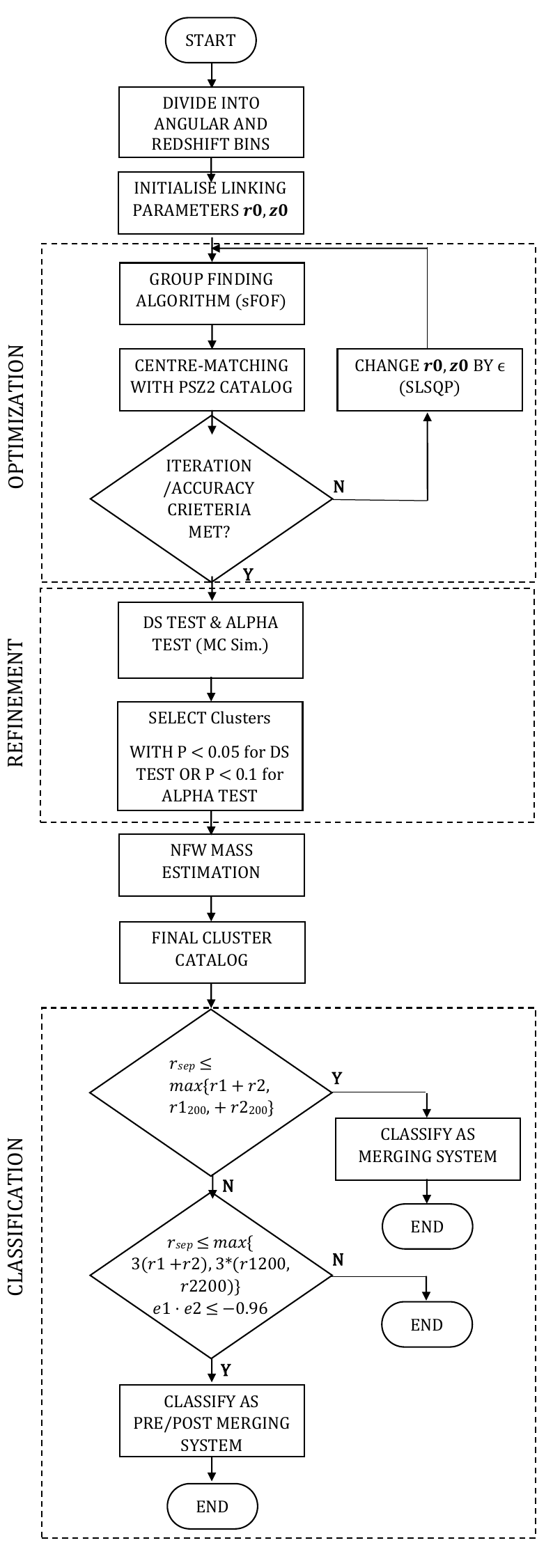}
    \caption{Flowchart of the algorithm used}
    \label{fig:flowchart}
\end{figure}

In a broader sense, when a system of multiple (at least a pair of) galaxy clusters come so nearby that their individual medium gets accreted and starts mixing up, the system would be called in an interacting state \citep[e.g.,][]{Int_Clust_Example}. Naturally, these systems would pass through pre-merging, merging and post-merging phases as it evolves in time. However, because of their high dynamical activity and complicated mass and energy distribution, it would be very difficult to identify their exact phase of mergers. One of the early attempts for a large dedicated algorithmic search for merging clusters was made by \citet{TempleMerger}. In our case, The challenge in predicting clusters in various merging stages is threefold. The first is reliably extracting clusters from the survey data. Once the clusters are extracted, further refinement needs to be performed, as the interaction of subsystems within massive clusters is the most likely candidate for a merger. The ultimate challenge lies in appropriately deciphering the merging phase of the system. Therefore it demands a proper scheme which we summarized in Fig.~\ref{fig:flowchart} and elaborated in the further sections.

\subsection{Primary data set for this study}\label{data}

Since the optical galaxy surveys provide the deepest and widest sky coverage as well as the best redshift information, we chose to work with the optical galaxy survey data. Currently, the latest galaxy data available is from the Sloan Digital Sky Survey data release 17 (SDSS DR17; \citep{SDSS17} and naturally we chose this as the primary data set for this study. The galaxy data has been downloaded from the SDSS sky server using the appropriate SQL query. We only select the spectroscopic class of `galaxies' and restricted the search to the northern sky up to redshift $z = 0.2$. The selected sample contains over 7,00,000 galaxies with spectroscopically or photometrically determined redshifts.

\subsection{Our group finding algorithm}
\label{fof}

Upon procuring the data from the SDSS archive, we further formulate a group-finding scheme based on the Friend-of-Friend algorithm. The central idea behind a group-finding algorithm is to group all the galaxies which lie within a predefined distance (linking length) from each other. Here, the direct friends are the ones who directly come within this predefined distance from the galaxy under consideration and friend-of-friends are those who are similarly connected via a direct fried. The very first of such algorithms were developed by \citet{huchra1982groups} on cFA redshift surveys where the authors used the linking length based on projected separation and line-of-sight velocity difference of galaxies in the examined volume.

For this work, we have used the open-source sFOF group finding algorithm by \citet{sfof}. This package fits to our plan well, as there is significant room for customization, also, there is a scope for automation. This enabled us to incorporate the determination of physically motivated linking length for identifying the structures as discussed in the following section.

\subsection{Choice of initial linking parameters}\label{LLopt}

Since the FoF algorithm gathers elements in the searched region that are close to each other to form a group, defining a connecting radius to give a numerical value to the closeness is essential. Such a distance is called the linking length and is usually taken as a fractional multiple of the mean separation of the elements in the locally searched volume \citep[e.g.,][]{tempel_LL_fraction}.
The exact choice of linking parameters is normally motivated by the scientific goal of the study, and it is known that small changes tend to cause significant differences in the formed groups \citep[For a review, refer][]{duarte_LL_review}. For example, the linking parameters tuned for a large-volume ($\sim100$s of Mpc$^3$) redshift survey sample from mock catalogues may not be sensitive enough to delineate interacting systems lying close by ($\sim10$~Mpc), or dynamical substructures within a cluster, which is a key area of this study. Ideally, we want the algorithm to be sensitive towards interacting systems (which corresponds to finer linking parameters), but we also wish to preserve the purity of halos avoiding spurious detections (corresponding to slightly coarser linking parameters). Thus, we need to explore the linking parameter space to optimize the parameters that best pick up the actual, physically interacting clusters. We choose to adapt our starting point based on the work done by \citet{tagoInitialLL} on SDSS DR 5, corresponding to $z_0$ = 250 km s$^{-1}$ and $r_0$ = 0.25 Mpc. In principle, any reasonable choice of initial linking parameters can be made, as we expect the optimizer to take care of the trade-offs and arrive at a global extremum. Hence, the Initial choice of linking length matters, although only marginally.   

\subsection {Scaling relation for linking length} 
As most surveys are flux-limited, only the brightest cluster members are visible at deeper redshifts due to the limit on the telescope sensitivity. These bright galaxies usually reside nearby the core of the clusters and may not account for the missing galaxies accurately. This would impact the measurement of the true size of the groups, especially in the line of sight \citep{Tempel_2014}. Hence, an appropriate scaling of linking parameters needs to be done separately in redshift space.
For the line of sight linking length, we used the scaling relation available in sFOF, as described below.

\begin{equation}
L(z) \propto L_0 \left(\frac{dN}{dZ} \times \frac{dz}{dV} \times \frac{1}{A_{sky}}\right)^{-0.5}   
\end{equation}

where $\frac{dN}{dZ}$  represents the surface number density of galaxies for the survey redshift range, $\frac{dz}{dV}$ is the differential comoving volume and  $A_{sky}$  is the fractional sky coverage of the sample. This surface number density-based scaling does not take in magnitude-based pre-selection (For detailed description, please see \citealt{sfof})

\subsection{Mass estimation of the detected clusters} \label{MassEst}
The large galaxy redshift surveys do not provide the mass information for each of the detected galaxies. However, mass is an important parameter in finding gravitational clustering \citep{Gupta_2023} and thus needs to be estimated from the information available in the survey data. Here, to estimate the masses of clusters, we undertook the single parametrized NFW profile approach similar to \citet{Tempel_2014}.

The NFW profile \citep{NFW_Profile} is given by,

\begin{equation}
\label{NFWProfile}
    \rho(r) = \frac{\delta_c~\rho_{\rm{crit}}}{\frac{r}{R_s} ~ \left( 1 + \frac{r}{R_s}  \right)^2}
\end{equation}

where $R_s$ represents a scaled radius and $\rho_{crit}$ is the critical density of the universe. This profile can be entirely parametrized on the cluster's virial mass ($M_{200}$). We start by making an initial guess about the virial mass of the cluster. In our case, for simplicity, we assumed a constant mass for each galaxy, so that the cluster mass is proportional to the richness of the cluster. Any reasonable initial guess can be made. Assuming that clusters tend to be virialised in regions where the density is 200 times the critical density of the Universe ($\rho_{crit}$) at that redshift, the virial radius corresponding to this mass was calculated as 

\begin{equation}
    \label{r200}
    r_{200} = \left(\frac{3~M_{200}}{4\pi~200~\rho_{\rm{crit}}}\right)^{1/3}
\end{equation}

Furthermore, the $\delta_c$ of Eq.~\ref{NFWProfile} has been parameterized using concentration parameter (i.e. $c_{200}$ in Eq.~\ref{c200}) and scale radius (i.e. $R_s$~\ref{Rs}) which are in turn a derivative of the initial guess mass.

\begin{equation}
    \label{c200}
    \log({c_{200}}) = 0.83 - 0.098~\log{\left( \frac{M_{200}}{10^{12}~h^{-1}~M_{\odot} }    \right)} 
\end{equation}
\begin{equation}
    \label{Rs}
    R_s = \frac{c_{200}}{r_{200}}
\end{equation}

Finally, the $\delta_c$ can be computed from

\begin{equation}
    \label{deltaC}
    \delta_c = \frac{200}{3}~\frac{c_{200}^3}{\ln{1 + c_{200}} - \frac{c_{200}}{1 + c_{200}}} 
\end{equation}

Therefore, from Eq. (\ref{r200},\ref{c200},\ref{Rs},\ref{deltaC} ), the NFW profile (Eq.~\ref{NFWProfile}) is shown to have sole dependence on the initial guess of $M_{200}$ and thus a method needs to be implemented to achieve more accurate value of the mass.

The refinement of guess mass has been achieved through an iterative process considering a simple hydrostatic equilibrium of the system, $2T=U$ i.e., virialization. Where $T$ is kinetic energy and $U$ is the potential energy of the identified clusters. The potential energy associated with the cluster can be estimated as 

\begin{equation}
    U = G\frac{M_{200}}{R_g} 
\end{equation}

Where $R_g$ is the gravitational radius and $G$ is the constant of gravitation. By equating the above-stated potential energy to potential energy obtained by integrating the NFW profile \citep{Binney08}, 

\begin{equation}
    U = 4\pi G~ \int_{0}^{r_{200}} \frac{M(r)}{r} ~\rho(r)~r^2~dr    
\end{equation}

we computed out $R_g$. Finally, a convergent solution was obtained by updating $M_{200}$ iteratively as $M_{200}^{\rm{new}}$ in 

\begin{equation}
    M_{200}^{\rm{new}} = 2.3525 \times 10^{12} ~ R_g ~ \left(\frac{\sigma_v}{100}\right)^2 \rm \frac {M_{\odot}}{Mpc~km^2 ~ s^{-2}}
\end{equation}
The exit condition can be imposed by monitoring the rate of convergence of this algorithm. As an estimate, We found that the difference of computed mass becomes smaller than 0.1\% per iteration, after an average of 15 iterations for each cluster, which we chose as a reasonable exit condition for the mass computation loop in our algorithm.

\subsection{Performance of LL-Optimized Cluster Finder}

\begin{figure*}
	\includegraphics[width=\textwidth]{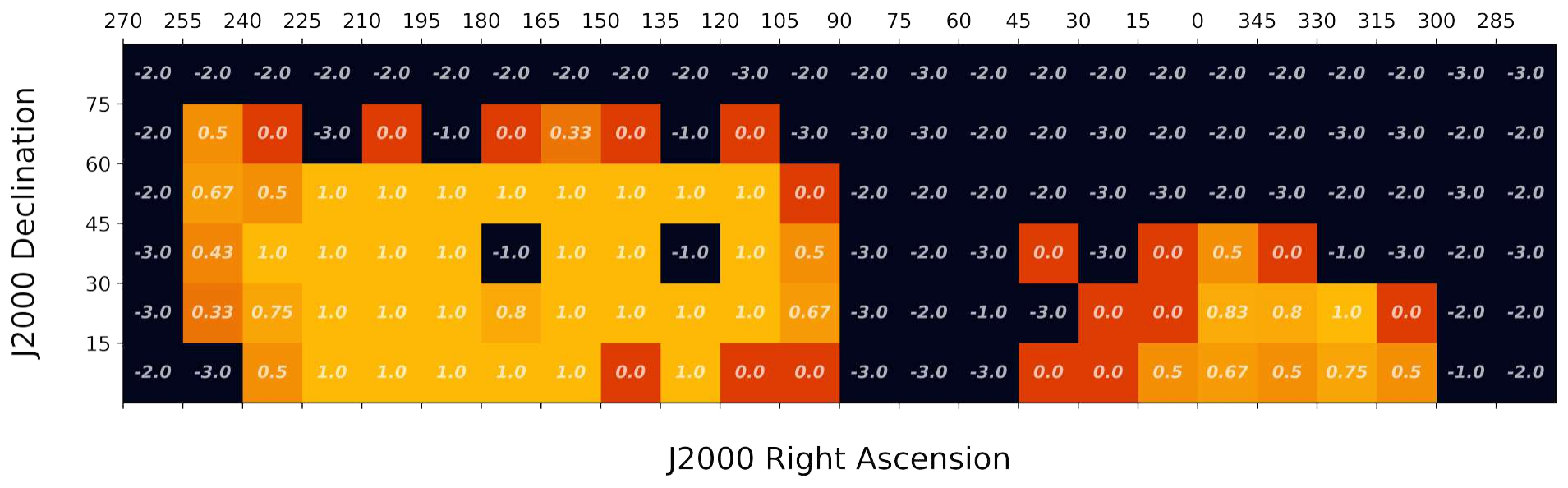}
    \caption{Sector-wise accuracy map of the optimization process. Each box is a $ 15 ^{\circ}\times 15 ^{\circ}$ sector in the sky. The fractions in each sector represent the accuracy achieved from 0.0 to 1.0. (-1) indicates no clusters found as a result of poor galaxy data. (-2) and (-3) indicates missing SDSS and PSZ2 data respectively. }
    \label{fig:accmap}
\end{figure*}

The optimization has been obtained by defining the objective function as the accuracy of centre-matching with the Planck SZ cluster centres \citep{PSZ2}. The Sunyaev-Zeldovich (SZ) effect \citep{SZ_original} is a mainstream method in astronomy to reliably detect galaxy clusters. Planck SZ cluster survey data release 2 (Planck SZ2) has identified over 1000 confirmed clusters using the SZ signal, providing a statistically relevant data set for implementing an optimization algorithm. The optimization metric here is simply the ratio of the number of FOF cluster centres matched with SZ cluster centres in the search bin to the total number of SZ clusters present. This matching accuracy is a measure of completeness. Ideally, we want the linking lengths to be tuned on physical information such that a complete SZ cluster sample is detected via the FOF algorithm. We maximize this ratio by varying the linking parameters, by starting with the initial conditions mentioned in section~\ref{LLopt}.
In order to ensure that the optimization process is free from biases and performs uniform clustering in the entire search space, we assess the accuracy of cluster-finding across the sectors with the help of a heatmap. Once the final groups are formed, their centres are compared with the PSZ2 catalogue within some tolerance. Based on this, we can define the sector-wise accuracy, as the ratio of centre-matched PSZ2 clusters to the total PSZ2 clusters present in the sector under consideration. The final sector-wise accuracy is shown in Fig.~\ref{fig:accmap}.

The fractions in each sector represent the accuracy achieved from 0.0 to 1.0. A negative number indicates the absence of data, either SDSS (-2) or Planck-SZ2 (-3). In some sectors, the data is available, however, the number is too less to perform clustering. Such sectors are indicated by (-1). Using this, a mean accuracy of close to 83\% has been achieved in centre-matching of SZ clusters to FOF clusters in the entire search region, excluding the negative grids in Fig.\ref{fig:accmap}.

In quite a few cluster-finding studies, the ratio of radial to transverse linking lengths is assumed empirically, based on their specific scientific goals \citep{duarte_LL_review}. In these studies, the actual values are then determined by running the cluster-finder on a mock catalogue with known outcome \citep[e.g.,][]{Berlind_fof}. Such studies essentially model the linking parameters without considering any physical pointers, such as SZ clusters used in this work, and are thus radically different from our work. The flux-limited catalogue prepared by \citet{Tempel_2014} (Hereafter referred to as Tempel14) is a well-cited, reliable work which also takes care of finger-of-god effect and redshift-based dimming effects. There, the ratio of linking lengths was taken as 10. 

As a measure of relative performance, we analyse the position offsets in locating centres of PSZ2 clusters (which serves as observed ground truth) to this independent study by Tempel and our work. 
\begin{figure}
    \includegraphics[width=0.5\textwidth]{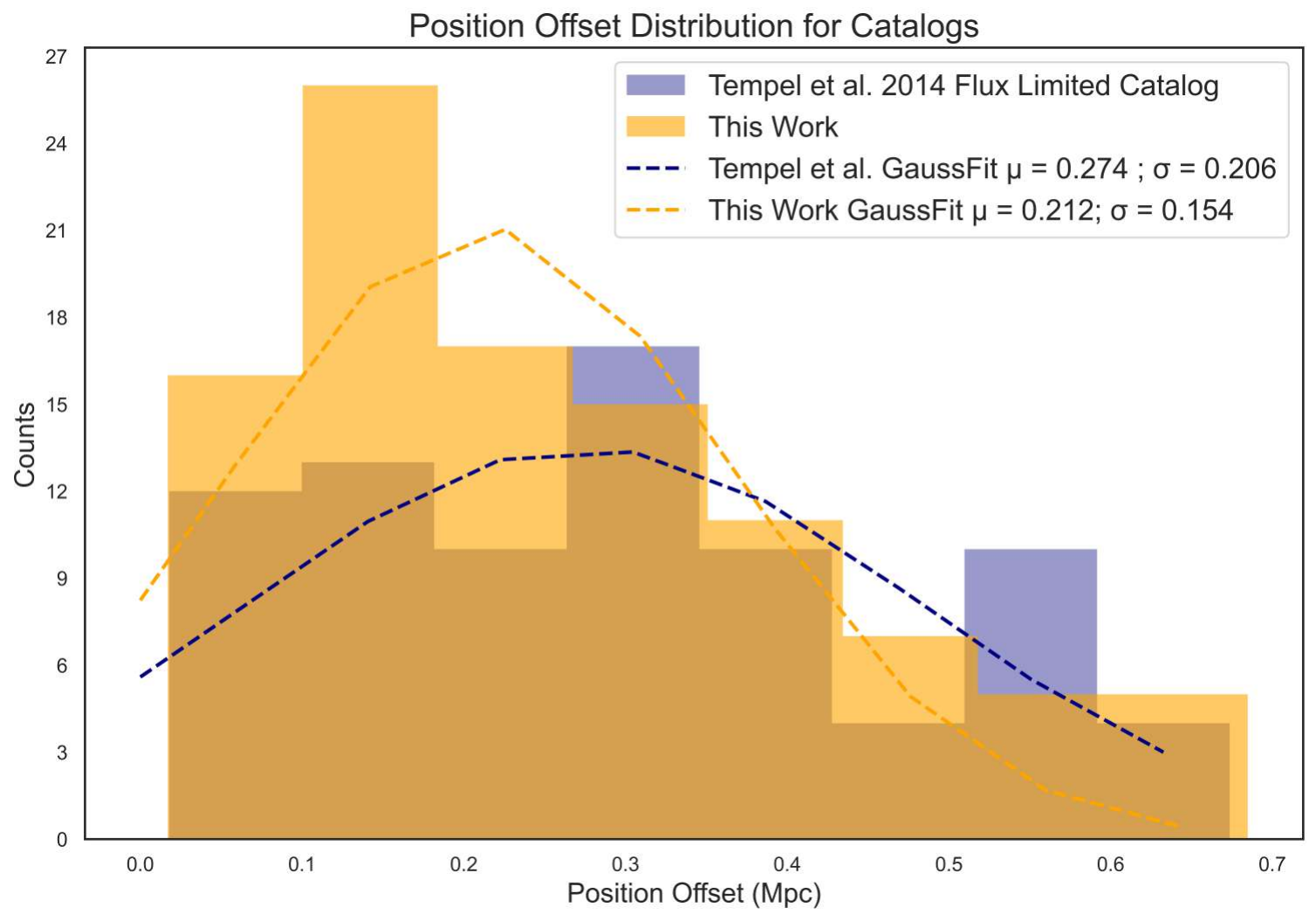}
    \caption{Distribution of position offsets for PSZ2 clusters for \citet{Tempel_2014} and this work}
    \label{fig:poff_comparison}
\end{figure}
The distribution of offsets between centres determined by the two methods mentioned above has been shown in
Fig. \ref{fig:poff_comparison}. The dashed lines show the best-fit Gaussians for both distributions. By comparing the means and standard deviations for both catalogues, it can be inferred that the cluster centres are more accurately determined, with a comparatively tighter spread in this current work. Additionally, the expected match with a larger number of PSZ2 clusters ensured the detection of more actual clusters.

As a final metric of reliability, we checked for the existence of clusters from robust, well-known catalogues (such as Abell) into clusters identified with our method. For Abell clusters, 191 out of 379 objects could be centre-matched using the same linking parameters in the same search volume. While about 2400 of 12000 WHL clusters could be centre-matched, this percentage may not be representative as accurate spectroscopic redshifts are not available with most of the WHL clusters.

\subsubsection{Quantitative analysis of the constructed group catalogue}

\begin{figure}
    \centering
        \includegraphics[width=0.85\linewidth]{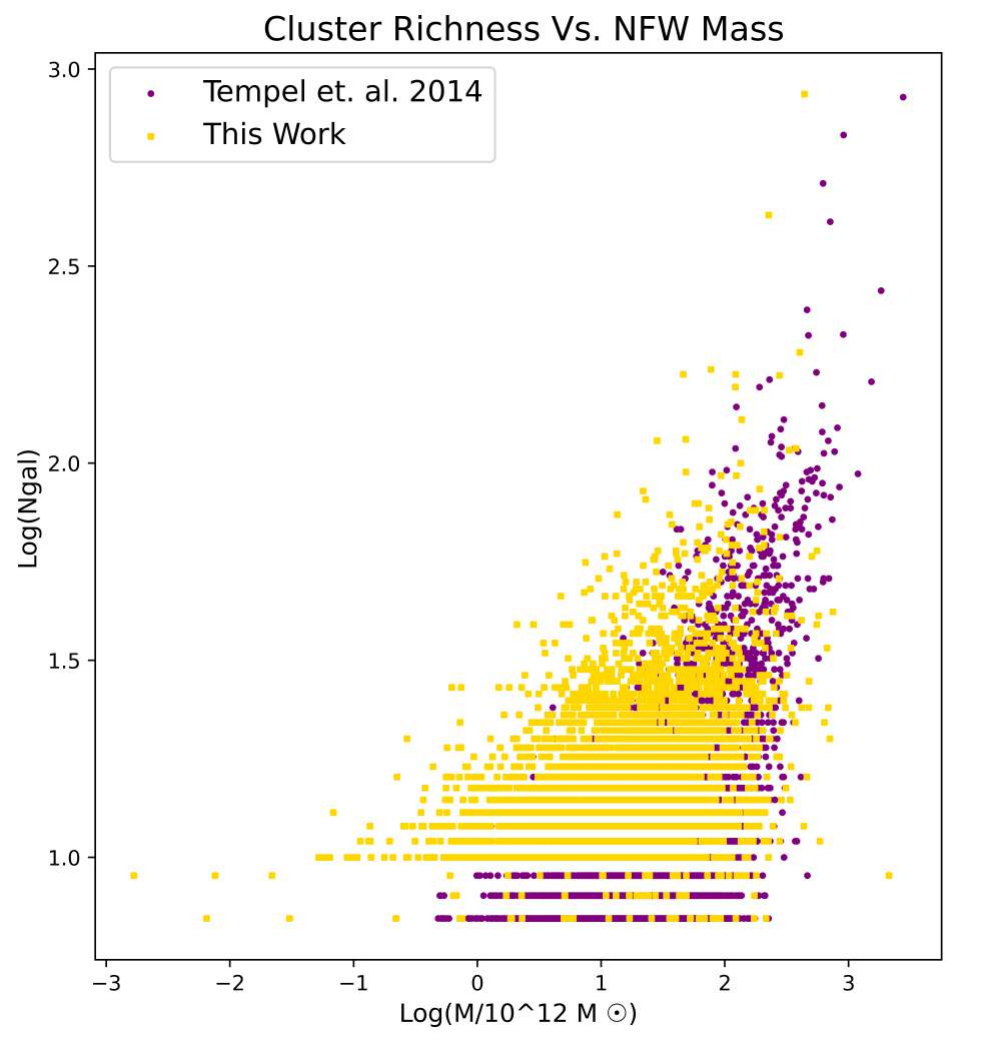}
        \caption{ Cluster Richness Vs. Cluster mass plotted for Tempel et. al. 2014 (Purple) and this work (Gold). A cutoff of Ngal = 7 is applied to ensure uniformity between both the datasets i.e. the richness threshold of this work. }
        \label{fig:clustprops}   
\end{figure}

To ensure that clusters generated by our clustering algorithm remain largely analogous to other cluster-finding studies, we can analyse the properties of clusters as a whole. Fig~\ref{fig:clustprops} shows the comparison of Cluster mass and cluster richness of this work with flux-limited group catalogue of \citet{Tempel_2014}. This comparison provides one way to show that the physical properties of clusters do not show reasonable deviation from an acceptable catalogue validated by calculating a mass function. However, for a stricter and more robust means of evaluation, a completeness study with a mock catalogue is needed. For this, an n-body cosmological simulation based on an identical DR17 sample will be required, which can be the topic of a separate independent study. However, it is beyond the scope of the current work.

\subsection{Refining substructures and galaxy groups}
\label{Refinement}
After identifying the structures using the optimized search parameters, our further task would be to determine the merging states. However, empirically, we expect merging systems to lie in close proximity to each other. Therefore, even though optimized, an average linking length used for the entire sky may not always be suitable to distinctly identify all those close-by structures. Consequently, our grouping algorithm may classify them as a single cluster and hence we may miss out on those interacting systems. 
We thus need a robust way to pick up substructures from the prepared cluster catalogue.
In this study, a "substructure" is essentially a signature of incomplete relaxation in the identified clusters. The sub-structures may be found either in the positional distribution or in the velocity dispersion of the member galaxies, or in both. In practice, various statistical tests have been formulated to pick up such substructures \citep[]{West_Tests,Bird_substructure},\citep[For a review, refer][]{Pinkley_Tests}. In this work, we have utilised a couple of such tests, the DS-Test (Dressler - Shectman or delta test) by \citet{DSTest} and the Alpha test by \citet{AlphaTest}.    

\subsubsection{Dressler - Shectman (DS) Test}

\begin{figure}
    \centering
        \includegraphics[width=0.85\linewidth]{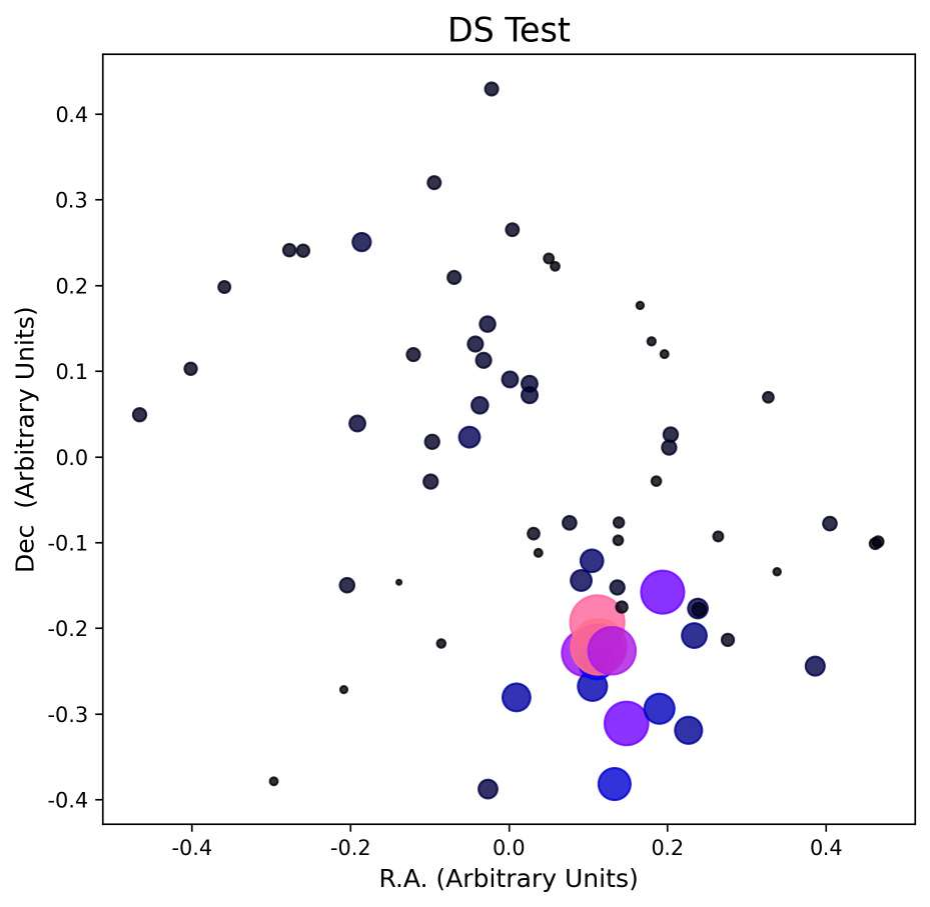}
        \caption{ An example bubble plot generated after DS test. The radii of circles and colours are proportional to $e^{\delta}$. A substructure can be seen to the south.}\label{fig:bplot}   
\end{figure}

In the DS Test, a delta statistic for each member galaxy of a cluster under consideration is computed as,
\begin{equation}
    \delta_i = \left( \frac{NN_n}{\sigma_{\rm{global}}} ~ (\bar{V}_{\rm{local}} - \bar{V}_{\rm{global}})^2 + (\sigma_{\rm{local}} - \sigma_{\rm{global}})^2  \right)^{1/2}
\end{equation}
where $NN_n$ refers to the number of spatially nearest neighbours for the galaxy (typically $N_n$ is taken as the square root of cluster richness $\sqrt{N_{gal}}$), the local quantities refer to quantities associated with the local group, while global quantities are of the entire cluster. This delta statistic is a measure of how different the local galaxy environment is from the global cluster; and is sensitive to both plane-of-sky distribution as well as the velocity dispersion. The delta statistic for the cluster can be then computed by taking the sum of individual galaxy statistics, $\Delta = \sum \delta_i$.
The results of the delta test performed on a chosen cluster from our catalogue can be visualised in the bubble plots as shown in Figure~\ref{fig:bplot}. Each galaxy is represented as a circle with a radius proportional to the delta statistic. A clear evidence of substructure can be observed in the plot.\\

\subsubsection{Alpha Test}

There may be situations where substructures are superimposed in plane-of-sky or have similar dispersion means. In such cases, the DS test does not provide desired results.
As a countermeasure, we perform the Alpha test, also known as the centroid shift test. The advantage of the alpha test is that the galaxies are weighted in inverse proportion to the local group's line-of-sight velocity dispersion, and hence, galaxies with smaller velocity dispersion dominate the statistic, which can be suppressed in the DS test. To perform the alpha test, we first calculate the centroids of plane-of-sky distribution,

\begin{equation}
    (\Bar{x},\Bar{y}) = (\frac{1}{N_{gal}}~ \sum{{x_i}} , \frac{1}{N_{gal}}~ \sum{{y_i}})
\end{equation}

Thereafter, for each galaxy, N nearest neighbours are selected to compute velocity dispersion, $\sigma_{local}$ and each galaxy is assigned a weight, $w = \frac{1}{\sigma_{\rm{local}}}$. Further, for $N_v$ nearest neighbours in \textit{velocity} space, a new centroid is calculated as,

\begin{equation}
    (\Bar{x}^\prime,\Bar{y}^\prime) =  (\frac{\sum_{i = 1}^{N_v + 1} x_i~w_i}{\sum_{i = 1}^{N_v + 1}w_i},\frac{\sum_{i = 1}^{N_v + 1} y_i~w_i}{\sum_{i = 1}^{N_v + 1}w_i})
\end{equation}

The unweighted centroid shift for each galaxy is then calculated as, 

\begin{equation}
    \gamma_i = [(\Bar{x}^\prime - \Bar{x})^2 + (\Bar{y}^\prime - \Bar{y})^2]^{1/2}
\end{equation}

Finally, the alpha statistic for the cluster is measured as the mean of $\gamma_i $ values,

\begin{equation}
    \alpha = \frac{1}{N_{gal}} ~ \sum_{i} \gamma_i
\end{equation}

 \citet{Pinkley_Tests} did not find the superimposition to be a major obstacle for this test. Hence, we can combine this with the DS test to obtain a robust mean of evaluating substructures. 

\subsubsection{Test normalization and hypothesis testing}
The risk with any statistical substructure test is the false-positive arising due to correlated positions and velocities of member galaxies. To tackle this issue, we randomly shuffle the velocities of galaxies without changing their positions and perform the tests. This ensures that any correlation, if exists, is broken and the true substructure is tested for. Based on this logic, a systematic hypothesis testing study was also carried out in the following way. We performed 100 Montecarlo simulations for each cluster and computed the level of significance as the fraction of Montecarlo iterations where the resulting statistic exceeded the observed statistic to the total Monte Carlo iterations (100 in this case). Such as, for delta test,  ($p = \frac{N_{\Delta > \Delta_{obs}}}{N_{\rm{total}}}$).  We then selected the clusters with $p < 0.05$ for either of the tests, split the cluster into substructures as indicated by the tests, and merged them into the main catalogue.

\subsection{Criterion for cluster merging phases}
\label{MergeCrit}
Once we adequately delineated the substructures, a further inspection was done to identify merger phases. We kept our criterion for defining the dynamical phases as simple as possible, however, made sure that the criteria were physically motivated.

Studies related to cluster mass estimation using top-hat filtered spherical overdensities \citep{Peebles1993} have shown that spherically collapsing cosmic structures are seen to be virialized in regions where the mean mass density is roughly 200 times the critical density of the universe. This radius is called $r_{200}$ and is considered to be the characteristic radius of galaxy clusters (For a review see \citet{clust_rad_review}). It is thus a natural choice to consider this radius to determine the interaction between two clusters. If this characteristic radius overlaps for two clusters, it would be safe to conclude that their member galaxies and intergalactic medium are within the interaction range of each other, indicating a merger. Beyond the virial radius, most of the interesting physics associated with a cluster lies roughly within the region of  $\sim3 \times r_{200}$, as this range contains the first turnaround radius \citep{turnaround_radius}. This is also the region where extreme peripheral radio relics can also be seen \citep[]{relic_range,Bagchi_2011_doublerelic}. At scales larger than this, physics goes beyond the collapsing objects and goes to the filamentary region made up of a warm-hot-intergalactic medium. It is therefore logical to assume  $3 \times r_{200}$ as the outer interaction limit for the cluster medium. Theoretical and computational works predict some interesting observable effects of cluster interaction in this region \citep{Gu_premerger_shock,Bonafede_2022_coma}. A few observational evidences have also been reported recently \citep{Radio_filament_example,vazza_radio_ridge}. Thus the above understanding of the cluster systems provides us with the clue to define the merging phases.

We, first calculated the value of the usual virial radius, commonly taken as $r_{200}$, for each group assuming spherical symmetry taking centres as the centre of mass of the distribution. The radius of the cluster ($r$) can also be estimated from the spatial dispersion of its members. We used the following criterion to declare the system as merging. 

\begin{equation}
\begin{aligned}
 & r_{\rm{sep}} \le \rm{max}\{r1 + r2,r1_{200} + r2_{200}\} \\
\end{aligned}
\end{equation}

 Here, r1 and r2 are the radii estimated by spatial dispersion of members, $r1_{200},r2_{200}$ are the radii corresponding to overdensity radius of 200 times the critical density of the Universe at the cluster redshift; usually considered as the virial radius of the systems (procedure as described in Section \ref{MassEst}), and $r_{\rm{sep}}$ is the separation between the clusters. For clusters which are in a pre or post-merging state, we expect them to be gravitationally influenced by each other. This interaction is expected to elongate the cluster along the direction of attraction. An estimate for this elongation vector can be obtained by taking the difference between the centre of mass and the centre of the spatial extent of the galaxy distribution inside the identified cluster. We, therefore, construct an elongation vector as, 

\begin{equation}
\vec{e} = \sum_{i} (\mu_{i} - m_{i}) \hat{e_{i}}   
\end{equation}

where $\mu$ is the centroid, $m$ is the median of the distribution in a given dimension, and $`i'$ is an index which runs over the dimensions. e.g, the line of sight component of this vector is ($\mu_{z} - m_{z}) \hat{e_{z}}$, which measures the difference in redshift centroid and redshift median of the distribution of galaxies in the cluster. Using the elongation vector defined above, we further defined the criteria for pre-merging and post-merging systems as:

\begin{equation}
\begin{aligned}
 & r_{\rm{sep}} > \rm{max}\{r1 + r2,r1_{200} + r2_{200}\} \\
 & r_{\rm{sep}} \le  \rm{max}\{3(r1+r2),3(r1_{200}+r2_{200})\} \\
 & \vec{e_1} \cdot \vec{e_2} \le -0.966 
\end{aligned}
\end{equation}

Where $r_{\rm{sep}}$ is the haversine separation. The first criterion ensures that a merging system is not reclassified as a pre-merging or post-merging system. The second and third criteria form the definition based on separation and anti-parallel orientation of elongation vectors, respectively.

\section{Expected multi-band tracers for the identified interacting galaxy clusters}
\label{results}
The observations revealed the web-like distribution of galaxies at large scales \citep{LSS_SDSS}. 
These structures are understood to be initiated from the primordial density fluctuation and mediated and influenced by non-linear growth at late times due to rapid merger and accretion processes \citep{coswebevo}.
Naturally, the nodes of this cosmic web are expected to host several interacting clusters, and occasionally multiple nearby interacting nodes along the large filaments form wall-like great attractors called the superclusters \citep[e.g.,][]{firstsupclust,Bagchi_2017_Saraswati,supclrex1} While, galaxy distribution unravels the filamentary structure in the Universe, except in the nodes, the diffuse intergalactic plasma is yet to be conclusively detected \citep{Fang_WHIM_OBS}. Since cosmological simulations have already indicated the presence of an intergalactic medium in filaments in the form of the warm-hot intergalactic medium (WHIM) \citep{dave2001baryons}, observational detection became essential for its confirmation. Nevertheless, WHIM being not enough hot ($<<10^7$~K), to become detectable, needs dynamical activity in the medium for additional energization. Such as the activity in the inter-cluster spaces during pre and post-merger phases. We thus expect to observe a few specific characteristic diffuse emission features from these interacting systems linked to their dynamical activities.

Since galaxy clusters form at the nodes of the cosmic web, they are massive and hot, more so when they interact with each other \citep[e.g.,][]{merger_temp_ex}. The hot plasma of the ICM, through bremsstrahlung emits diffuse X-rays \citep{sarazin_x_ray}. However, at present, there are no highly sensitive all-sky X-ray surveys in the public domain. Thus, we resort to the ROSAT survey which was done in 1990-91 \citep{rosat_main}. For a few well-studied systems, detailed X-ray maps are available. In such cases, we used public data from NASA Chandra X-ray observatory \citep{Chandra_Main}, or XMM-Newton mission archives \citep{xmm_main}.

Furthermore, it is a well-established fact that the diffuse radio emission seen in galaxy clusters is an outcome of the hierarchical merging process \citep{Ferreti_2008,van_Weeren_2019}. This diffuse emission is of synchrotron origin and arises due to accelerating relativistic electrons as they move through cluster-scale magnetic fields. The cause of the acceleration of the particles can be connected to merger activity \citep{vanWeeren_acc_mech,Cassano_2010}, i.e. a shock front, or scattering across regions of magneto-hydrodynamic turbulence. The extended diffuse radio emission is sometimes visible as a bridge connecting two massive clusters. \citet{A399_401_Deep_Study} have recently reported that the acceleration mechanism for such emission is similar, where particles scatter from magnetic inhomogeneities, sometimes called the second-order Fermi acceleration.
Another significant effect of massive interacting clusters can be found on the member galaxies of these structures. As the member galaxies get attracted towards the massive core the clusters or super-clusters, due to their rapid motion through the hot ICM, possibly during the interacting phase, they experience ram pressure. The magnitude of this ram pressure can be enough to strip gravitationally bound gas from these galaxies, often rendering them with a long tail of diffuse radio emission \citep{Lal_2022_RPS,venturi_RPS}. Moreover, this ram pressure stripping (RPS) is thought to have a profound impact on star formation and galaxy evolution as a whole. Since these features of cluster galaxies are the imprints of large mass distribution, may act as the tracers of interacting galaxy clusters.

The ideal instrument to observe this diffuse emission must have a high sensitivity to low surface brightness, as well as a sufficiently small beam size to distinguish point sources \citep{Feretti_2012}. Since the diffuse sources are expected to have a steep spectrum \citep[e.g.,][]{steep_spectra}, low-frequency observations are suited well. Keeping all of this in view, we select LOw-Frequency ARray (LOFAR) Two-metre Sky Survey (LoTSS) \citep{lotss2} images as the primary source to check for diffuse radio emission around the identified clusters.

\newcommand{\summarytable}{\multirow{2}{*}{M61*}&146.421&54.645&0.0465&3.11877\\&146.796&54.429&0.0462&4.17813\\\hline
\multirow{2}{*}{M137*}&227.502&33.464&0.1128&34.26554\\&227.548&33.515&0.1127&55.27979\\\hline
\multirow{2}{*}{PPM93*}&163.580&54.849&0.0718&5.37769\\&163.652&54.862&0.0727&3.91902\\\hline
\multirow{5}{*}{PPM95*}&167.338&41.559&0.0763&9.13971\\&167.850&39.602&0.076&2.15536\\&167.928&40.854&0.0755&34.25835\\&168.013&40.538&0.0752&1.0386\\&168.392&40.284&0.0753&5.30148\\\hline
\multirow{2}{*}{PPM144*}&215.274&48.309&0.0742&2.0842\\&215.552&48.542&0.0726&17.90491\\\hline
\multirow{2}{*}{PPM164*}&225.138&47.274&0.0874&9.01206\\&225.506&47.291&0.0888&0.36345\\\hline
\multirow{2}{*}{PPM169*}&230.222&30.539&0.0779&1.41221\\&230.412&30.677&0.0788&2.19787\\\hline

}
\begin{table}
\centering
\caption{Summary of interacting systems which show diffuse emission as tracers described in this work. [The full table for merging and pre/post-merging can be found as Supplementary Table~1\&2]}
\label{tab:summaryTable}
\begin{tabular}{|l|cccc|}
\hline
System & R.A. (Deg) & Dec. (Deg) & z & M ($\times 10^{13} \rm{M_\odot} $) \\
\hline
\multirow{2}{*}{M61*}&146.421&54.645&0.0465&3.11877\\&146.796&54.429&0.0462&4.17813\\\hline
\multirow{2}{*}{M137*}&227.502&33.464&0.1128&34.26554\\&227.548&33.515&0.1127&55.27979\\\hline
\multirow{2}{*}{PPM93*}&163.580&54.849&0.0718&5.37769\\&163.652&54.862&0.0727&3.91902\\\hline
\multirow{5}{*}{PPM95*}&167.338&41.559&0.0763&9.13971\\&167.850&39.602&0.076&2.15536\\&167.928&40.854&0.0755&34.25835\\&168.013&40.538&0.0752&1.0386\\&168.392&40.284&0.0753&5.30148\\\hline
\multirow{2}{*}{PPM144*}&215.274&48.309&0.0742&2.0842\\&215.552&48.542&0.0726&17.90491\\\hline
\multirow{2}{*}{PPM164*}&225.138&47.274&0.0874&9.01206\\&225.506&47.291&0.0888&0.36345\\\hline
\multirow{2}{*}{PPM169*}&230.222&30.539&0.0779&1.41221\\&230.412&30.677&0.0788&2.19787\\\hline

\end{tabular}
\end{table}

 Here, in Table~\ref{tab:summaryTable} we listed out and briefly mentioned about only the interacting systems that are detected with diffuse radio emission in LoTSS-2 images. However, a comprehensive list of all interacting systems detected by our algorithm is presented as a supplementary table attached to this paper. The detailed properties of each of the objects mentioned in Table~\ref{tab:summaryTable} are presented in further texts. In each of the image panels describing these sources, we plotted the SDSS galaxies in that particular region (e.g., Fig.~\ref{fig:Abell 1190}a). To construct these images, an appropriate angular and redshift limit has been chosen adaptively so as to enhance and visualize the galaxy filaments surrounding the interacting system. The FoF-identified clusters are slightly exaggerated by choosing larger points and different colours to represent the members of each galaxy group. The dashed circles are shown at different overdensities, the inner circle represent an estimate of $\rm{r_{200}}$, while, the outer circle shows the estimate of $3 \times \rm{r_{200}}$. An additional redshift-based transparency was also added to background galaxies to distinctly project the interacting systems. In the remaining parts of this section, we present only a subset few systems which exhibit characteristic multi-band signatures and interesting dynamical scenarios; while details of other systems exhibiting diffuse emission can be found in Appendix \ref{Remaining_Systems}.

\subsection{Systems detected as a part of supercluster}
Although there are several accounts of superclusters being identified in optical wavelengths \citep{supclrex1,supclrex2,supclrex3}, there is very limited literature available for radio and X-ray wavelengths. When the clusters undergo merging and accretion, the ICM may get spewed across the filament which forms during merger \citep{Paul_2011_ICM_Shocks}. Thus, the filamentary material connecting them can also exhibit traces of diffuse radio and X-ray emission \citep{Radio_filament_example}, although it tends to be extremely faint, and currently lies beyond the sensitivity of most of the surveys. Recently, \citet{A399_401_Deep_Study} reported a radio bridge between clusters Abell 399 and Abell 401. Such filamentary connections have also been observed in X-ray wavelengths \citep{X_ray_filament_ex}. \citet{Planck_Filaments} is a first and remarkable example of observation of filaments using the SZ effect as well. 

Depending on the choice of linking parameter, the algorithm may be able to detect parts of a much larger interacting system. In this section, we report systems which are detected as part of optical superclusters which also show diffuse radio and X-ray emission nearby cluster centres. The locations of these systems may prove to be interesting regions to spot radio filaments when even more sensitive all-sky surveys are carried out in future.

\subsubsection{Abell 1190 - Abell 1203 (PPM95)}

\begin{figure*}
\centering
	 	\begin{subfigure}{0.485\textwidth}
	 	\centering
                    \includegraphics[width=\linewidth]{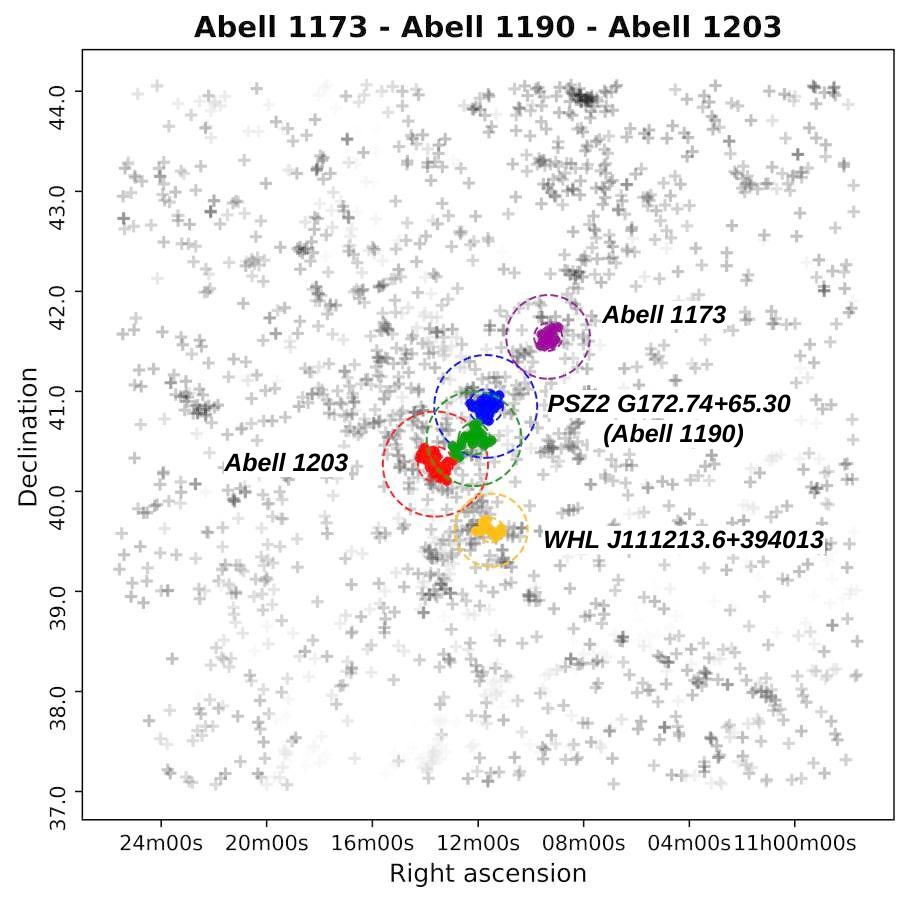}
                 \caption{}
        \end{subfigure}
       \begin{subfigure}{0.485\textwidth}
       \centering
                \includegraphics[width=\linewidth]{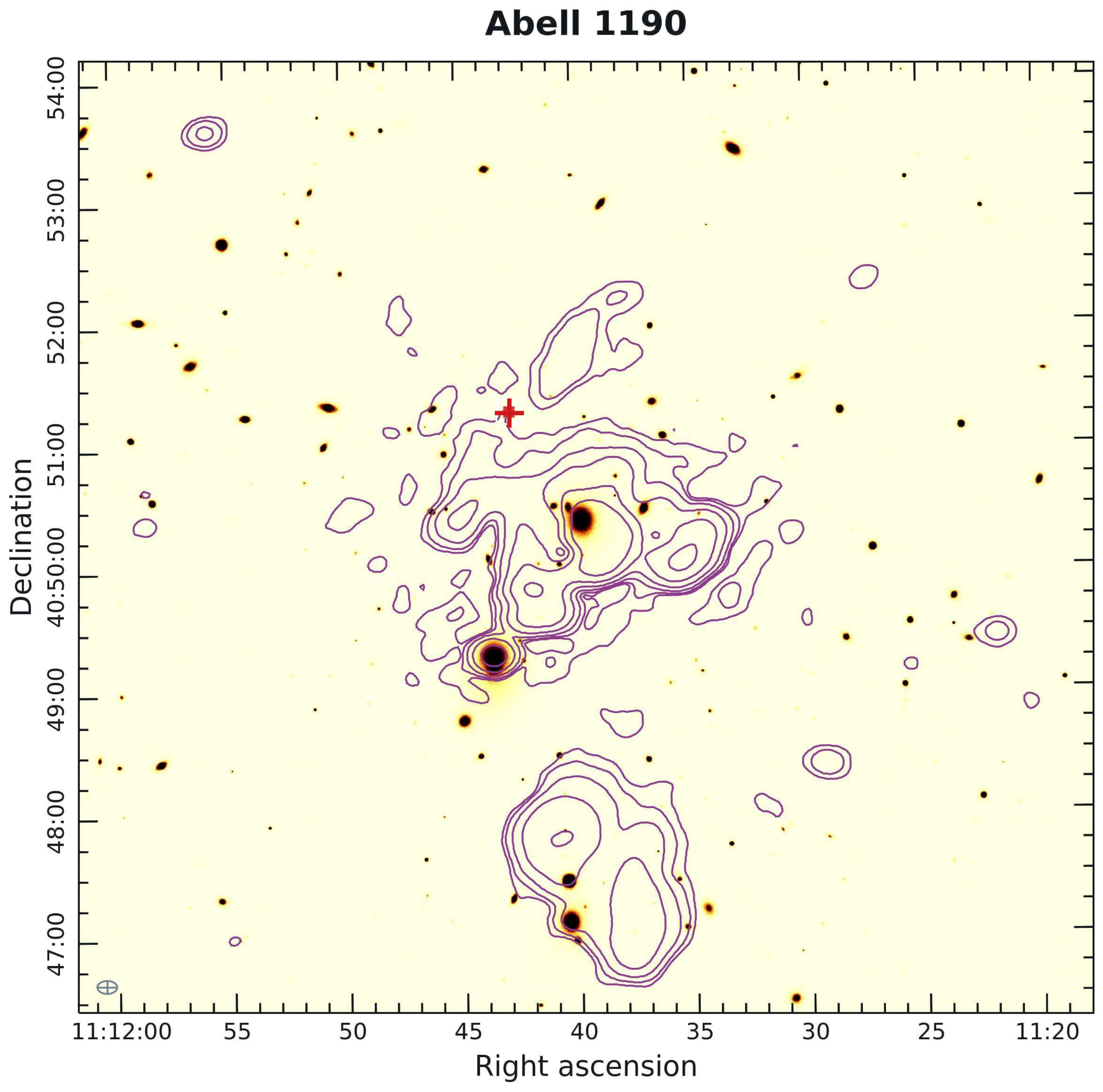}
                 \caption{}
        \end{subfigure}
       	\begin{subfigure}{0.485\textwidth}
       	\centering
                 \includegraphics[width=\linewidth]{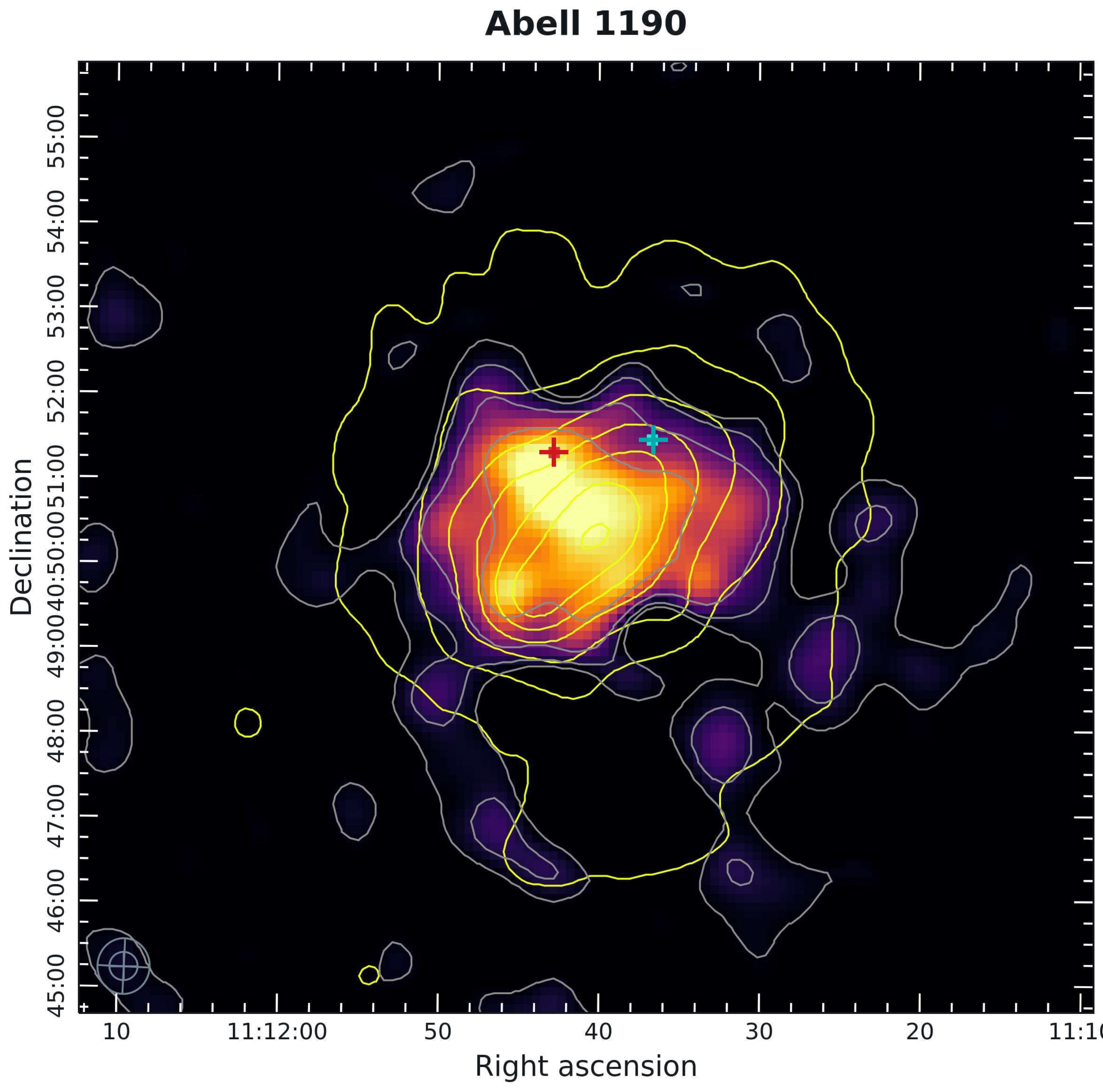}
                 \caption{}
        \end{subfigure}
       \begin{subfigure}{0.485\textwidth}
       \centering
                \includegraphics[width=\linewidth]{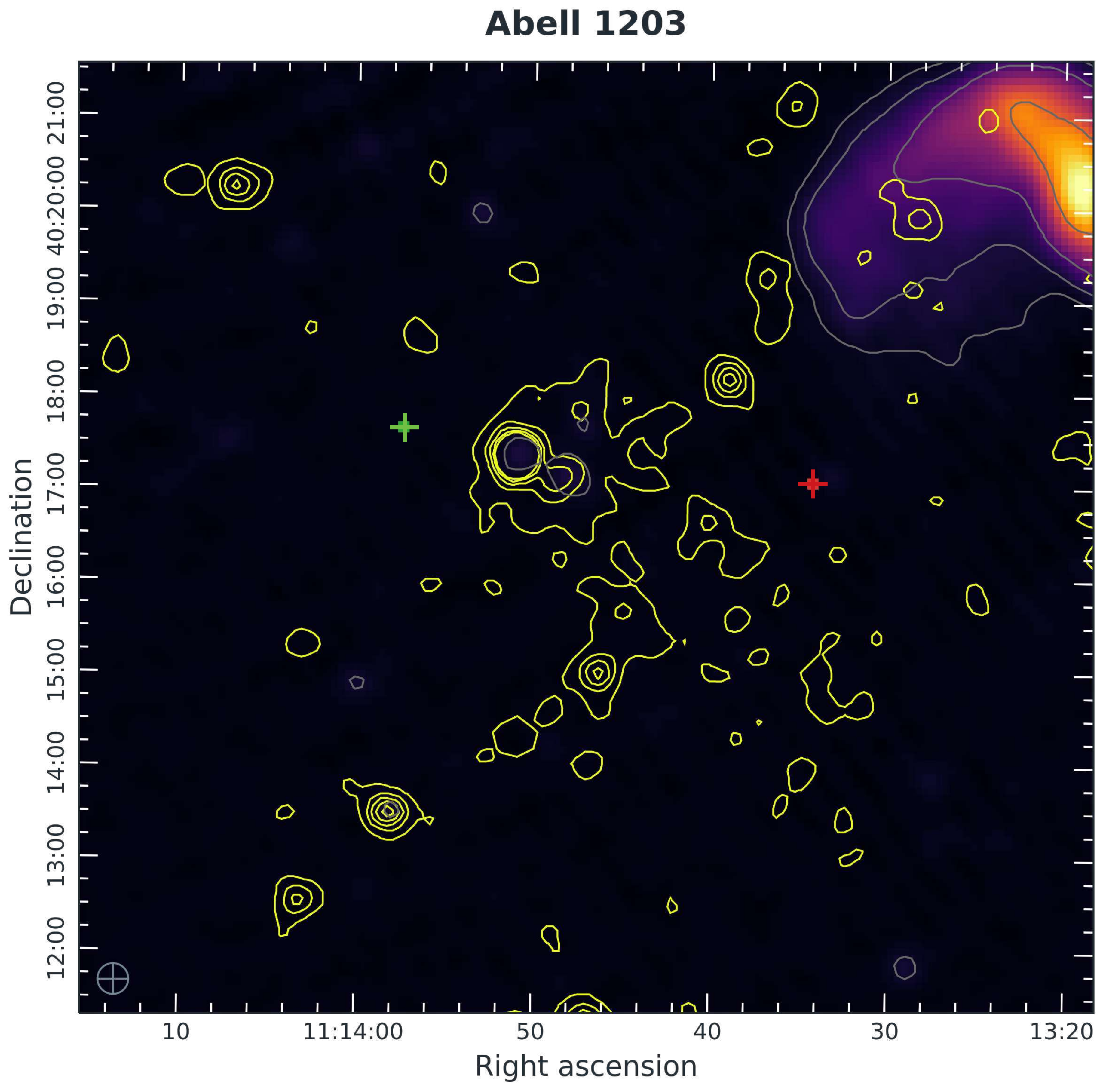}
                 \caption{}
        \end{subfigure}

        \caption{\textbf{Abell~1190 - Abell 1203}: Panel~\textbf{(a)}: Plot of SDSS~DR~17 galaxies. The dotted circles represent the bounds of $r_{200}$ and $3r_{200}$ respectively. Panel \textbf{(b)}: Pan-STARRS i-band optical image overlaid by LoTSS-2 low-resolution image contours [at 3, 9, 27, 81, 243] of $\sigma$ where $\sigma_{\rm{rms}} = 150 \rm{\mu Jy/beam^{-1}}$.  Panel~\textbf{(c)}: LoTSS-2 point source subtracted image with same contours (grey) overlaid with XMM-Newton (39~ks exposure data) contours (yellow) for Abell~1190. Panel~\textbf{(d)} Abell~1203 field LoTSS-2 low resolution image with contours (grey)[3,9,27,81,243] of $\sigma$ where $\sigma_{\rm{rms}} = 250 \rm{\mu Jy/beam^{-1}}$ overlaid with XMM-Newton contours (yellow; 15.9~ks exposure data). }\label{fig:Abell 1190}
       
\end{figure*}

\begin{figure}
    \centering
     \includegraphics[width=\linewidth]{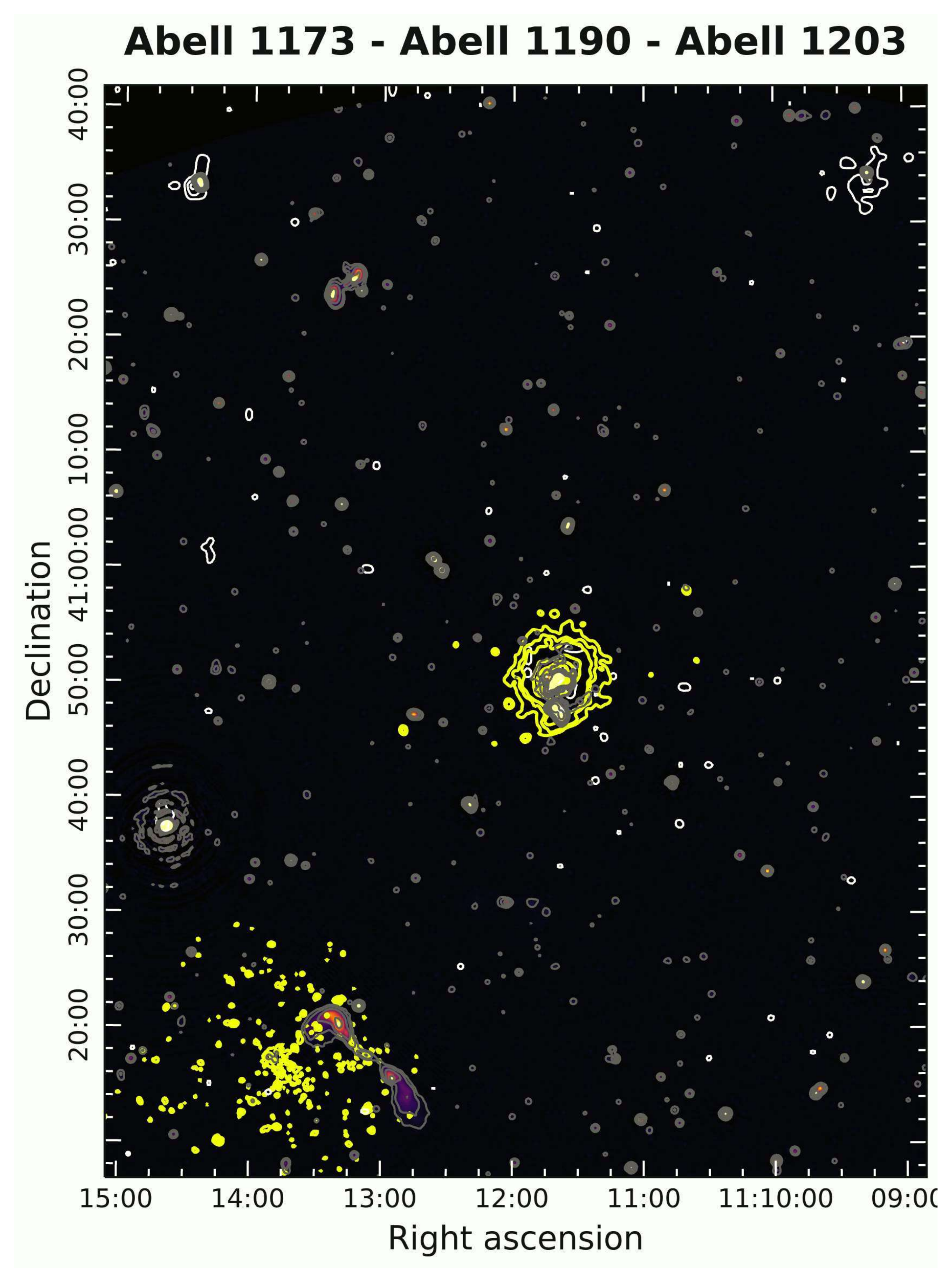}
    \caption{From top to bottom: Abell~1173, Abell~1190 and Abell~1203. XMM X-ray contours (Yellow) along with ROSAT (white) contours are overlaid onto LoTSS-2 low-resolution image with contour levels (grey) at [3,9,27,81,243] of $\sigma$ where $\sigma_{\rm{rms}} = 250 \rm{\mu Jy/beam^{-1}}$ .}
    \label{fig:Abell 1190_Big}
\end{figure}

Figure~\ref{fig:Abell 1190}a shows the plot of SDSS~DR~17 galaxies in nearby fields where a system of five clusters can be seen all together. Of these, the northernmost cluster corresponds to Abell~1173, the central clusters correspond to Abell~1203/Abell~1190 and the southernmost cluster lies close to WHL~J111213.6+394013. These systems are located at the mean redshift of $z = 0.076$. The identified systems are seen to fall on a filamentary extension spanning almost 10 Mpc. Previously, Abell~1190 and 1203 were detected as a part of the SCL~38 supercluster by \citet{scl38supclstr} in optical wavelengths based on data from SDSS~DR~7, which is consistent with the present work. Here, Abell~1190 can be identified with diffuse emission both in X-ray \citep[e.g,][]{clust_her_project} and Radio wavelengths \citep{PlanckLotssdr2}. The extension of the emission of the source at the centre of Abell~1190 is around 400~kpc (3$\sigma$). Abell~1203 does not show any distinct emission in this image, however, detailed \textit{Chandra} data indeed shows diffuse X-ray emission originating from the ICM. The detection of this radio emission could have been hampered by the presence of a bright radio galaxy in the foreground. The southernmost cluster is devoid of any emission at the identified centre. In Figure~\ref{fig:Abell 1190_Big}, XMM X-ray contours are overlaid onto a LoTSS-2 low-resolution image. Abell~1203 shows irregular morphology possibly indicating that the cluster may be in a dynamically active state. \citet{PSZ2} report the mass of Abell~1190 as $M_{500}$ of $2.45\times 10^{14} \rm{M_\odot}$. Our estimated mass of Abell~1190 is around $7.7\times 10^{13} \rm{M_\odot}$. In X-rays, this cluster is found to be moderately luminous ($L_x = 1.048 \times 10^{44} \rm{erg  s^{-1}}$) having a temperature of $4.061^{+0.344}_{-0.336}\, \rm{keV}$ \citep{xraygclustsample}.        

\subsubsection{Abell 2061 (PPM169)}
\begin{figure*}
\centering
        
	 	\begin{subfigure}{0.49\textwidth}
	 	\centering
                \includegraphics[width=\linewidth]{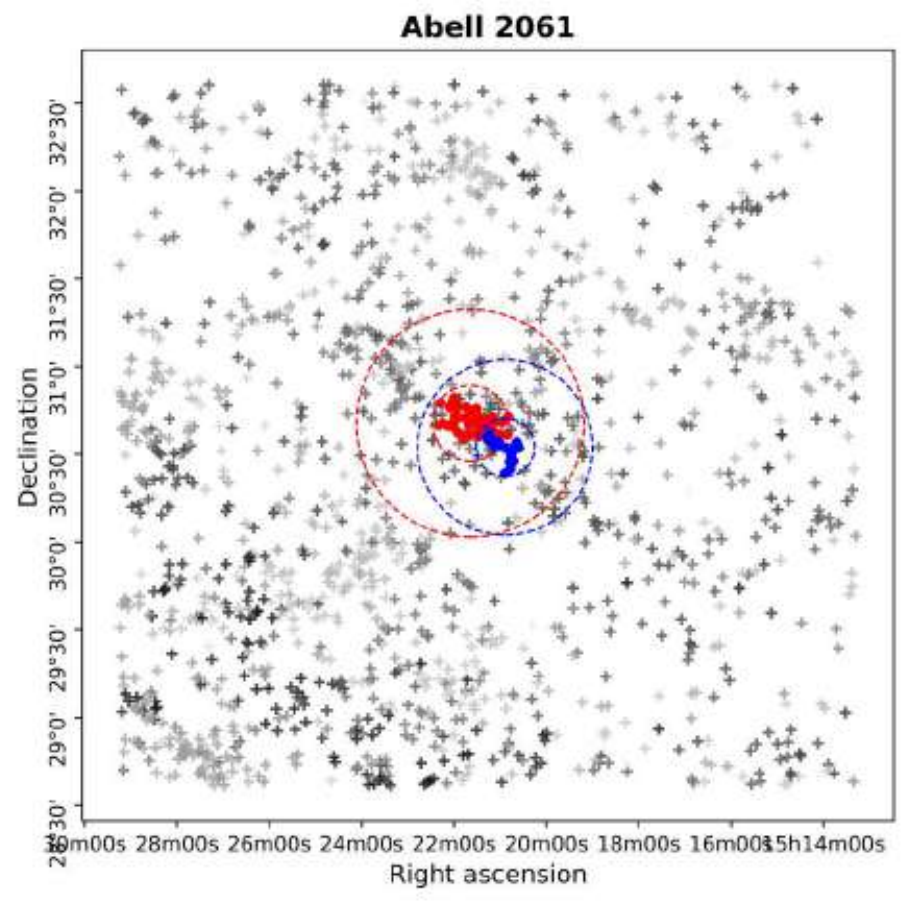}
                \caption{}
        \end{subfigure}
       \begin{subfigure}{0.49\textwidth}
       \centering
                \includegraphics[width=\linewidth]{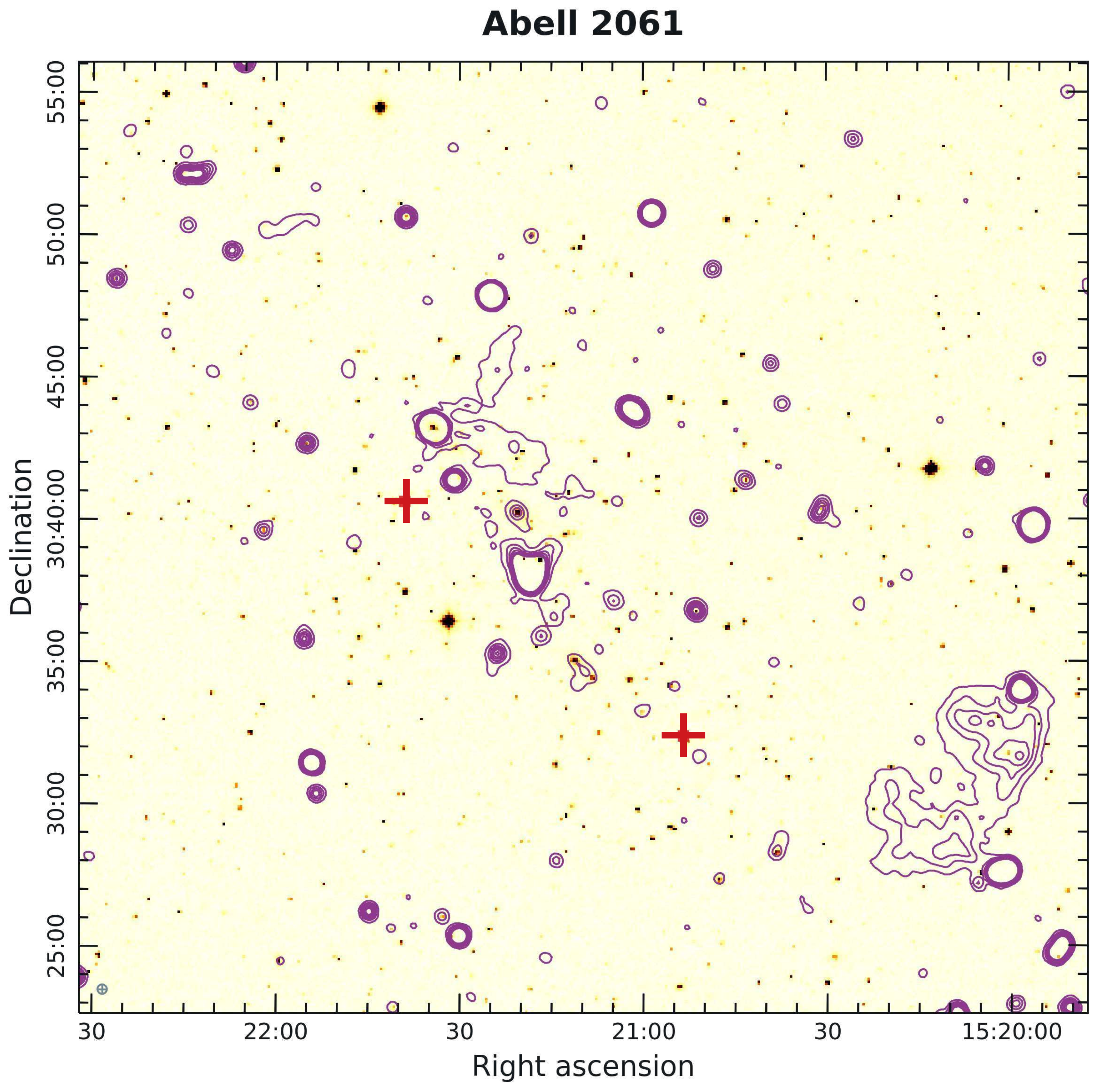}
                \caption{}
        \end{subfigure}
        \begin{subfigure}{0.49\textwidth}
        	\centering
                \includegraphics[width=\linewidth]{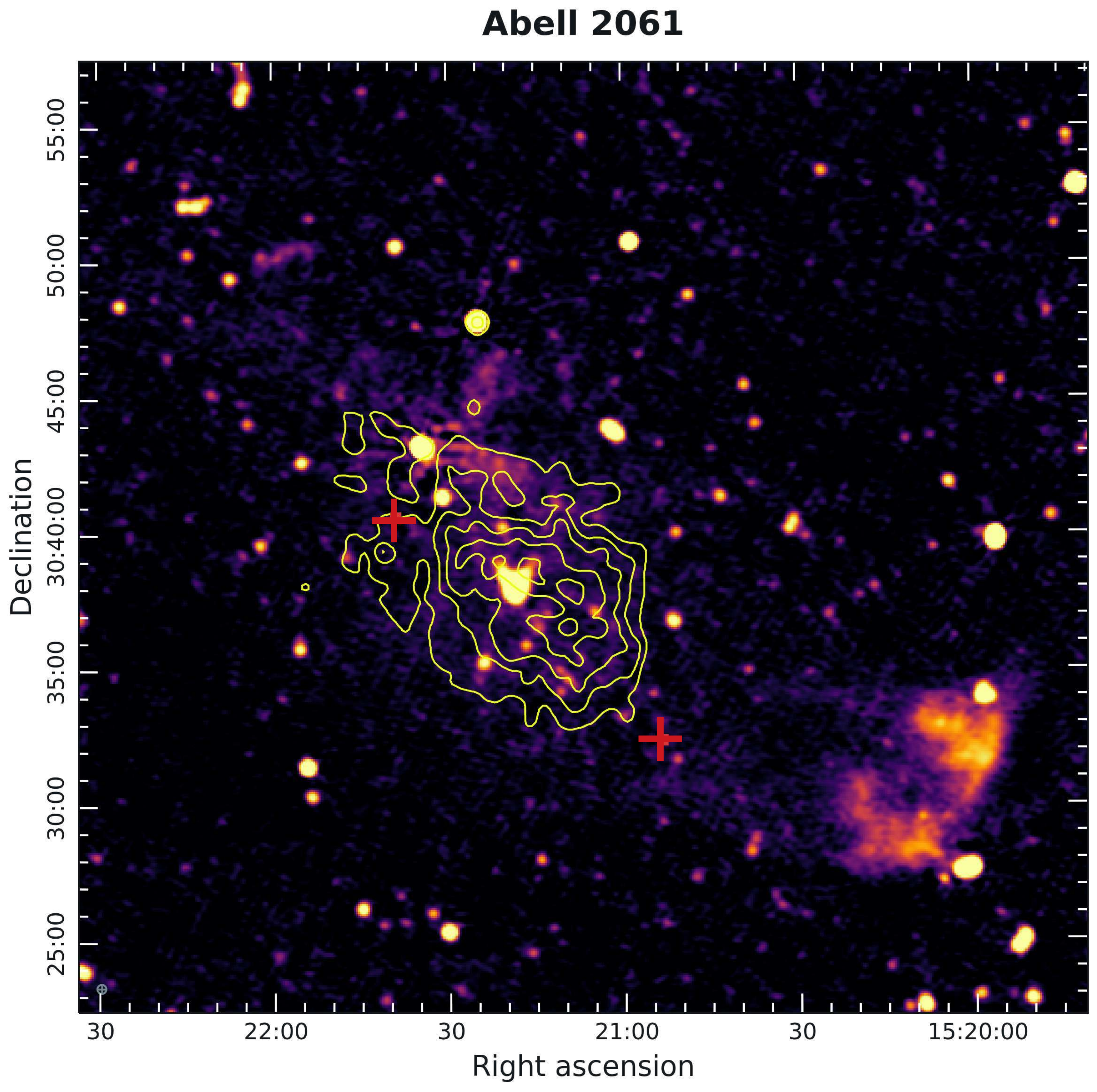}
                \caption{}
        \end{subfigure}
        \begin{subfigure}{0.49\textwidth}
        	\centering
                \includegraphics[width=\linewidth]{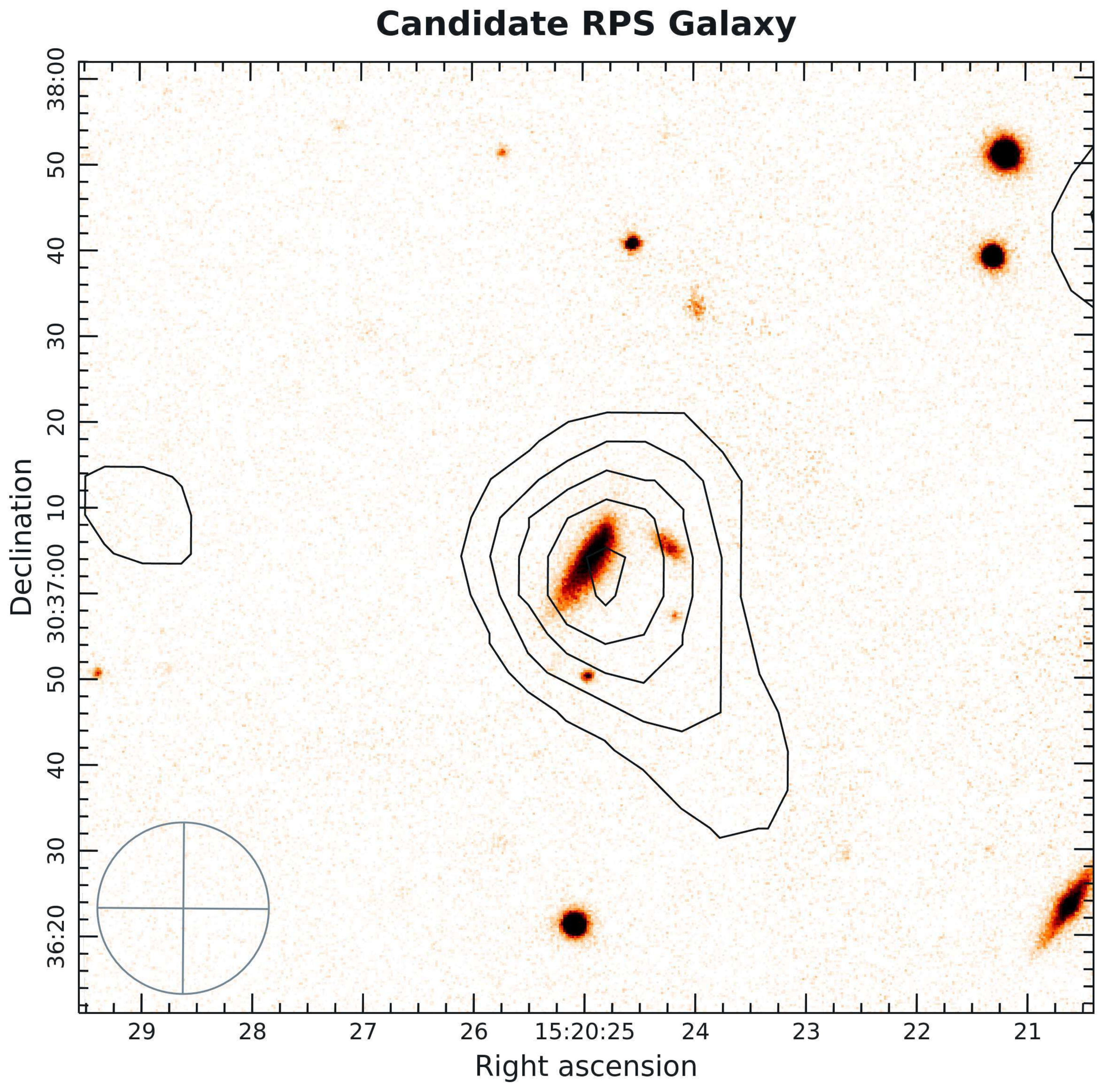}
                \caption{}
        \end{subfigure}
       
       \caption{ \textbf{Abell 2061} Panel~\textbf{(a)}: A plot of SDSS galaxies. Panel~\textbf{(b)}: DSS r band optical image overlaid with LoTSS-2 low resolution image contours at [3,6,9,12,15,18] of $\sigma$ where $\sigma_{\rm{rms}} = 200 ~\rm{\mu Jy/beam}$. Panel~\textbf{(c)}: Chandra X-ray contours (yellow) with exposure of 33~ks overlaid on LoTSS-2 point source subtracted image with contours (grey) with the same levels as in Panel Panel~\textbf{(b)}. Panel~\textbf{(d)}: A candidate RPS galaxy in the field of A2061; Pan-STARRS i-band optical image overlaid with contours (grey) at [3,6,9,12] of local $\sigma_{\rm{rms}}$ where $\sigma_{\rm{rms}} = 80~\rm{\mu Jy/beam}$. }\label{fig:PPM68}

\end{figure*}

The corona borealis supercluster is one of the most prominent superclusters in the northern sky \citep{corborsupclstr}. In literature, it is noted that more than five Abell clusters are in the process of infall and merger in this system. Our algorithm has been able to detect a merging substructure in Abell 2061, which is a part of the larger Corona Borealis supercluster. This interacting system lies at a mean redshift of z = 0.078. Fig. \ref{fig:PPM68}\textbf{a} displays the SDSS galaxy distribution in the field nearby the clusters. A very prominent filament can be seen that emerges from NE-SW which appears to host this interacting system. \citet{PSZ2} report a mass of $M_{500}=3.59 \times 10^{14} ~ \rm{M_\odot}$. \citet{StrubleVelDisp} have shown that this cluster exhibits a velocity dispersion of $ 1020 ~ \rm{km s^{-1}} $ which is high for a cluster of this mass, and thus indicative of dynamical activity in the ICM. In Fig.~\ref{fig:PPM68}\textbf{c}, it can be seen that the overall X-ray morphology is irregular, and elongated along the galaxy filament seen in Fig.~\ref{fig:PPM68}\textbf{a}. \citet{xraygclustsample} have reported the X-ray temperature of $4.668^{+0.141}_{-0.159}~\rm{keV}$ and a X-ray luminosity of $L_x = 1.873 \times 10^{44} ~ \rm{erg}^{-1}$. This suggests that the cluster is moderately hot and luminous.

\citet{A2061ppm} reported a possibility of radio filament between Abell 2061 and Abell 2067 which lies to the NE along the same filament. In the LoTSS 2 map (see Fig. \ref{fig:PPM68} \textbf{c}), we found an elongated diffuse emission starting from A2061 and traces of diffuse radio emission along the filament. A bow shock feature can also be seen ahead of cluster A2061 towards the south-west. The angular extent for radio emission from relic is $\sim 475~\rm{kpc}$ with radio flux density $342 \pm 34$ mJy. For the halo, the angular extension is about $1020$ kpc with the diffuse radio flux density of $407 \pm 41$~mJy. For this central diffuse emission, \citet{A2061ppm} reported an integrated spectrum of $\alpha^{1.4}_{0.3} = 1.8 \pm{0.3}$ using the Green Bank Telescope, making this an ultra-steep spectrum source. Owing to these radio features, this system becomes a potential site for observing inter-cluster radio bridges and WHIM filament using even sensitive interferometers like SKA in future.

\subsection{Planck Clusters in LoTSS-2 Region}
The Planck SZ2 catalogue contains more than 1200 confirmed galaxy clusters found by the virtue of SZ effect. SZ-identified galaxy clusters are considered to be reliable as they are independent of redshift \citep{SZ_detailed_review}, and therefore not hampered by any selection or projection effects. The second data release of the Planck SZ catalogue has reliability close to or upwards of 90\%. \citep{PSZ2}

In this section, we describe the list of interacting clusters with at least one confirmed PSZ2  cluster member which lies in the field of LoTSS-2. These clusters are independently identified with diffuse radio emission in the work by \citet{PlanckLotssdr2}, however, the dynamical states of these sources and their multi-wave-band features have not been explored well. The other additional properties relevant to the interaction scenario are also presented here in the further sections.

\subsubsection{PSZ2~G053.53+59.52 (Abell~2034/M137)}
\begin{figure*}
\centering
        
	 	\begin{subfigure}{0.50\textwidth}
	 	\centering
                \includegraphics[width=\linewidth]{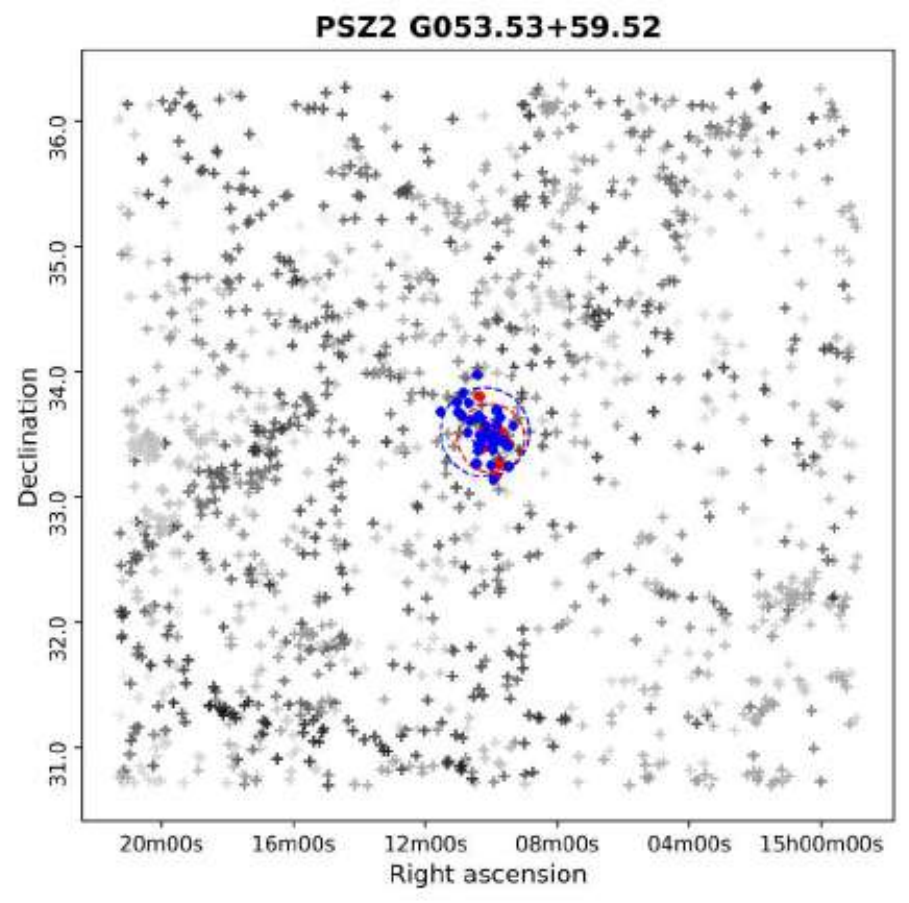}
                \caption{}
        \end{subfigure}
       \begin{subfigure}{0.49\textwidth}
       \centering
                \includegraphics[width=\linewidth]{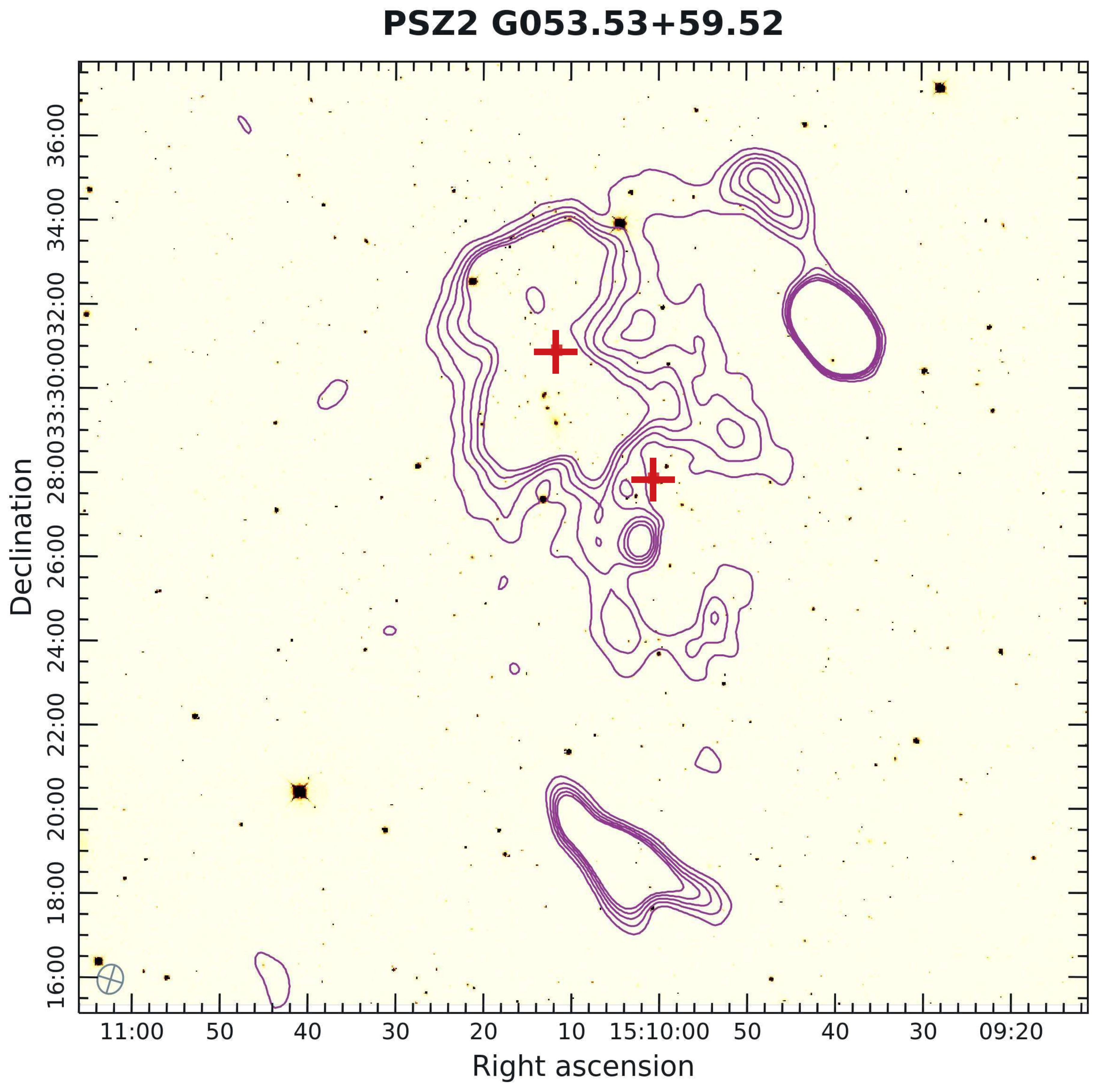}
                \caption{}
        \end{subfigure}
        \begin{subfigure}{0.49\textwidth}
        	\centering
                \includegraphics[width=\linewidth]{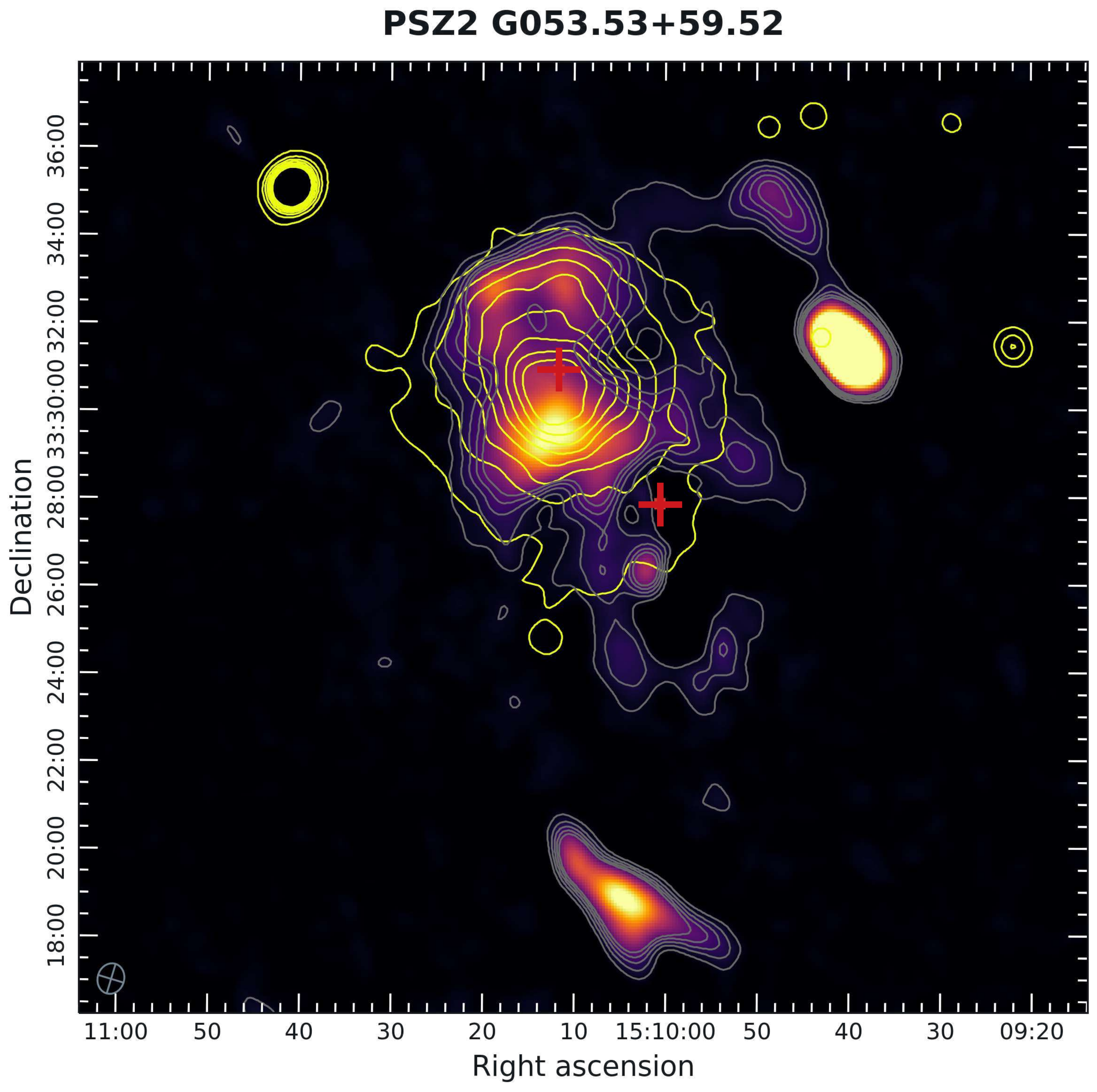}
                \caption{}
        \end{subfigure}
        \begin{subfigure}{0.49\textwidth}
        	\centering
                \includegraphics[width=\linewidth]{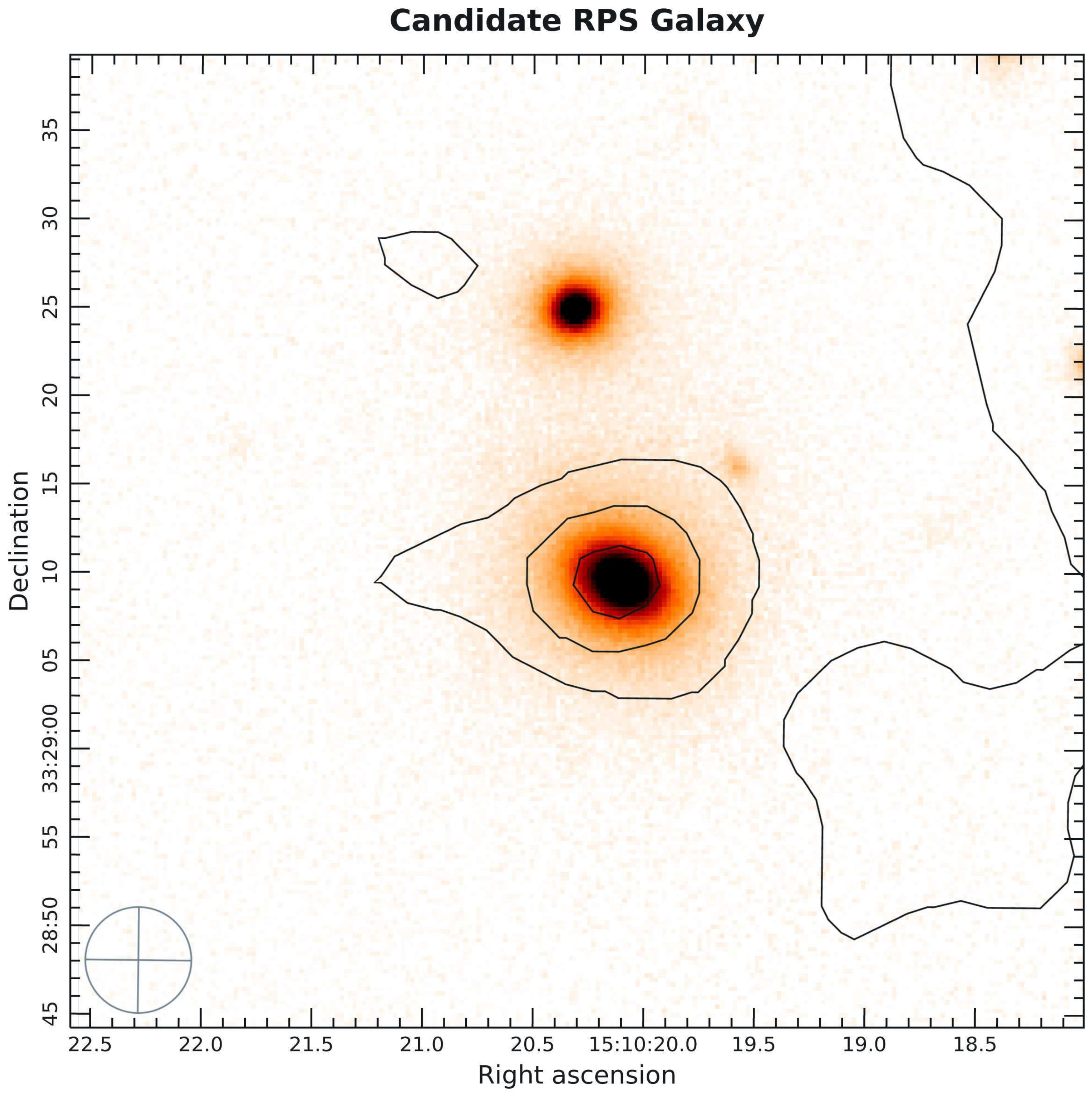}
                \caption{}
        \end{subfigure}
       
        \caption{ \textbf{PSZ2~G053.53+59.52} Panel~\textbf{(a)}: A plot of SDSS galaxies. Panel~\textbf{(b)}: DSS optical image overlaid with LoTSS-2 low resolution image contours at [3,6,12,24,48..] of $\sigma$ where $\sigma_{\rm{rms}} = 100 ~\rm{\mu Jy/beam}$. Panel~\textbf{(c)}: XMM-Newton X-ray contours (yellow) with exposure of 13~ks overlaid on LoTSS-2 point source subtracted image with contours (grey) at [3,6,9,12,15] of local $\sigma_{\rm{rms}}$ where $\sigma_{\rm{rms}} = 300 ~\rm{\mu Jy/beam}$ Panel~\textbf{(d)}: A candidate RPS galaxy in the field of PSZ2~G053.53+59.52; Pan-STARRS i-band optical image overlaid with contours (grey) at [3,6,9,12] of local $\sigma_{\rm{rms}}$ where $\sigma_{\rm{rms}} = 95~\rm{\mu Jy/beam}$. }\label{fig:Abell 2034}
       
\end{figure*}

Fig.~\ref{fig:Abell 2034}a shows a plot of SDSS galaxies in the field of Abell~2034 clearly highlighting the identified merging system in colour. This system is well known and studied in great detail by many authors reporting many multi-band signatures of interaction. This system is located at the mean redshift of $z = 0.113$. A prominent filament can be seen just northwards of the cluster spanning from west to east. \citet{PSZ2} have reported the cluster mass of $M_{500}$ of $5.85\times 10^{14} \rm{M_\odot}$,while we estimate an upper limit on mass as $M\sim 8\times 10^{14} \rm{M_\odot}$ . From the weak-lensing maps, the bimodal structure (north-south) of this cluster can clearly be identified, along with additional mass peaks revealing a highly irregular structure \citep{OkabeA2034}. A comparatively recent study by \citet{OlivieraA2034} related to the numerical density distribution of the red sequence galaxies, reveals a third structure towards the west, along with historically identified north and south substructures. This relatively new insight is consistent with the centres identified in this work. Early shallower X-ray observations of this cluster suggested the presence of a cold front, however, \citet{A2034MergerShock} proved with deeper observations that the signature was a shock front with a Mach number $M = 1.59^{+0.06}_{-0.07}$. Critically, the authors also confirmed the existence of a substructure just ahead of the shock which is a remnant of the merging substructure. \citet{xraygclustsample} have reported the cluster to be hot, with an X-ray temperature of $7.8^{+0.1}_{-0.1}~\rm{keV}$, using the data from NASA Chandra observatory; while the X-ray luminosity found to be $L_x = 3.5 \times 10^{44}~\rm{erg\,s^{-1}}$. 

\citet{GiovanniniA2034} and \citet{A2034Merger} were a few preliminary works that explored diffuse radio emission in this cluster. The interpretation of the 144~MHz radio image shows the presence of several large-scale diffuse radio sources. The region near the centres shows diffuse radio emission and a structure resembling the radio filaments appears to be connecting the two centres. A distinct relic can also be clearly seen in the southern part of the image. The extent of the central diffuse radio emission source is $\sim1.4$~Mpc with a total flux density of $2.0\pm0.2$~Jy. The extent of the nearby southern relic is $\sim600$~kpc with a total flux density of $104\pm10$~mJy.  Previously, \citet{A2034Detailed} have performed a detailed study on all the prominent diffuse sources present in the nearby field of this system. In addition to these, the detection of a candidate RPS source (see Fig.~\ref{fig:Abell 2034}d) in the LoTSS~2 high-resolution map provides a clue to the presence of strong dynamical activities in the system. The stretched-out radio tail in the galaxy was found to point away from the cluster centre possibly indicating strong ram pressure striping of its gas arising from its motion through the hot ICM.

\subsubsection{PSZ2~G080.16+57.65 (Abell~2018/PPM164)}
\begin{figure*}
\centering
        
	 	\begin{subfigure}{0.495\textwidth}
	 	\centering
                \includegraphics[width=\linewidth]{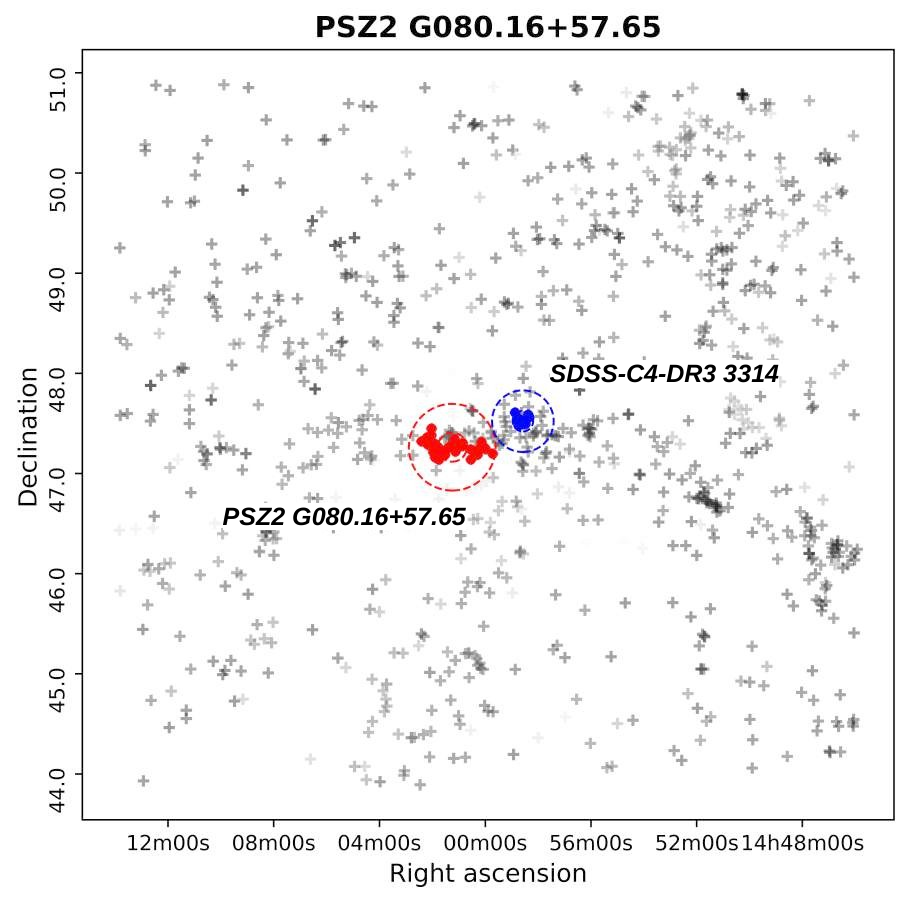}
                \caption{}
        \end{subfigure}
       \begin{subfigure}{0.485\textwidth}
       \centering
                \includegraphics[width=\linewidth]{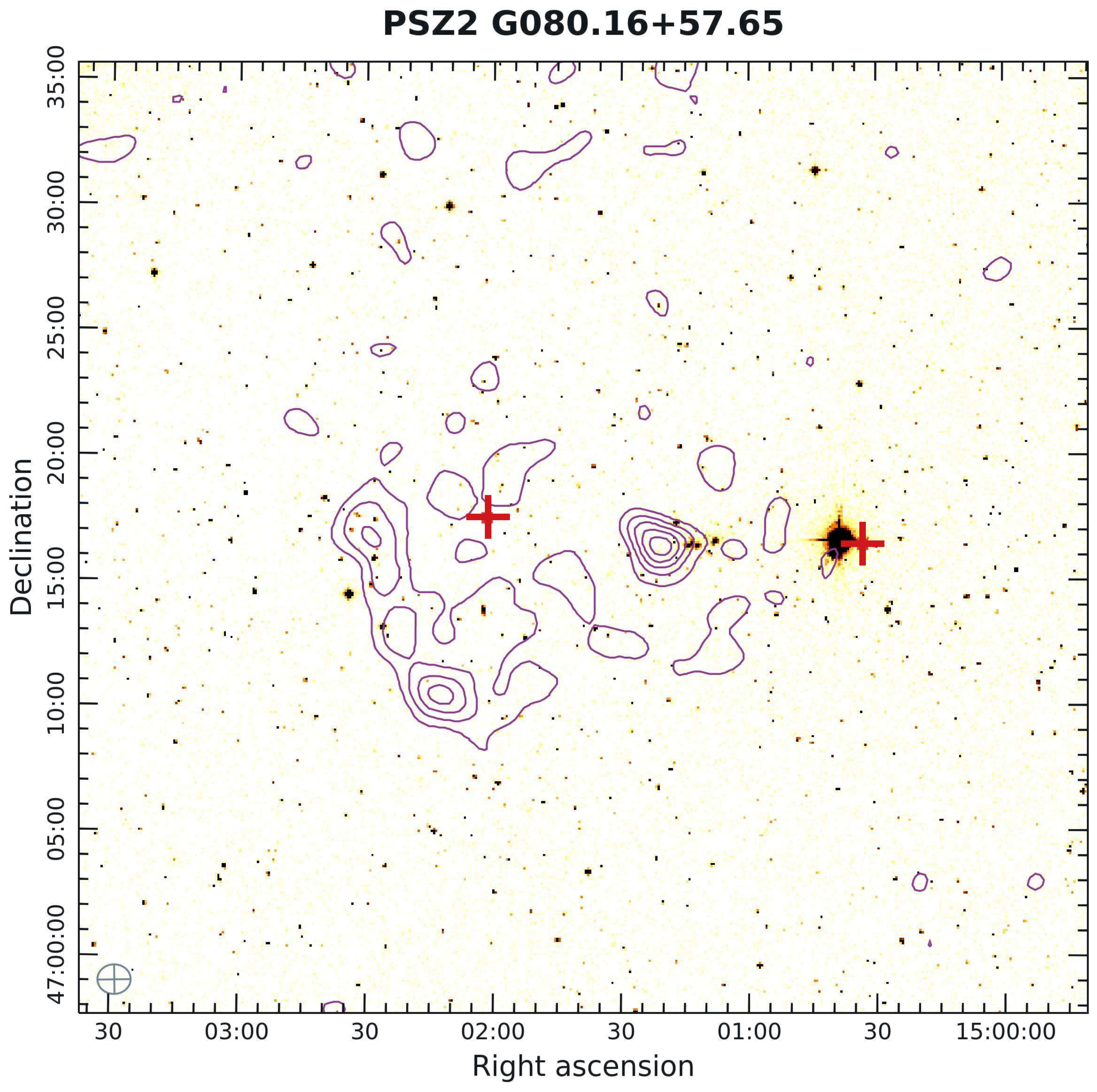}
                \caption{}
        \end{subfigure}
    	\begin{subfigure}{0.485\textwidth}
        	\centering
                \includegraphics[width=\linewidth]{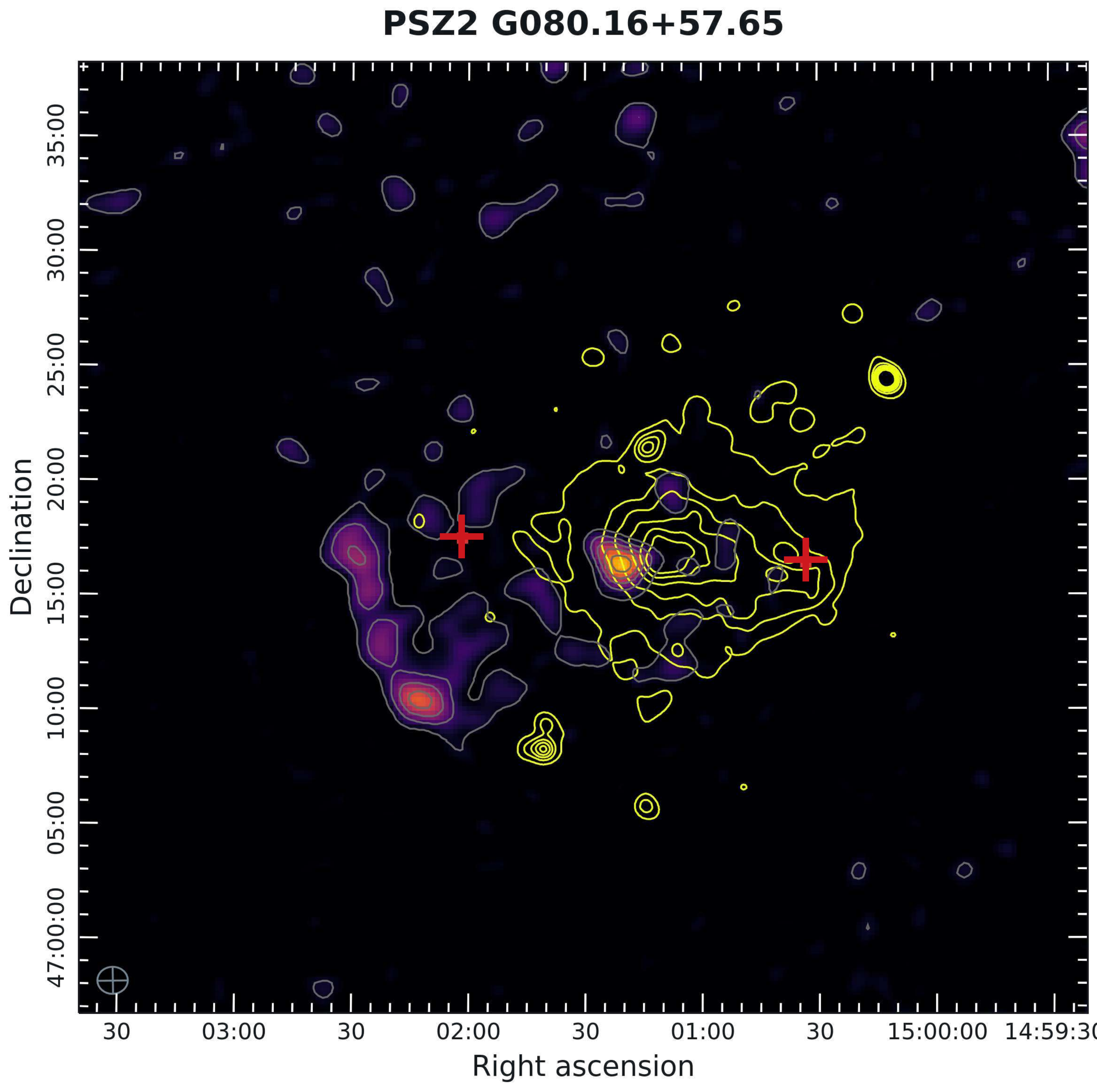}
                \caption{}
        \end{subfigure}
        \begin{subfigure}{0.485\textwidth}
        	\centering
                \includegraphics[width=\linewidth]{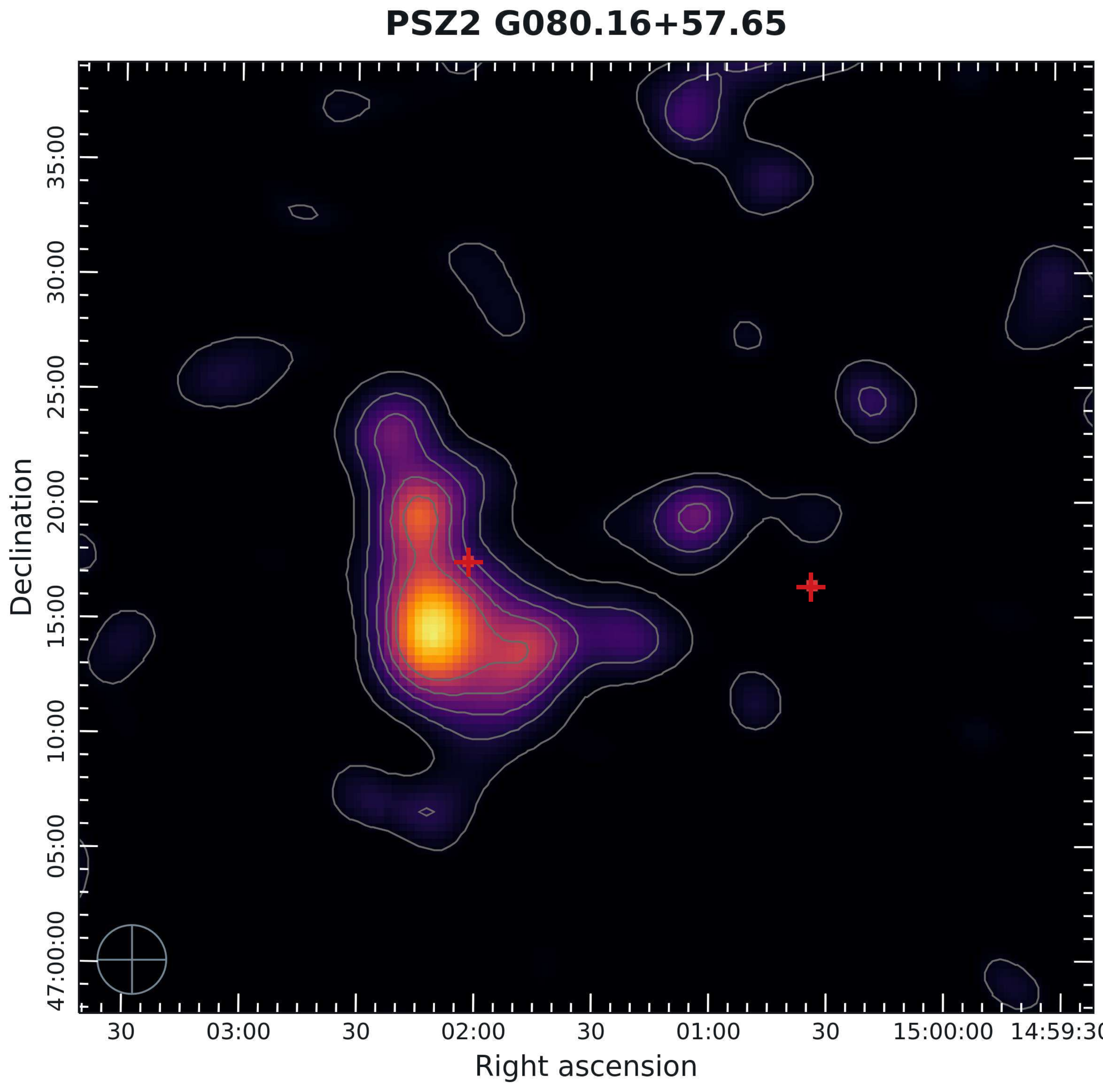}
                \caption{}
        \end{subfigure}

        \caption{ \textbf{PSZ2~G080.16+57.65}  \textbf{(a)}: A plot of SDSS galaxies. Panel~\textbf{(b)}: DSS red band optical image overlaid with LoTSS-2 low-resolution image contours at [3,6,9,12,15,18] of $\sigma$ where $\sigma_{\rm{rms}} = 150 ~\rm{\mu Jy/beam}$. Panel~\textbf{(c)}: XMM-Newton X-ray contours (yellow) with exposure of 35.50 ks overlaid on LoTSS-2 point source subtracted image with contour levels (grey) at [3,6,9,12,15] of $\sigma$ where $\sigma_{\rm{rms}} = 390 ~\rm{\mu Jy/beam}$. Panel~\textbf{(c)} LoFAR LBA Sky Survey low-resolution image overlaid with radio contours at [3,6,9,12,15] of local $\sigma_{\rm{rms}}$ where $\sigma_{\rm{rms}} = 4.0 ~\rm{ mJy/beam}$ }\label{fig:p61}
       
\end{figure*}

The region surrounding this cluster is seen to be rich in the number of galaxies, as shown in \ref{fig:p61}a. A couple of filamentary structures can be spotted around the located centre, approximately from north-south and east-west. It appears that the system may lie on the node (intersection) of filaments, As a result, this system may have undergone several interactions in the past and may do so in the future, thus making it difficult to reliably associate the merger signatures with specific clusters. We identify PSZ2~G080.16+57.65, also known as Abell~2018, in an interacting state with nearby cluster SDSS-C4-DR3~3314, located at the mean redshift of $z = 0.088$. \citet{PSZ2} have reported cluster mass $M_{500}$ of $2.51\times 10^{14} ~\rm{M_\odot}$. We have estimated a mass of $\sim 9\times 10^{13} \rm{M_\odot}$ for Abell~2018 and lower limit of $ 3 \times 10^{12} \rm{M_\odot}$ for SDSS-C4-DR3~3314. The X-ray morphology of Abell~2018 is seen to be mildly elongated and clumpy, indicative of an unrelaxed state. \citet{ChecXMateMorphology} have shown that this cluster possesses an extremely dynamic morphology and is the 8th most dynamically active cluster amongst their sample of 118 clusters. The authors have also reported an asymmetry parameter of $1.5 \pm{0.4}$ which is a measure of departure from the spherically distributed luminosity profile. \citet{Paul_low_mass_clusters} have calculated the X-ray luminosity of $\rm{L_x} = 8.14^{+0.04}_{-0.03} \times 10^{43} \rm{erg s^{-1}}$ which tells that the cluster is  moderately luminous.

LoTSS-2 low-resolution images reveal the presence of several diffuse structures, most prominently a cluster radio shock to the southeast of the Abell~2018 centre, as well as a diffuse emission, just off-centre. The larger elongation of the radio shock is roughly perpendicular to the estimated merger axis, with its extent being $\sim$ 1.0 Mpc. We also calculated the radio flux density of $38\pm5$~mJy. The off-centre diffuse emission has an ultra-steep spectral index of $\alpha^{400}_{144} = -1.6$; however, as it is neither located exactly at the centre, nor at the outskirts, the identity of this emission remains doubted \citep{Paul_low_mass_clusters}. The LoFAR LBA images (see Fig.~\ref{fig:p61}d) show diffuse radio sources along the relic as well as a small source near the cluster centre. We have computed the flux density of $735 \pm 84 ~\rm{mJy}$ for the relic, and $755 \pm 18~\rm{mJy}$ for the diffuse patch. This computes to  $\alpha^{54}_{144} = -1.1$. Thus, it can further strengthen our argument with this additional information from the LBA data that the emission is a steep spectrum, possibly indicating a relic. With the insight from this work, it may be possible that the emission arises due to a historical shock seen in interaction with the halo. The other cluster, SDSS-C4-DR3~3314 does not show any well-defined diffuse radio signatures, which may be due to its extremely low mass as estimated by us.         

\subsubsection{PSZ2 G089.52+62.34 (PPM144)}
\begin{figure}
\centering
        
	 	\begin{subfigure}{0.8\linewidth}
	 	\centering
                \includegraphics[width=\linewidth]{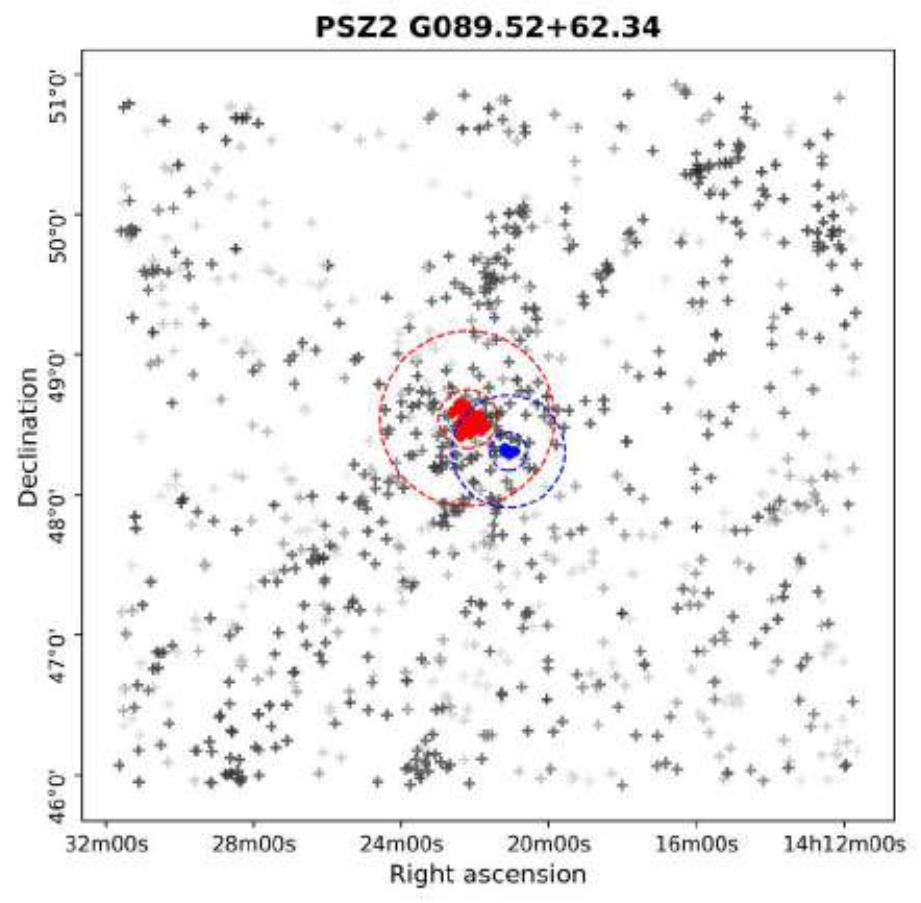}
                \caption{}
        \end{subfigure}
      \begin{subfigure}{0.8\linewidth}
      \centering
                \includegraphics[width=\linewidth]{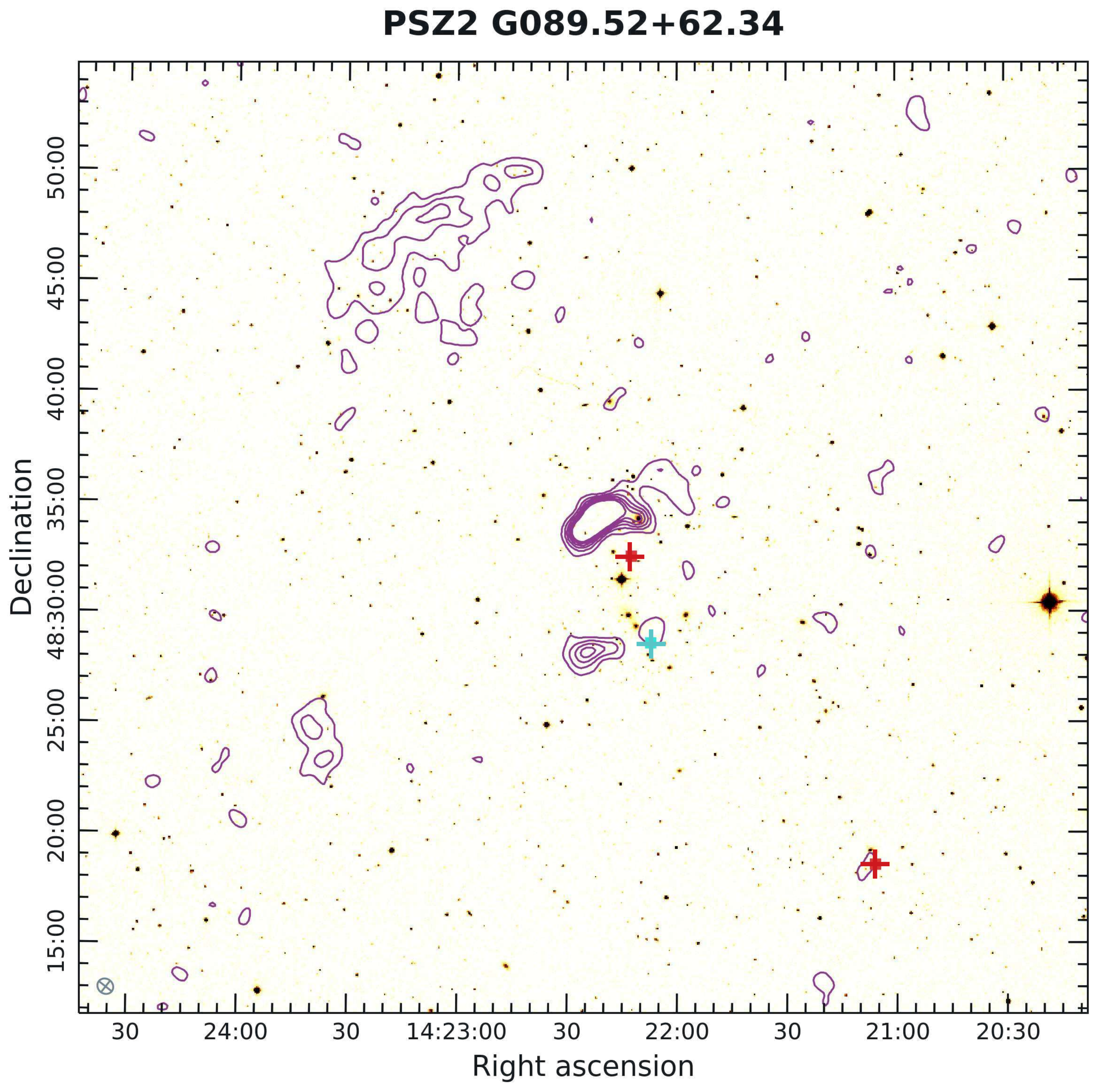}
                \caption{}
        \end{subfigure}
    	\begin{subfigure}{0.8\linewidth}
        	\centering
                \includegraphics[width=\linewidth]{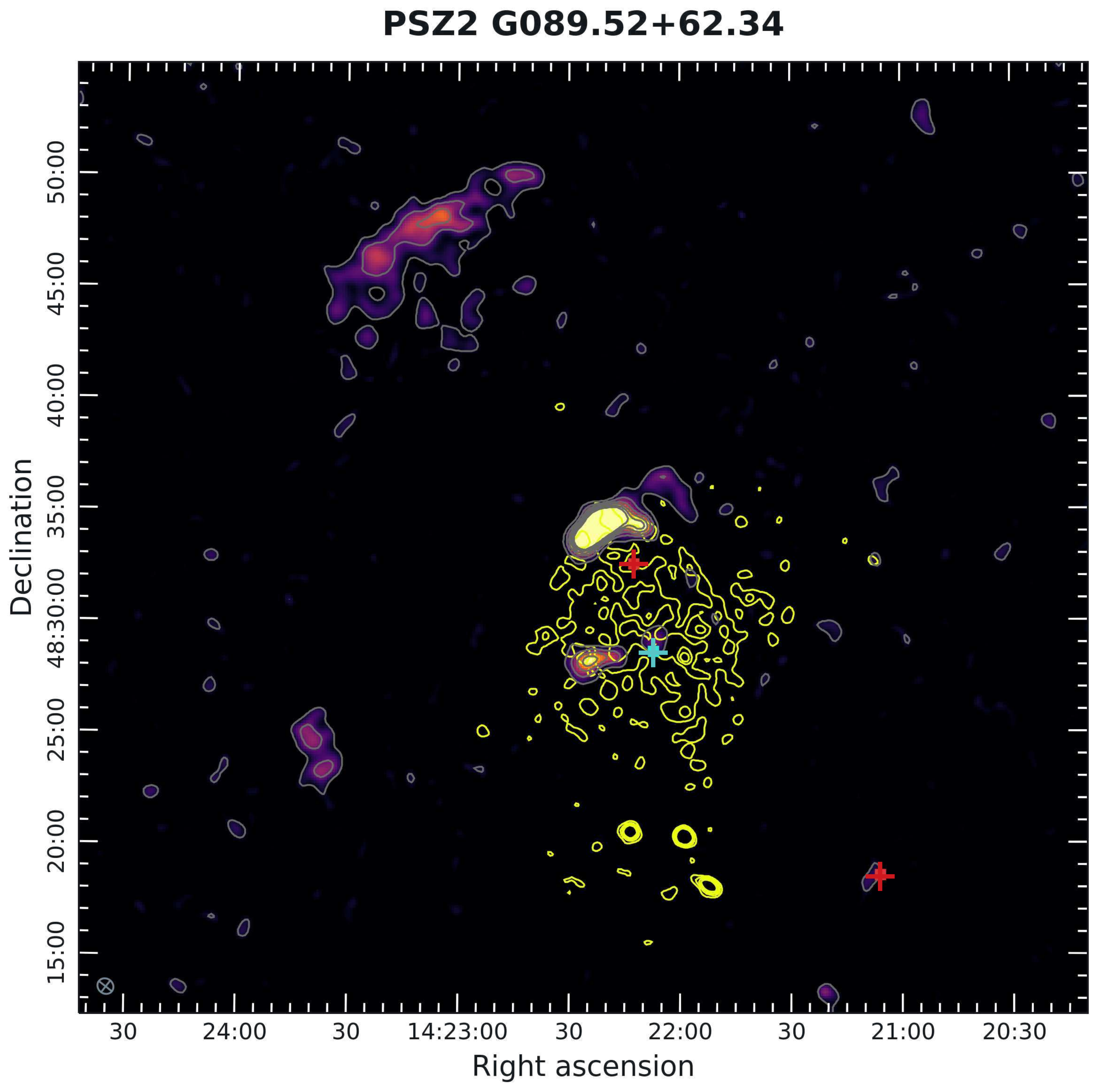}
                \caption{}
        \end{subfigure}

        \caption{\textbf{PSZ2 G089.52+62.34} Panel~\textbf{(a)}: A plot of SDSS galaxies. Panel~\textbf{(b)}: DSS r-band optical image overlaid with LoTSS-2 low resolution image contours at [3,6,9,12,15,18] of $\sigma$ where $\sigma_{\rm{rms}} = 250~\rm{\mu Jy/beam}$. Panel~\textbf{(c)}: XMM - Newton X-ray contours (yellow) with exposure of 19~ks overlaid on LoTSS-2 point source subtracted image with same contour levels (grey)}\label{fig:PPM64}
       
\end{figure}

 This couplet of interacting system is located at a mean redshift of  $ z = 0.07 $. The region surrounding this system is galaxy-rich, and a filamentary structure of SDSS galaxies can be seen originating from NE-SW in Fig. \ref{fig:PPM64}.  We compute the mass of the entire interacting system to be $1.9 \times 10^{14} M_\odot$. \citet{StrubleVelDisp} have reported a velocity dispersion of this cluster to be $ \sim 803 \rm{km s^{-1}}$ which is relatively higher for the cluster of this mass. \citet{FlinSubstructures} find a substructure in this system with an orientation similar to that we found in this work (NE-SW). In Fig. \ref{fig:PPM64} \textbf{c}, the X-ray morphology appears to be extremely clumpy and irregular, strongly suggesting a highly disturbed cluster medium. The cluster is also moderately luminous with $L_x = 2.54 \times 10^{44} \rm{erg}^{-1}$. \citep{Paul_low_mass_clusters}.

\citet{van_Weeren_2019} reported the presence of radio relics in the system for the first time. Apart from the Northern prominent relic, the system is also known to host a second inner relic \citep{Paul_low_mass_clusters}; also along NE closer to the cluster core. We compute the flux density of the outer relic to be $ 58.2 \pm 20.4$~mJy. In the LoTSS low-resolution image \ref{fig:PPM64} \textbf{c}, another diffuse signature can be observed near the cluster centre. Prima-facie, it appears to be a tailed galaxy, however, the optical signature seems to be missing in \ref{fig:PPM64} \textbf{b}.     

\subsection{Non-Planck Clusters in LoTSS-2 Region}

The SZ signal,

is strongly related to the mass of a cluster, mainly the luminous mass in the ICM  \citep{sz_mass_rel}. For a few low-mass systems, it is possible that the signature was beyond the reach of PSZ2 survey sensitivity. Such clusters may still continue to exhibit diffuse radio and X-ray sources due to their high dynamical activities. This section represents such clusters which are centre-matched from any of the SDSS, WHL or NSC cluster catalogues. These catalogues are based on observations in optical wavelengths. The utility of the catalogues is that they have a very large coverage, and low false detection rates \citep{WHL_Catalog}.

\subsubsection{RXCJ1053.7+5452 (PPM193)}
\begin{figure*}
\centering
        
	 	\begin{subfigure}{0.49\textwidth}
	 	\centering
                \includegraphics[width=\linewidth]{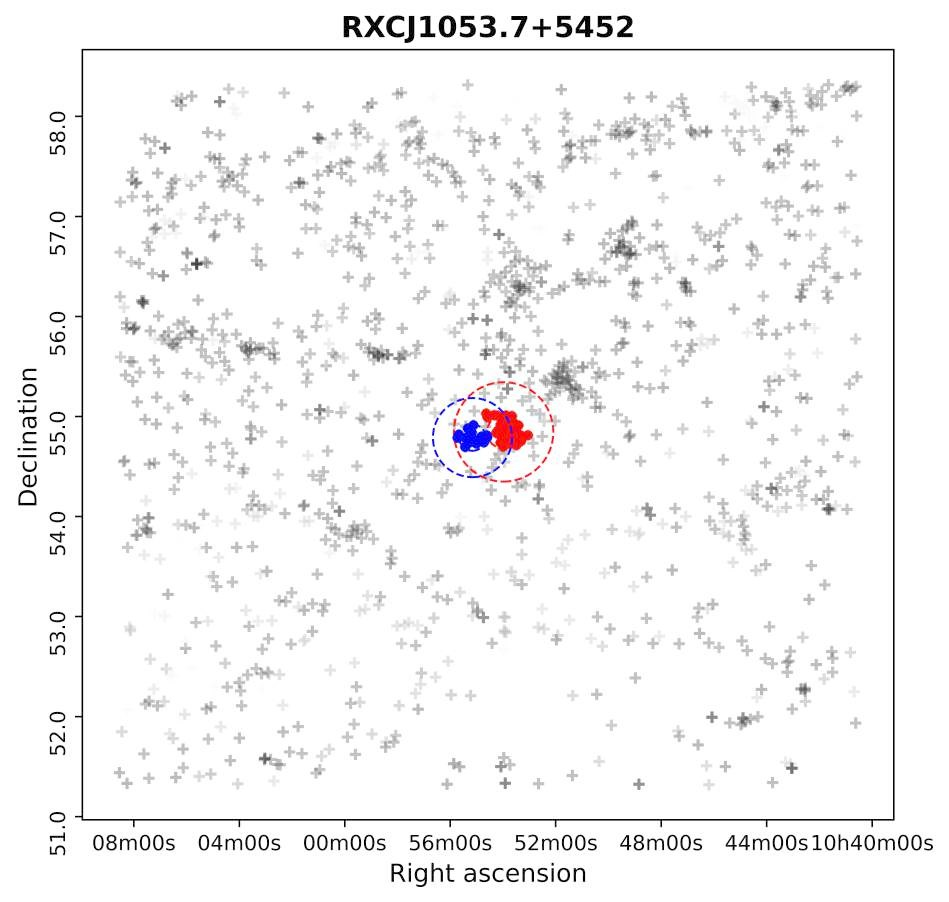}
                \caption{}
        \end{subfigure}
       \begin{subfigure}{0.49\textwidth}
       \centering
                \includegraphics[width=\linewidth]{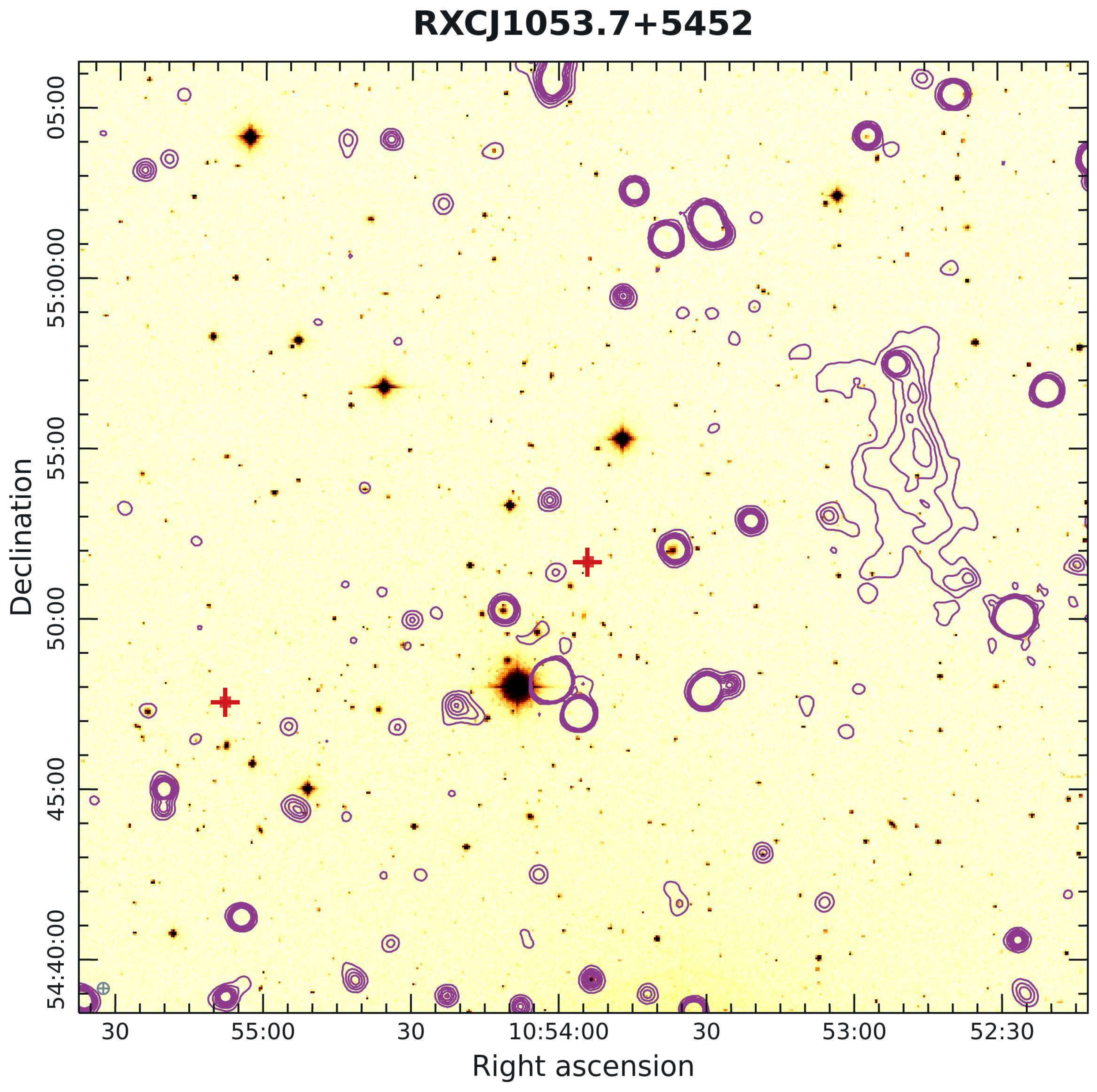}
                \caption{}
        \end{subfigure}
    	\begin{subfigure}{0.49\textwidth}
        	\centering
                \includegraphics[width=\linewidth]{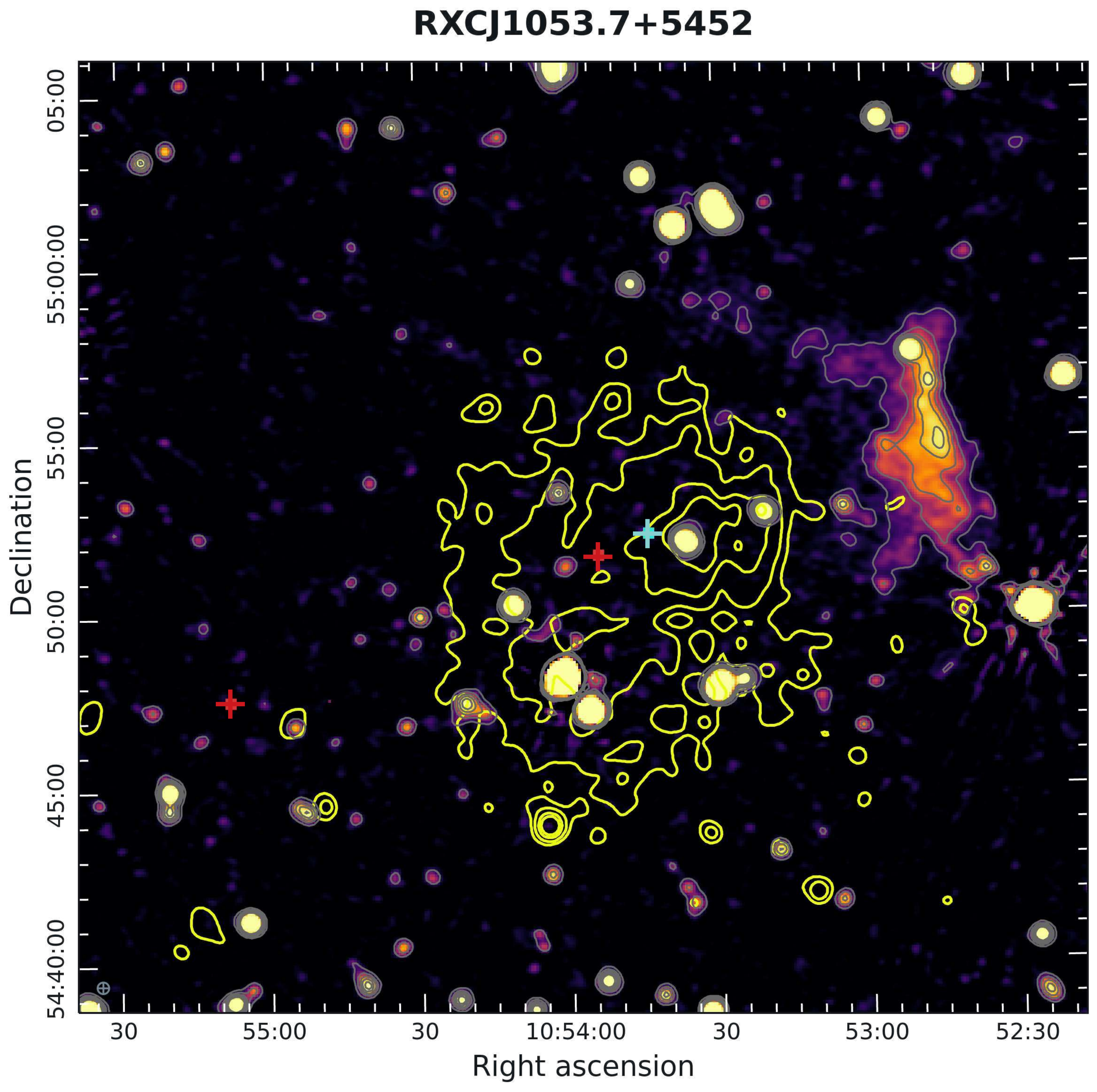}
                \caption{}
        \end{subfigure}
        \begin{subfigure}{0.49\textwidth}
        	\centering
                \includegraphics[width=\linewidth]{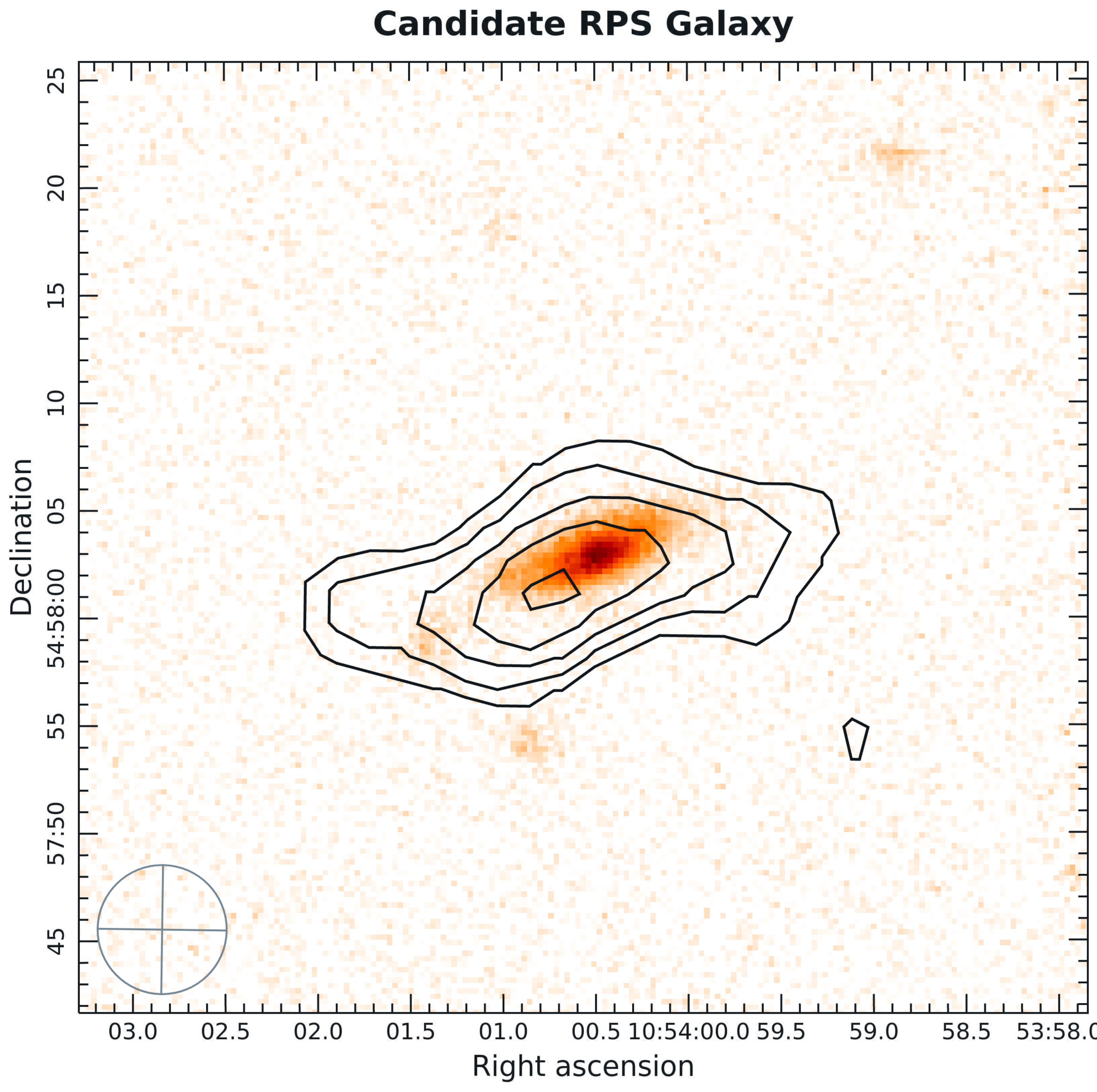}
                \caption{}
        \end{subfigure}
  
        \caption{\textbf{RXC~J1053.7+5452} panel\textbf{(a)}: A plot of SDSS galaxies in the nearby field. Panel~\textbf{(b)}: LoTSS-2 low resolution image contours at [3,6,9,12,15,18] of $\sigma$ where $\sigma = 200 ~\rm{\mu Jy/beam}$ overlaid on DSS optical image. Panel~\textbf{(c)}: Chandra X-ray contours (yellow) with exposure 24.76~ks overlaid on LoTSS-2 low-resolution image at the same contour levels (grey) with $\sigma = 175 \rm{\mu Jy/beam}$. Panel~\textbf{(d)}: A candidate RPS galaxy in the field of RXC~J1053.7+5452. LoTSS high-resolution contours at [3,4,5,6,7] $\sigma$ where $\sigma = 25~\rm{\mu Jy/beam}$ overlaid on PanSTARRS r-band optical image. }\label{fig:p19}
       
\end{figure*}

Located at the mean redshift of $z = 0.070$, our algorithm has flagged this system as a pre-merging/post-merging system. Although this system shows up in the SDSS cluster catalogue, it has been studied in little detail until now. The neighbourhood of this system shows the presence of a few other clusters and a subtle filament-like structure roughly extending from Northwest towards the centre as evident in Fig.~\ref{fig:p19}a. The total mass of this cluster obtained in this work is $\sim 1.0\times 10^{14}~\rm{M_\odot}$. The central velocity dispersion of galaxies of $665^{+51}_{-45}$, reported by \citet{sdssVeldisp2007}, does not indicate any extreme dynamical activity. However, \citet{Merging_Clust_Collab} have previously identified this as a merging system and report a general bi-modality present in the central region based on X-ray peaks. The X-ray morphology shows X-ray centre lies slightly ahead of the optical centre identified in this work, which is a possible indication of the independent evolution of dark and baryonic matter in a galaxy cluster as it undergoes interaction. A thorough X-ray study using Suzaku and {\it Chandra} telescopes has been done by \citet{RXCJ10_X_ray_Study}, where, a particular focus has been given to the region near the relic that this system is known to host. They report a temperature of $3.04^{+1.08}_{-1.08} \rm{KeV}$ for the cluster. Considering the mass of this system, this average temperature is significantly lower which may have resulted because of high cooling flows in the cluster centre of the cool-core cluster \citet{RXCJ10_X_ray_Study}. Nevertheless, the authors also suggested that the low temperature has been aided by the adiabatic expansion which follows a recent collision. In addition, the authors reported a sudden drop in temperature and density outwards of the surface brightness edge, indicating a cold front. If true, the presence of a cold front will solidify the interaction scenario presented in this work along the East-West direction.

A significantly large radio shock can be seen to the right of the identified centres in the LoTSS low-resolution images (see Fig.~ \ref{fig:p19}c). The larger elongation of this diffuse emission is roughly perpendicular to the merger axis. The extent of this radio shock is $\sim 770$~kpc with a radio flux density $193\pm20$~mJy. Until now, due to a lack of multi-band radio observations, the spectral index of the relic remained undetermined. Using the 1382~MHz observations carried out by \citet{RXCJ_van_Weeren} we calculate the spectral index of the diffuse emission source to be $ \alpha ^{1382}_{144} \sim -1.1 $; which is steep as expected for a relic. The absence of any optical source in Fig.~\ref{fig:p19}b, and the steep spectral index is clear evidence for a radio-relic nature of this diffuse radio source. As a tracer of interaction in this system, we identified an RPS galaxy candidate in a LoTSS-2 high-resolution map (see Fig.~\ref{fig:p19}d). The contours show stripping of gas towards the northeast, with the cluster centre located roughly to the southwest.
 
\subsubsection{SDSS-C4-DR3-3088 (M61)}
\begin{figure}
\centering
        
	 	\begin{subfigure}{0.8\linewidth}
	 	\centering
                \includegraphics[width=\linewidth]{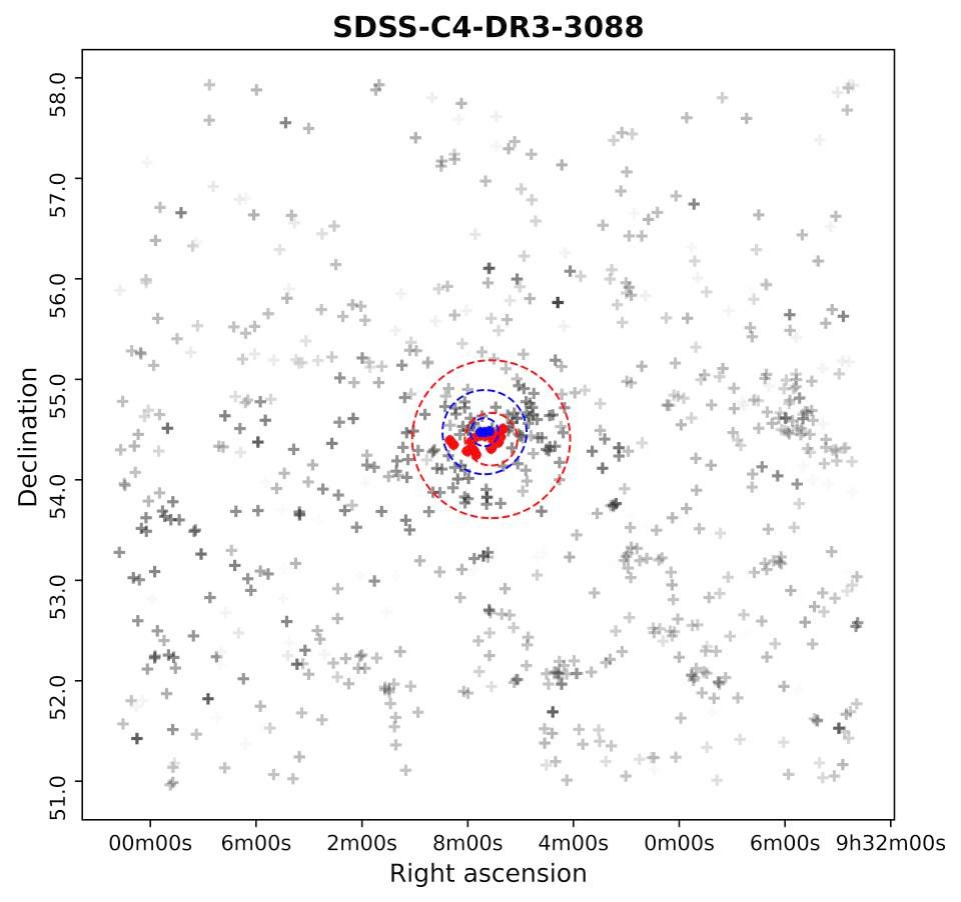}
                \caption{}
        \end{subfigure}
       \begin{subfigure}{0.8\linewidth}
       \centering
                \includegraphics[width=\linewidth]{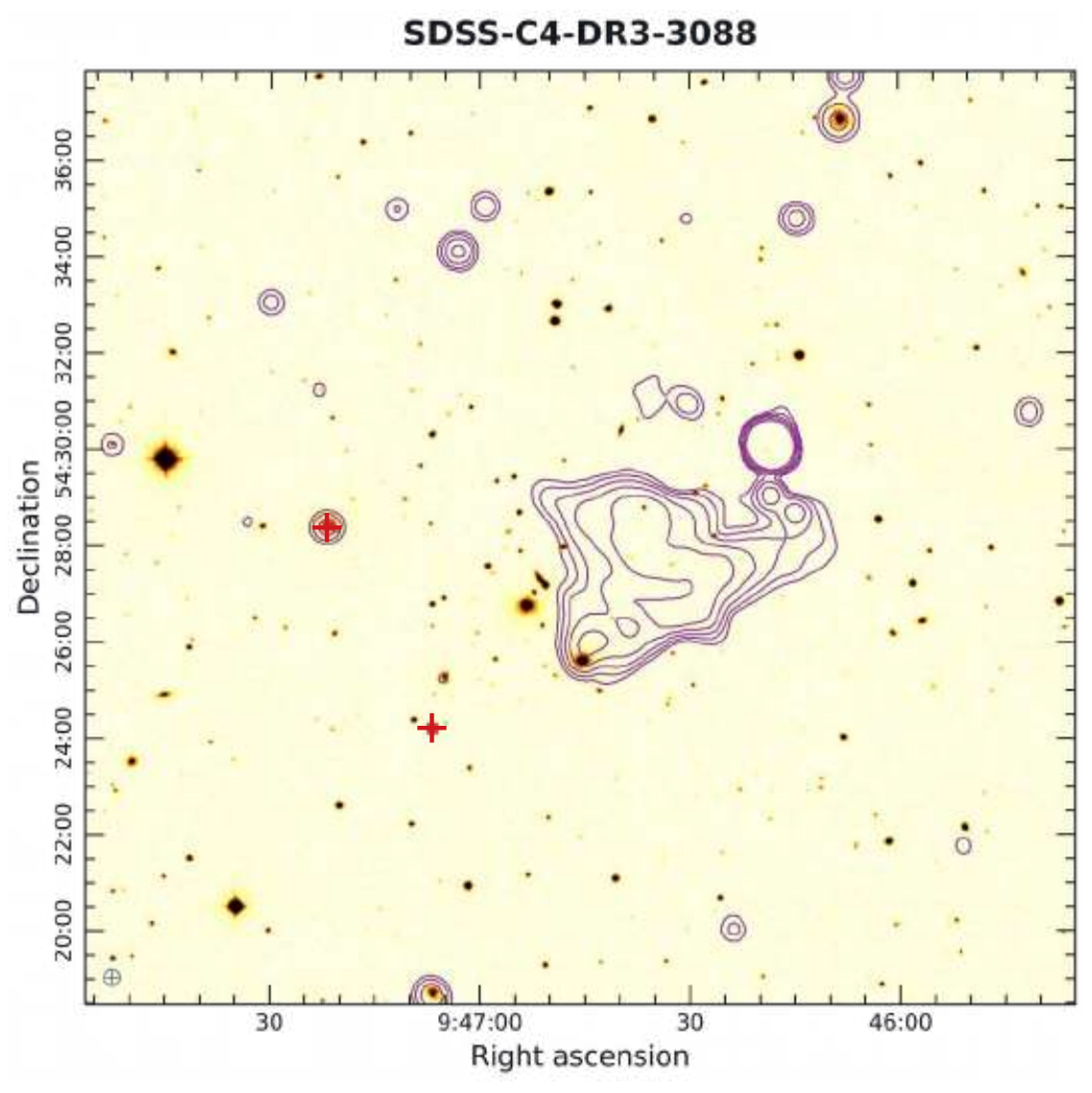}
                \caption{}
        \end{subfigure}
    	\begin{subfigure}{0.8\linewidth}
        	\centering
                \includegraphics[width=\linewidth]{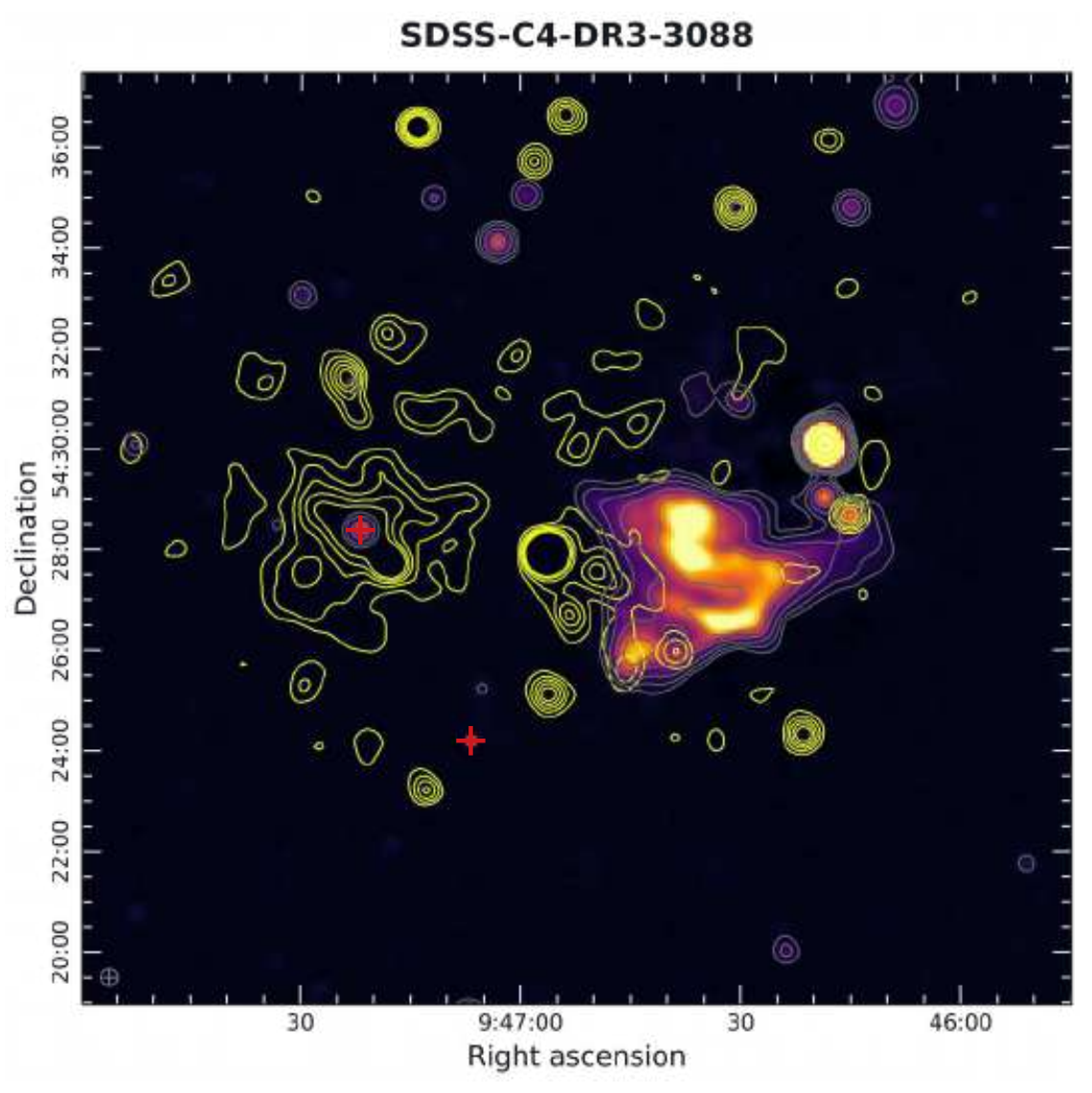}
                \caption{}
        \end{subfigure}
      \caption{\textbf{SDSS-C4-DR3-3088} panel\textbf{(a)}: A plot of SDSS galaxies in the nearby field. Panel~\textbf{(b)}: LoTSS-2 low-resolution image contours at [3, 6, 12, 24, 48] of $\sigma$ where $\sigma = 450 \rm{\mu Jy/beam}$ overlaid on DSS optical image. Panel~\textbf{(c)}: Chandra X-ray contours overlaid on LoTSS-2 low-resolution image with same contour levels (grey).}\label{fig:SDSS-C4-DR3-3088}
       
\end{figure}

In Fig.~\ref{fig:SDSS-C4-DR3-3088}a,  a faint filament-like structure is visible from top to bottom left. The neighbourhood of this galaxy cluster is rich in number. Our algorithm has classified this as a merging system comprising of two members. This system lies at the mean redshift of $z=0.046$. The 61 member galaxies of this cluster were identified by \citet{p4old_simard}, who also report the velocity dispersion of the member galaxies to be $\sigma = 556 \pm{61} $. The authors reported a mass of $2.88 \times 10^{14} M_\odot$, while, \citet{MandalFossilPlasma} reported M$_{500}$ to be $1.6^{+0.39}_{-0.36} \times 10^{14}~\rm M_\odot$, derived from X-ray observations. In this work, we estimate the mass of this cluster to be a lower limit of $\sim 7 \times 10^{13} ~\rm M_\odot$. \citet{MandalFossilPlasma} have studied the cluster in both X-ray and radio wavelengths and reported an X-ray temperature of $3.1 \pm 0.4 ~\rm keV$ which is an indication that the cluster is moderately hot considering that the cluster has a low mass. The centre identified in this work overlaps with the X-ray peak reported by the authors; however, it appears to be originating from a point source, possibly a BCG with signatures in optical and radio counterparts. The overall X-ray morphology of this cluster is highly irregular and extremely clumpy which points to high dynamical activity. A general bimodal structure can be seen in X-ray contours where the eastern lobe encompasses the BCG. The western lobe contains several optically bright galaxies and the centre is marked by the C4 cluster catalogue \citep{C4_cat}. 

This cluster is known to host an ultra-steep spectrum source with $\alpha_{150-610} = -1.74 \pm 0.23$ \citep{MandalFossilPlasma}. The authors also argued that a strong morphological link exists between bright radio galaxy in the field of this cluster and this diffuse emission. An additional remark that the authors make is that the kinematic time of energized electrons is much larger than the radiative lifetimes in this case which points to the revival of fossil electrons. An explanation for this revival may be the interaction described in this work. LoTSS-2 low-resolution images show the mentioned revived plasma source. We report the linear extension for the diffuse source to be $\sim330$~kpc, with the total $3\sigma$ flux density of $3.4\pm0.3$~Jy. The flux density of this source is relatively higher than the usual cluster peripheral relics suggestive of local re-energization of the plasma.    

\section{Summary and conclusions}\label{discussion}

In this work, we present an improved search algorithm for identifying interacting galaxy clusters from large optical galaxy survey data and interpret the interaction scenario traced by multi-band signatures of such systems. A comprehensive catalog of interacting systems has been prepared implementing an optimized Friends-of-Friends clustering algorithm on the latest data from SDSS~DR-17. This algorithm has been in use since the 1980s because of its simplicity of implementation and cheap computational cost. 
Here, we employed an open-source package of the FoF algorithm by Samuel Farrens \citep[sFoF; ][]{sfof} for our study. To account for selection effects, sFOF uses a scaling relation for the linking parameters. It has been observed that the output of the FOF algorithm is highly sensitive to the choice of this linking parameter which in turn depends on the goal of the study \citep{tagoInitialLL}. Moreover, the linking parameter significantly depends on various technical and astrophysical conditions and therefore varies over the sky patches. Unlike the usual method of determining of linking-length using a mock catalogue from simulated clusters, which lacks the real sky conditions, we use an optimization technique based on the physical verification of real galaxy clusters. To ensure that our algorithm is successful in locating actual clusters to its maximum possible accuracy, we for the first time, designed the algorithm that optimizes the linking length over the full survey volume based on the Planck~SZ galaxy cluster catalogue.  

Interacting clusters so far have largely been studied on an individual basis. Very few attempts have been made to systematically identify and catalogue interacting clusters. Most recently \citet{TempleMerger} have performed a somewhat similar study using SDSS DR-12 and concluded that due to the complicated nature of merging systems, automatic detection is not a straightforward procedure. In this work, by setting reasonable criteria, we defined the dynamical states of the cluster systems into two different categories, namely merging and pre-merging/post-merging. Our criteria are based on physical and geometrical parameters of clusters such as virial radius, elongation and haversine separation which has enabled us to remove much of the ambiguity associated with merging clusters. However, we should admit that distinctly defining pre and post-merging situations remains error-prone.

It is quite well known that galaxy clusters evolve through the process of hierarchical merging as well as continuous accretion. Cosmological simulations revealed that, at large scales, nodes of the cosmic web represent the region of high merging activity. Clusters which lie along the filaments are also likely to undergo merging or have undergone merging in the past. 

 Since cluster interactions and the presence of filaments are expected to correlate, we can think of galaxy filaments in redshift surveys as the tracers of interacting systems as flagged by the algorithm we have described. We also present the interaction scenario for each detected system from this viewpoint.

\subsection{Systems detected as parts of supercluster}

Our algorithm is so designed that it can detect a system with at least two interacting clusters to as many as objects that pass the criteria laid down by us to locate the interacting systems. Superclusters are the largest members in the hierarchical scheme of the cosmic web where even multiple massive clusters come closer and interact with each other to form a pattern of voids and filaments of galaxies and galaxy clusters \citep{LSS_SDSS}. In this work, we have identified such a system which is in fact a part of a much larger supercluster namely SCL-38 supercluster as identified by \citet{scl38supclstr}. Our algorithm identified a total of five galaxy clusters which include Abell 1173, Abell 1190 and Abell 1203 with a combined mass of around $ 5 \times 10^{14} ~\rm M_{\odot} $ This supercluster is also known to host two more Abell clusters, Abell~1155 and Abell~1187, which did not match our criteria of interaction. The largest separation among the five clusters in the system is at least 10 Mpc, possibly making the densest part of a much larger structure which also includes two more Abell clusters as mentioned above. Of these five clusters, Abell 1190 is the most massive system, and hence comes as the prominent detection in our algorithm as well in LoTSS diffuse radio maps, and two more have been also detected in X-rays. While, galaxy distribution clearly shows a filamentary connection, no tracers for the intergalactic medium be detected in radio or X-rays at the current sensitivity.
Along with Abell 1190, we have also detected the Abell 2061 system, which is part of the larger Corona borealis supercluster. The Corona Borealis supercluster is known to host more than five clusters which are supposed to be gravitationally bound and collapsing. Of these, we detect subgroups which fit our interaction criteria in the vicinity of A2061. The centres reported in this work suggest a dynamical scenario that is consistent with a highly characteristic diffuse radio emission spanning NW-SE hosted the system. The distribution of SDSS galaxies is also preferentially along similar directions. Although in the LoTSS-2 subtracted map, there are hints of faint radio emission originating all across this intergalactic filamentary region, no claim can be made with confidence at current sensitivity; but this makes Abell 2061 a rather interesting candidate for potential radio bridge observations using future telescopes.

\subsection{Interacting systems}

Apart from being connected by filaments, interacting clusters also show signs of strong dynamical activities through thermal (e.g., X-rays) and non-thermal (e.g., radio) emission features. Such emissions are diffuse in nature and therefore linked to tenuous intra and inter-cluster medium of these systems \citep{van_Weeren_2019,formanxrayICM}. Synchrotron radio emissions detected in these sources are in various forms, such as radio-halos or radio relics (see \citealt{Paul_2023_Diffuse_Emission,van_Weeren_2019} for the classifications), and are often linked to the galaxy-cluster formation dynamics, providing the strong evidence of the presence of cosmic magnetic fields. Moreover, these radio sources elucidate the astrophysical process of particle acceleration in these systems.

\subsubsection{Radio Halos} 
As we have already elaborated, due to the presence of magnetic fields, the electrons in the ICM that are accelerated to relativistic energies through cluster interactions, quickly lose energy by synchrotron radiation, peaking at radio wave bands. Turbulence being the dominant particle accelerator in the densest central part of the cluster, manifests itself as extended central diffuse radio sources that roughly follow the ICM distribution and are termed as radio halos. The previous works related to this topic have shown that there is a strong linkage between cluster mergers and radio halos \citet{cassanoHaloMergerLink}, making radio halos a good indicator for the interaction of massive systems.

In this work, many of the systems that are flagged as interacting systems show at least one of them hosts a radio halo. Abell~1190, part of the supercluster system SCL~38 hosts a halo-like emission with an extension of 400~kpc. 
A2061 also hosts a confirmed radio halo with the flux density of $407 \pm 41$, and its extent stretches as large as ~1 Mpc. Similarly, PSZ2~G053.53+59.52, a very closely interacting two-cluster system, was found to host a large diffuse radio emission permeating the entire X-ray halo. However, it seems the emission is contaminated with many radio galaxies. 

\subsubsection{Radio Shocks (Relics) and Revived Fossil Plasma Sources}

Radio relics or cluster radio shocks are usually considered as markers of shock generated during cluster mergers or due to adiabatic compression in the intermediate space between two interacting clusters \citep{Gu_premerger_shock}.  
These shocks are found to be supersonic (i.e., Mach number $M>1$) that can only be generated during violent major cluster mergers \citep{Paul_2011_ICM_Shocks} and therefore are the unambiguous signature of interacting systems. Since, astrophysical shocks can accelerate particles to relativistic energies through diffusive shock acceleration, under the influence of $\mu$G magnetic fields they emit synchrotron radiation \citep{ryurelicsdsa}. These are sources of diffuse radio emission with the usual extension of a few hundred kpc to Mpc \citep{van_Weeren_2019} and are mostly found near the cluster periphery. 

Many of the interacting systems identified by our algorithm show prominent radio relics confirming their dynamical activities. For example, RXC~J1053.7+5452 is a system with a characteristically large radio shock. This feature has an extension of about 770~kpc and exhibits a spectral index of $\sim -1.1$, as estimated by us using LoTSS-2 low-resolution maps. 
 
Furthermore, PSZ2~G080.16+57.65 and G089.52+62.34 are also detected with relic-like emissions.  PSZ2~G053.53+59.52 is a well-studied system with high dynamical activity \citep{A2034Detailed}. According to the criteria proposed in this work, the system is currently in the merging phase, however, the presence of unclassified relic-like features in the surrounding region indicates that other merger events may have also taken place in the past.

Another class of ultra-steep-spectrum cluster diffuse radio sources called the revived fossil plasma sources have been discovered in numbers in recent years \citep[see][for a review]{Paul_2023_Diffuse_Emission}.
It is likely, that the fossil electrons from tailed radio galaxies and lobes of radio galaxies which are deposited in the ICM get revived during merger and accretion processes \citep{Ensslin-krishna2001}. SDSS-C4-DR3-3088 represents such a case of a revived fossil plasma source. 

\subsubsection{Ram pressure stripped galaxies as the interaction marker}

The ICM tends to be hot ($\gtrsim10^7$K), and even more so in the case of interacting clusters due to gas sloshing that results from dynamical activities \citep{ZuHoneSloshHeating}. As the member galaxies of a cluster move through this hot ICM, they experience a ram pressure that strips the gravitationally bound gas from a galaxy, a process known as the ram pressure stripping (RPS) \citep{NulsonRPSEffect}. In the case of spiral galaxies, this effect is easier to observe in their radio features, since the radio tail can be seen to extend much beyond the galactic disk, which may itself appear bow-shaped. The tail emanating due to the stripping of gas is usually pointed away from the cluster centre as the galaxy falls towards it. RPS galaxies can thus be seen as an additional tracer of cluster interaction \citep{Ebling_RPS_Merger}.  Figure~\ref{fig:p55}d shows such a galaxy identified by \citet{PoggiantiRPSA21124}. Additionally, using the information from LoTSS-2 high-resolution maps, and the centre and merger axis location, we found a few more such RPS candidates, as shown in Fig.~\ref{fig:PPM68}, Fig.~\ref{fig:Abell 2034}d, Fig.~\ref{fig:p19}d, and Fig. \ref{fig:M72} It is to be noted that a detailed multiwavelength study is needed for confirming the RPS galaxies; which is beyond the scope of this paper.

Finally, we may conclude that, by implementing a physically relevant new optimization scheme for choosing linking length in the existing FoF algorithm, we successfully produced a comprehensive list of interacting galaxy cluster systems from the latest SDSS data release, DR-17. We have also proposed a set of physically motivated criteria and classified these interacting systems into two phases, 'merging' and 'pre-merging/post-merging'. Interestingly, we found that the bulk of cases show cluster interaction along the prominent cosmic filaments (as seen in SDSS optical galaxy distribution) with the most violent ones at their nodes. Important to mention here is, this is exactly what the cosmological simulations based on L-CDM cosmology predict. The imprint of interactions such as diffuse cluster emissions in radio or X-rays, has also been identified in many of these systems or in their proximity. 

While, these systems show the well-established multi-band emission features as a sign of interaction, such as radio halos, relics or revived fossil plasma as well as clumpy and elongated morphology in X-rays, in a handful of cases the nature of the emission features could not be characterised. Moreover, the non-detection of filamentary connections and ridges etc. in the available multi-band deep images and surveys made it difficult to show the direct evidence for interactions in these systems. Nevertheless, we are hopeful that with the upcoming extremely sensitive radio interferometers, such as SKA, or with the eROSITA telescope in X-rays, it may also be possible to detect the faint filamentary materials that span several Mpc connecting these interacting clusters. It would thus be useful to have a catalogue of interacting systems, such as the one we presented here, to hunt for such extremely faint signatures. Once a large number of such systems, undergoing various phases of a merger are catalogued, it will also be possible to stack them and make predictions for the required detection sensitivity for future surveys. 

\section*{Acknowledgements}

We would like to thank the editor and anonymous reviewer for their helpful comments for improving the quality of the manuscript.

Funding for the Sloan Digital Sky Survey IV has been provided by the Alfred P. Sloan Foundation, the U.S. Department of Energy Office of Science, and the Participating Institutions. SDSS acknowledges support and resources from the Center for High-Performance Computing at the University of Utah. The SDSS website is www.sdss4.org.

SDSS is managed by the Astrophysical Research Consortium for the Participating Institutions of the SDSS Collaboration including the Brazilian Participation Group, the Carnegie Institution for Science, Carnegie Mellon University, Center for Astrophysics | Harvard \& Smithsonian (CfA), the Chilean Participation Group, the French Participation Group, Instituto de Astrofísica de Canarias, The Johns Hopkins University, Kavli Institute for the Physics and Mathematics of the Universe (IPMU) / University of Tokyo, the Korean Participation Group, Lawrence Berkeley National Laboratory, Leibniz Institut für Astrophysik Potsdam (AIP), Max-Planck-Institut für Astronomie (MPIA Heidelberg), Max-Planck-Institut für Astrophysik (MPA Garching), Max-Planck-Institut für Extraterrestrische Physik (MPE), National Astronomical Observatories of China, New Mexico State University, New York University, University of Notre Dame, Observatório Nacional / MCTI, The Ohio State University, Pennsylvania State University, Shanghai Astronomical Observatory, United Kingdom Participation Group, Universidad Nacional Autónoma de México, University of Arizona, University of Colorado Boulder, University of Oxford, University of Portsmouth, University of Utah, University of Virginia, University of Washington, University of Wisconsin, Vanderbilt University, and Yale University.

LOFAR data products were provided by the LOFAR Surveys Key Science project (LSKSP; https://lofar-surveys.org/) and were derived from observations with the International LOFAR Telescope (ILT). LOFAR (van Haarlem et al. 2013) is the Low-Frequency Array designed and constructed by ASTRON. It has observing, data processing, and data storage facilities in several countries, which are owned by various parties (each with its own funding sources), and which are collectively operated by the ILT foundation under a joint scientific policy. The efforts of the LSKSP have benefited from funding from the European Research Council, NOVA, NWO, CNRS-INSU, the SURF Co-operative, the UK Science and Technology Funding Council and the Jülich Supercomputing Centre.

This research has made use of data obtained from the Chandra Data Archive and the Chandra Source Catalog, and software provided by the Chandra X-ray Center (CXC) in the application packages CIAO and Sherpa.

This research is based on observations obtained with XMM-Newton, an ESA science mission with instruments and contributions directly funded by ESA Member States and NASA.


\section*{Data Availability}
The optical SDSS data used in this work is publically available at \url{http://skyserver.sdss.org/dr17/SearchTools/sql}. The Planck-SZ2 clusters were obtained from \url{https://heasarc.gsfc.nasa.gov/W3Browse/planck/plancksz2.html}. The LoTSS-2 data is also publically available at \url{https://lofar-surveys.org/dr2_release.html} while the LOFAR-LBA data is available at \url{https://www.lofar-surveys.org/lolss.html}. Data from XMM-Newton was obtained from \url{http://nxsa.esac.esa.int/nxsa-web/#search}. Data related to Chandra X-ray observatory is available at \url{https://cda.harvard.edu/pop/dispatchQuick.do}. Analysed data can be made available upon reasonable request to the authors.


\bibliographystyle{mnras}
\bibliography{./refs} 



\appendix
\section{Multiwavelength Interaction Tracers of Systems }
\label{Remaining_Systems}

\subsubsection{PSZ2~G095.22+67.41 (PPM136)}

\begin{figure*}
\centering
        
	 	\begin{subfigure}{0.495\textwidth}
	 	\centering
                \includegraphics[width=\linewidth]{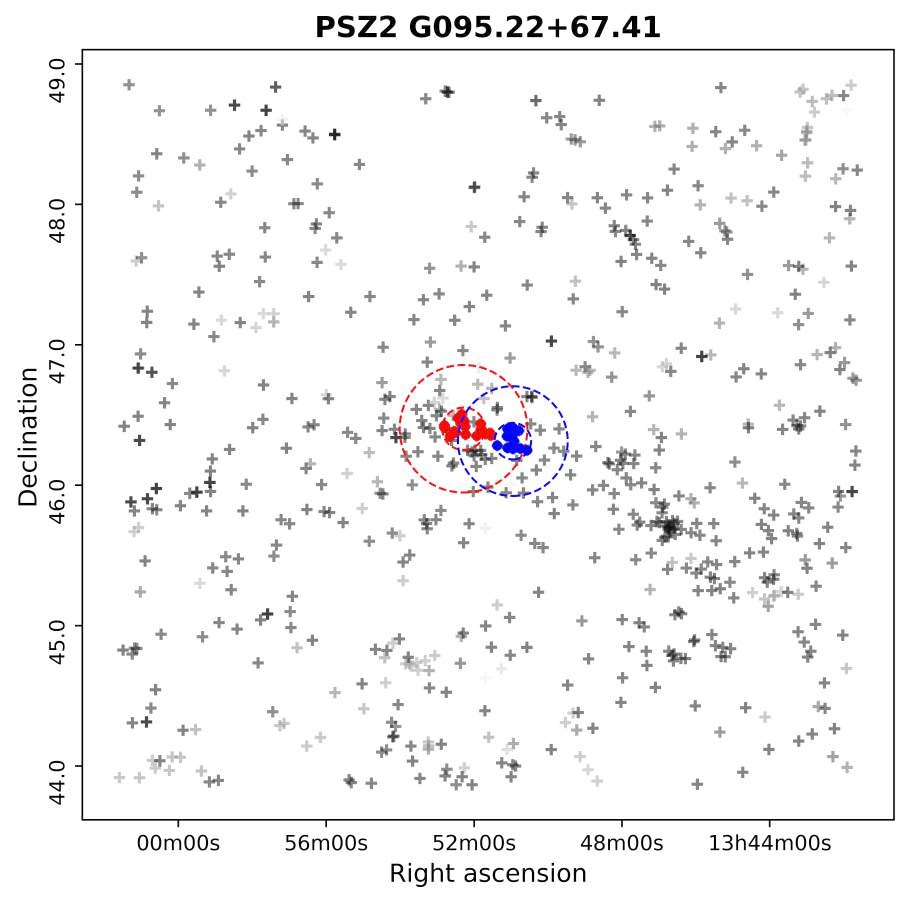}
                \caption{}
        \end{subfigure}
       \begin{subfigure}{0.485\textwidth}
       \centering
                \includegraphics[width=\linewidth]{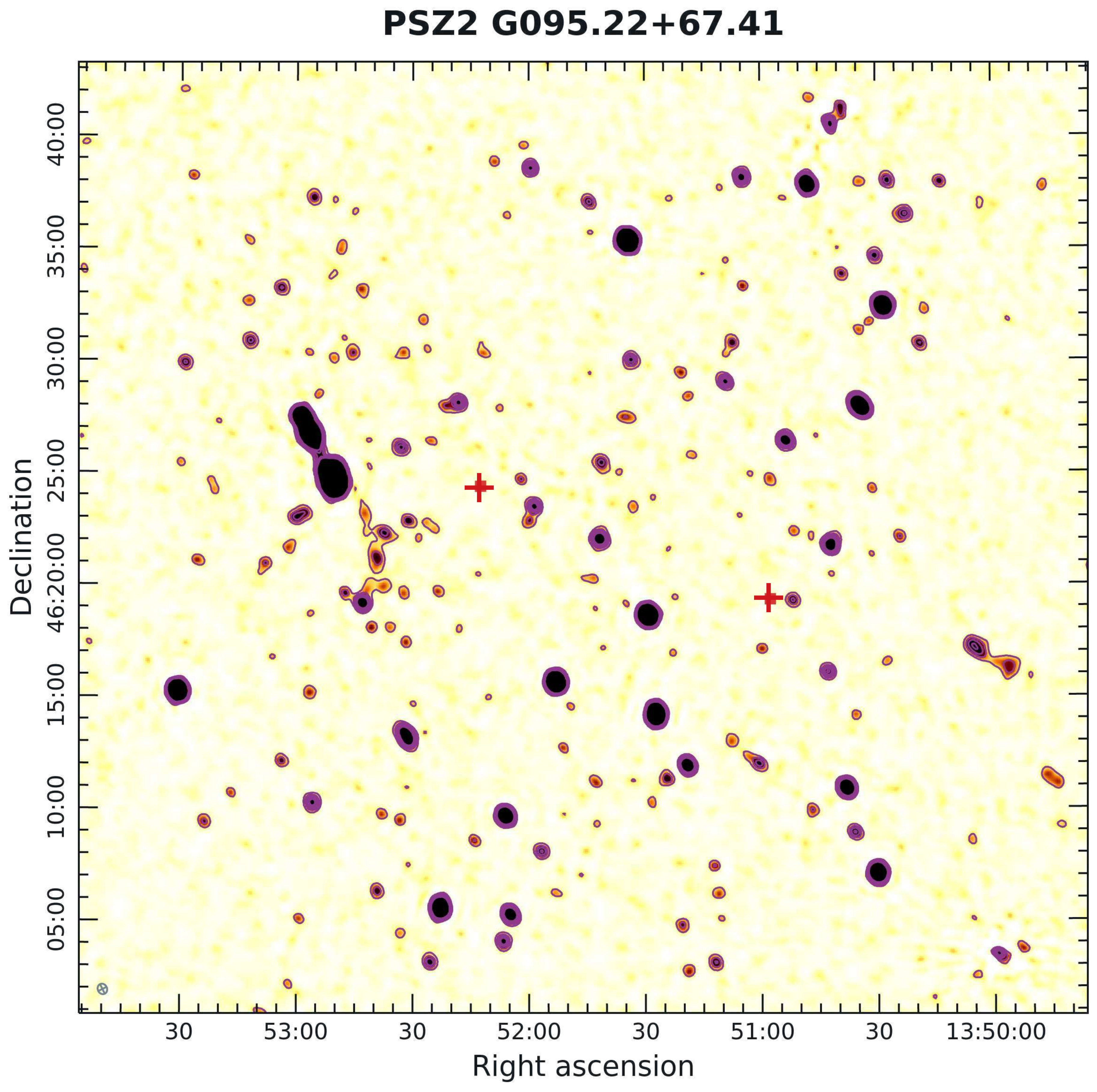}
                \caption{}
        \end{subfigure}
    	\begin{subfigure}{0.485\textwidth}
        	\centering
                \includegraphics[width=\linewidth]{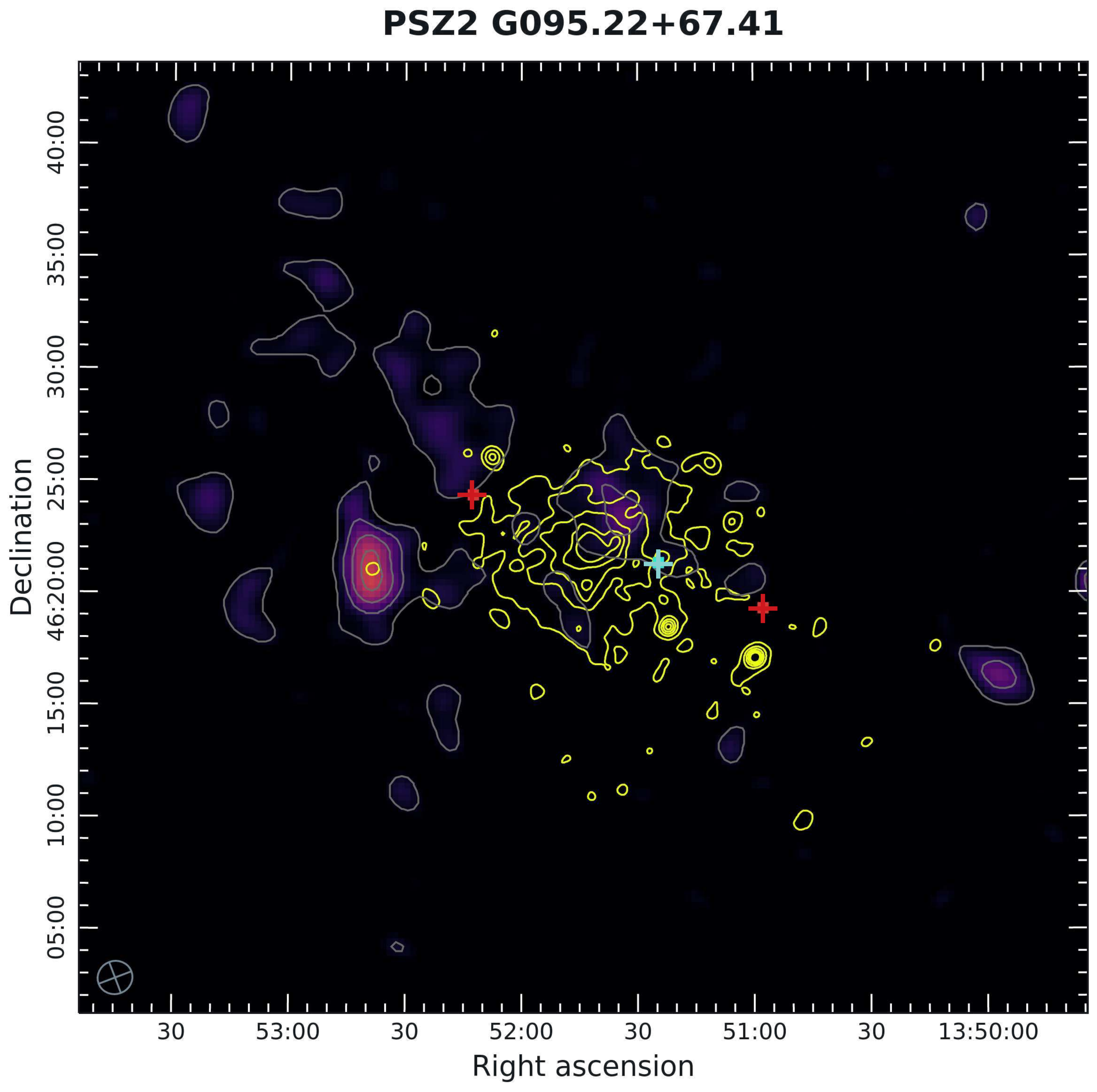}
                \caption{}
        \end{subfigure}
        \begin{subfigure}{0.485\textwidth}
        	\centering
                \includegraphics[width=\linewidth]{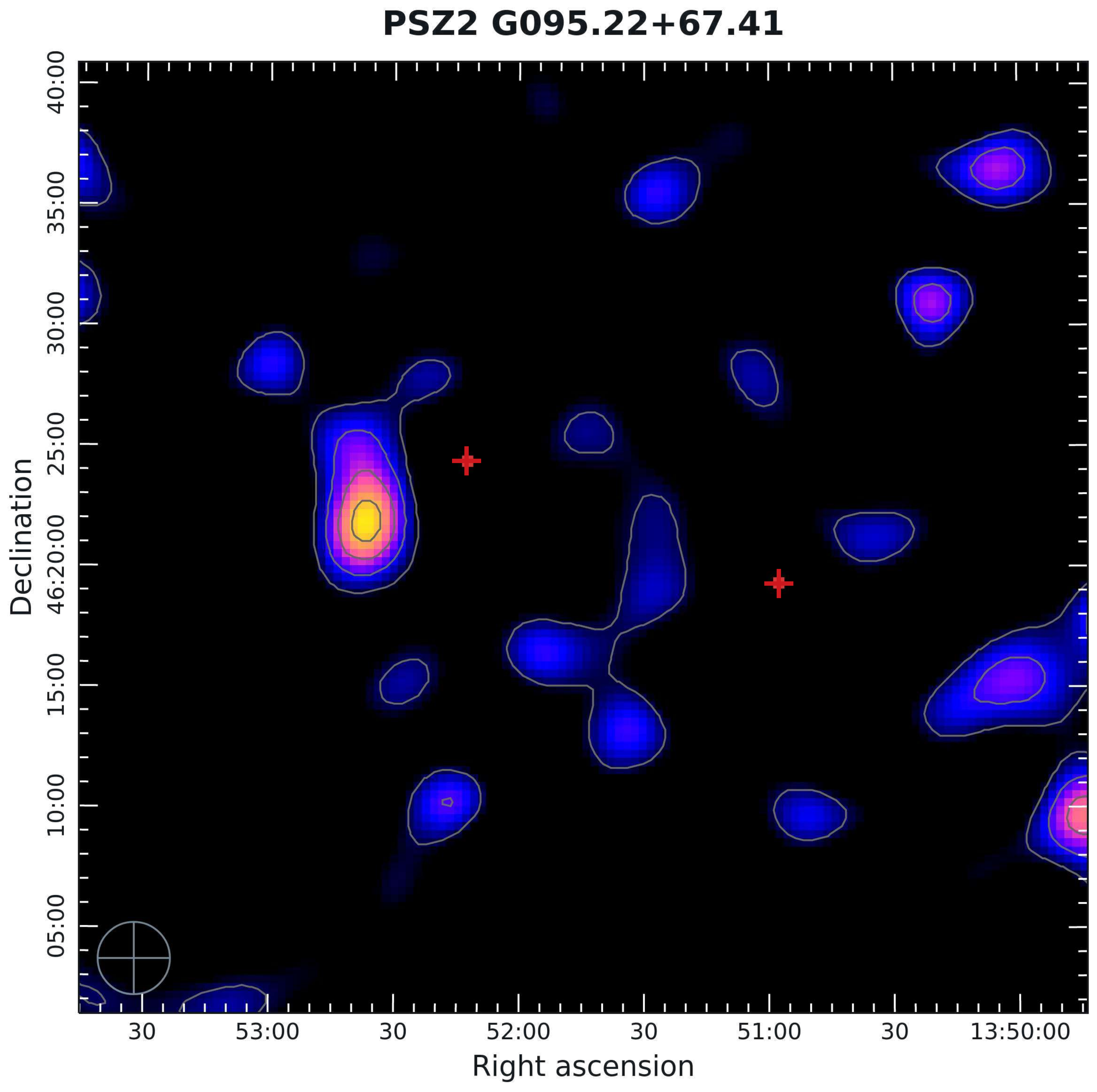}
                \caption{}
        \end{subfigure}

        \caption{ \textbf{PSZ2~G095.22+67.41} Panel \textbf{(a)}: A plot of SDSS galaxies. Panel~\textbf{(b)}: DSS red band optical image overlaid with LoTSS-2 low resolution image contours at  [3,6,9,12,15] of $\sigma$ where $\sigma_{\rm{rms}} = 200 ~\rm{\mu Jy/beam}$. Panel~\textbf{(c)}: XMM-Newton X-ray contours (yellow) with exposure of 24~ks overlaid on LoTSS-2 point source subtracted image with contour levels (grey) [3,6,9,12,15] of $\sigma$ where $\sigma_{\rm{rms}} = 300 ~\rm{\mu Jy/beam}$. Panel~\textbf{(d):} LoFAR LBA Sky Survey low-resolution image overlaid with radio contours at [3,6,9,12,15] of local $\sigma_{\rm{rms}}$ where $\sigma_{\rm{rms}} = 4.5 ~\rm{ mJy/beam}$ }\label{fig:p44}
       
\end{figure*}

The couplet of interacting clusters, PSZ2~G095.22+67.41 flagged by our algorithm can be seen in Fig \ref{fig:p44}c. The system lies in a galaxy-rich neighbourhood, with a subtle filamentary structure emanating from the southwest towards the centre of the image. The first studies of interaction associated with this cluster are very recent \citep{vanweerenHETDEX}, and the entire dynamical scenario is still not fully understood. This is a relatively smaller cluster with an estimated mass of only $\sim 5\times 10^{13} \rm{M_\odot}$, located at a mean redshift of $z = 0.063$. The X-ray morphology of the cluster is clumpy and irregular indicating that the cluster is not in a relaxed state, and is possibly undergoing a merger along the east-west direction which coincides with the apparent stretch of galactic filament.

LoTSS low-resolution images reveal an elongated diffuse structure to the east of centres. This was first claimed as a candidate relic by \citet{vanweerenHETDEX}. This diffuse emission has an extent of $\sim385$~kpc with a radio flux density of $15.7\pm2.9$~mJy. \citet{PlanckLotssdr2} do not report a well-observed diffuse emission in the central region of this cluster. The Planck centre is slightly off from the centre which can be identified from the X-ray contours, nicely encompassing the BCG. Interestingly, the point source subtracted image reveals a diffuse emission which appears to be in the vicinity of the centre, and the radio contours primarily indicate that the emission does not arise from the BCG. Although prima facie, it appears to be a halo. It exhibits a very low flux ($\sim6$~mJy) and is not properly aligned with the X-ray centre. Stronger evidence for the eastern relic-like emission can be found in LoFAR~LBA images (see Fig.~\ref{fig:p44}d). With a flux density of $130 \pm {19}$~mJy for the source in the LBA image, we estimated an upper limit on the spectral index as $\alpha^{56}_{144} -2.2$ which is indicative of an ultra-steep spectrum relic.

\subsubsection{PSZ2 G057.78+52.32 (Abell 2124/PPM175)}
\begin{figure*}
\centering
        
	 	\begin{subfigure}{0.49\textwidth}
	 	\centering
                \includegraphics[width=\linewidth]{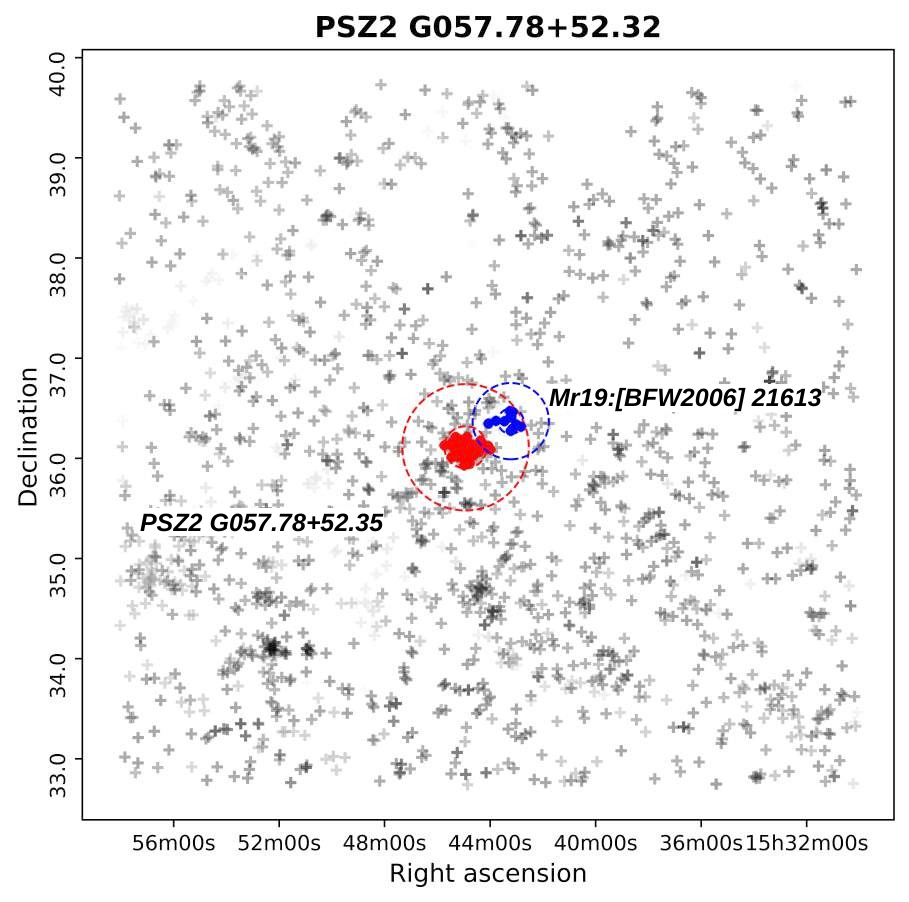}
                \caption{}
        \end{subfigure}
       \begin{subfigure}{0.49\textwidth}
       \centering
                \includegraphics[width=\linewidth]{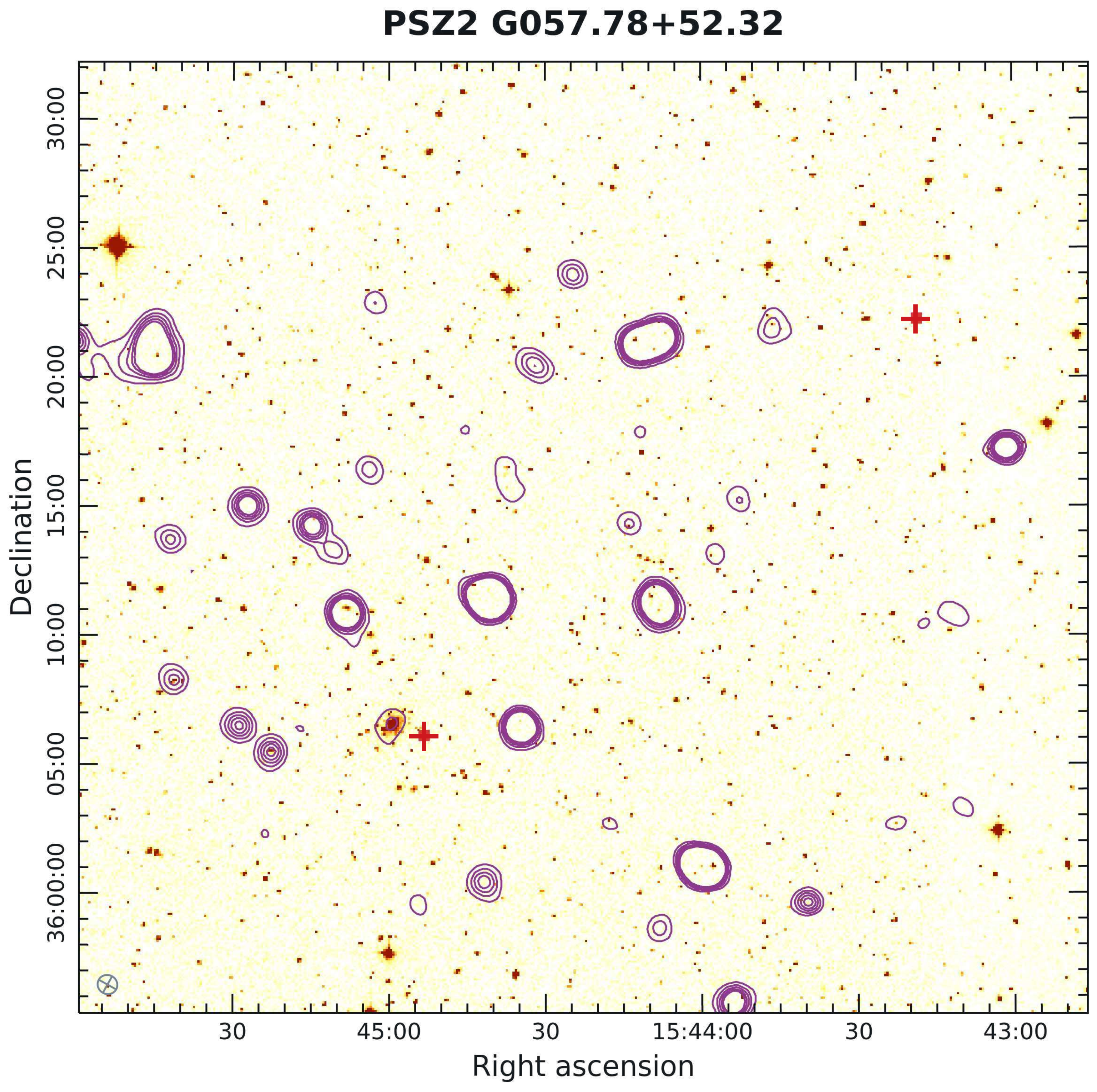}
                \caption{}
        \end{subfigure}
    	\begin{subfigure}{0.49\textwidth}
        	\centering
                \includegraphics[width=\linewidth]{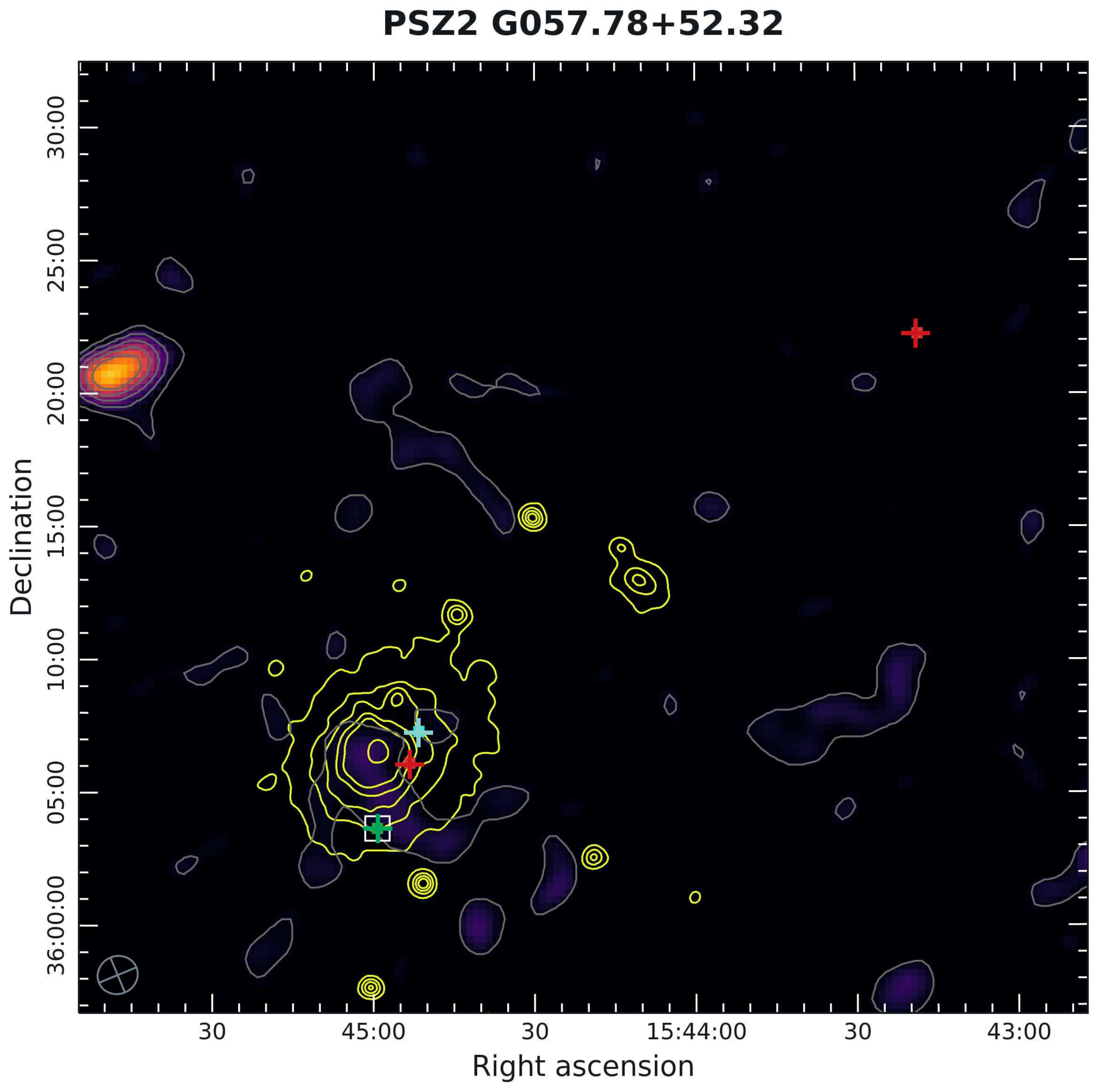}
                \caption{}
        \end{subfigure}
          \begin{subfigure}{0.49\textwidth}
        	\centering
                \includegraphics[width=\linewidth]{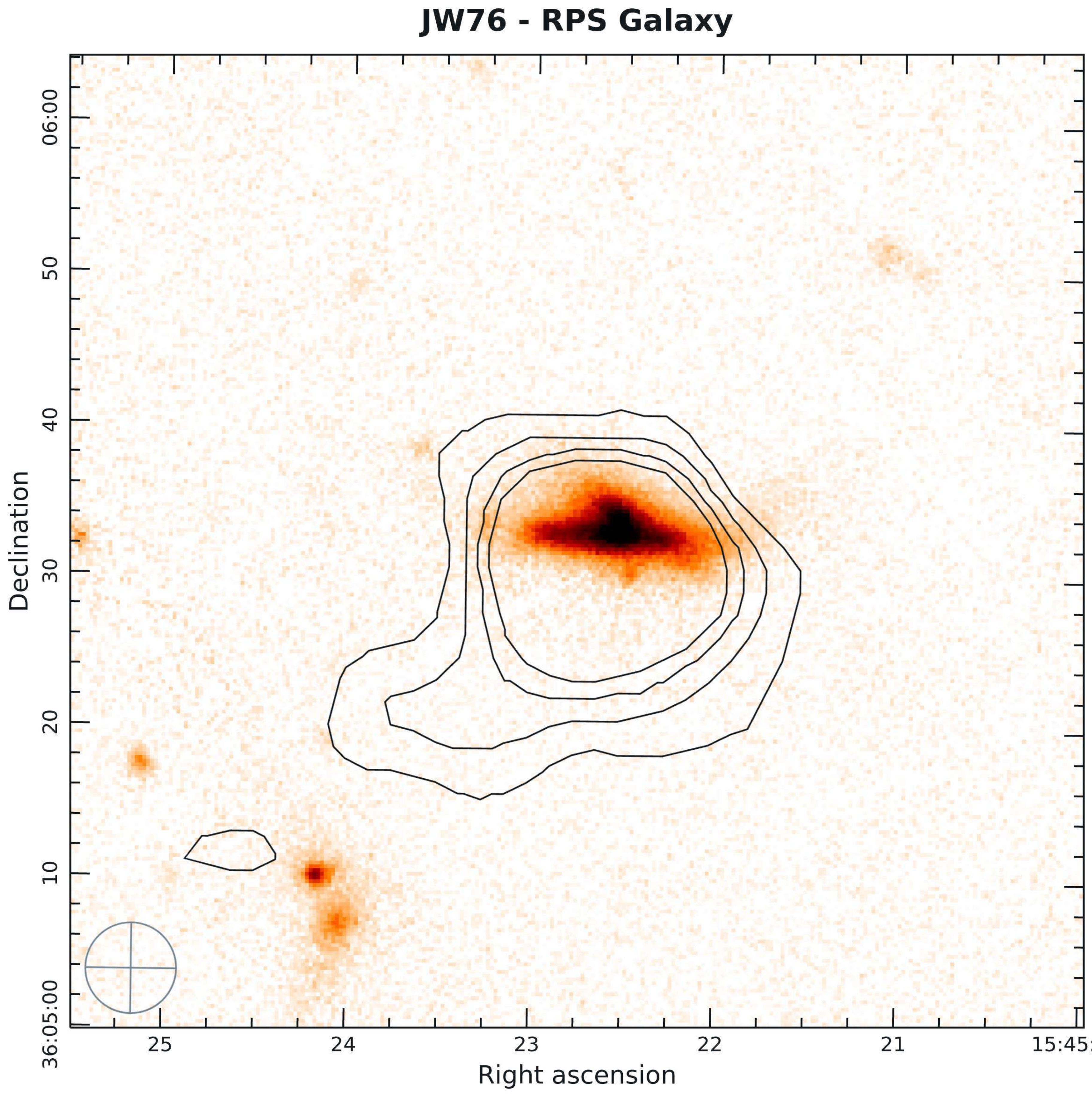}
                \caption{}
        \end{subfigure}

        \caption{ \textbf{PSZ2~G057.78+52.32} Panel~\textbf{(a)}: A plot of SDSS galaxies. Panel~\textbf{(b)}: DSS r-band optical image overlaid with LoTSS-2 low resolution image contours at [3,6,9,12,15] of $\sigma$ where $\sigma_{\rm{rms}} = 500~\rm{\mu Jy/beam}$. Panel~\textbf{(c)}: XMM-Newton X-ray contours (yellow) with exposure of 35~ks overlaid on LoTSS-2 point source subtracted image with same contour levels (grey) at $\sigma_{\rm{rms}} = 350 \rm{\mu Jy/beam}$ Panel~\textbf{(d)}: A candidate Ram-pressure stripped galaxy; Pan-STARRS r band optical image overlaid with LoTSS-2 high-resolution image with  contour levels of [3,6,9,12] at $\sigma_{\rm{rms}} = 80 ~\rm{\mu Jy/beam}$.}\label{fig:p55}
       
\end{figure*}

A prominent filamentary structure can be seen going from top-left to bottom-right in Fig.~\ref{fig:p55}a. PSZ2~G057.78+52.32, also known as Abell~2124 is seen to be interacting with a nearby cluster Mr19:[BWF2006]~21613 and is in the state of pre-merger/post-merger. This system is located at a mean redshift of $z = 0.065$. Abell~2124 is a fairly well-studied cluster. One of the first studies based on optical data shows the presence of two substructures, one roughly to the north of the centre, and one to the south at the scale of $\sim 130$~kpc \citep{FlinA2124}. This is a possible indicator that the cluster may have undergone interactions in the past. \citet{PSZ2} report a $M_{500}$ of $2.38\times 10^{14} \rm{M_\odot}$. \citet{BiltonA2124} have calculated the velocity dispersion of $751^{40}_{80}$ which points towards high dynamic activity. In addition to this, \citet{PoggiantiRPSA21124} have found six candidate jellyfish galaxies in the field of Abell~2124. Figure~\ref{fig:p55}d shows one of such ram pressure-stripped galaxies, JW76, where the centres and merger axis lie roughly to the northwest of this galaxy. \citet{xraygclustsample} have determined the X-ray temperature of $5.17^{+0.36}_{-0.29}~\rm{keV}$ using data from {\it Chandra} observatory, which can be considered to be hot for the mass the cluster has. It is also highly luminous with $0.5$–$7.0$~keV X-ray luminosity of $6.4 \pm{0.1} \times 10^{43}~\rm{erg s^{-1}}$. The overall X-ray morphology is regular, however, it is asymmetric and slightly elongated in the NW-SE direction \citep{LakhchauraA2124}. The authors also discover a mismatch between the X-ray peak and optical peak in the central region of this cluster pointing towards ongoing galactic merger. The other cluster, Mr19:[BWF2006]~21613 shows no signatures in Radio or X-rays, which may be accounted for its low mass.    

The first radio study of Abell~2124 was performed by \citet{HanischA2124} to search for halo using WSRT. The authors conclude that no significant halo-like emission can be associated with this cluster. However, the 144~MHz LoTSS point source subtracted image reveals an elongated diffuse emission nearby one of the centres along the merger axis. We calculate the size of this diffuse emission to be around $\sim890$~kpc with a radio flux density of $48\pm6$~mJy. Radio contours on the optical image reveal that this emission may be arising from the BCG.

\subsubsection{Abell 2249 (M72)}
\begin{figure*}
\centering
	 	\begin{subfigure}{0.49\textwidth}
	 	\centering
                \includegraphics[width=\linewidth]{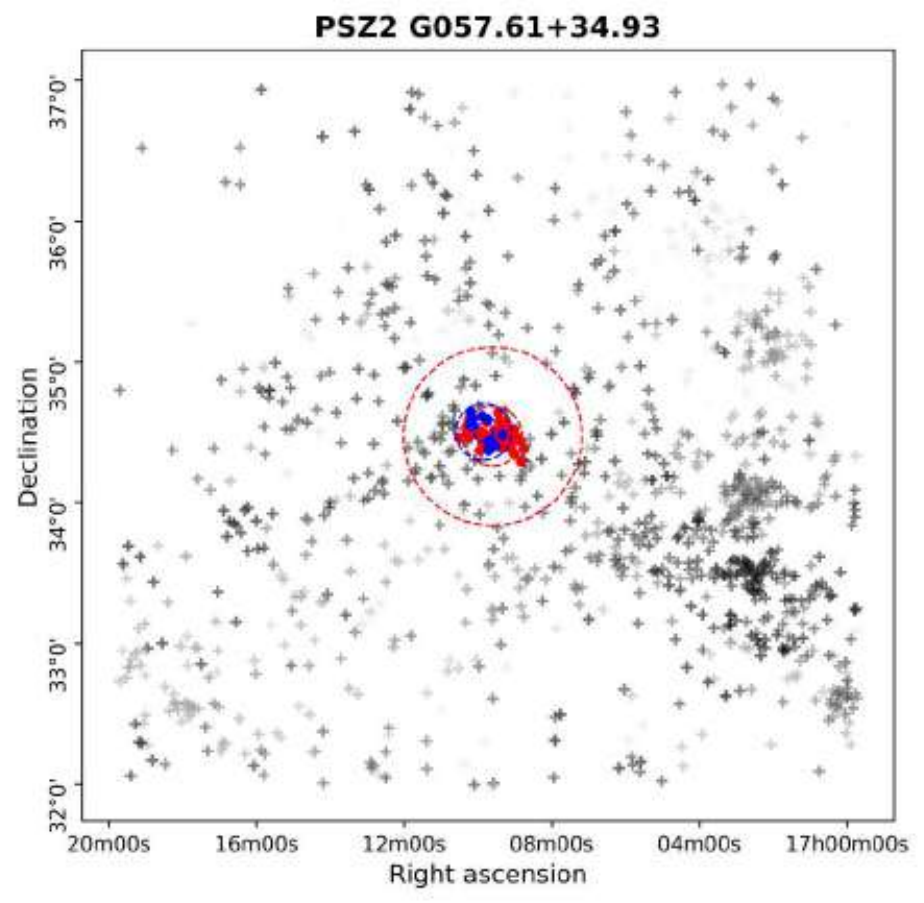}
                \caption{}
        \end{subfigure}
      \begin{subfigure}{0.49\textwidth}
      \centering
                \includegraphics[width=\linewidth]{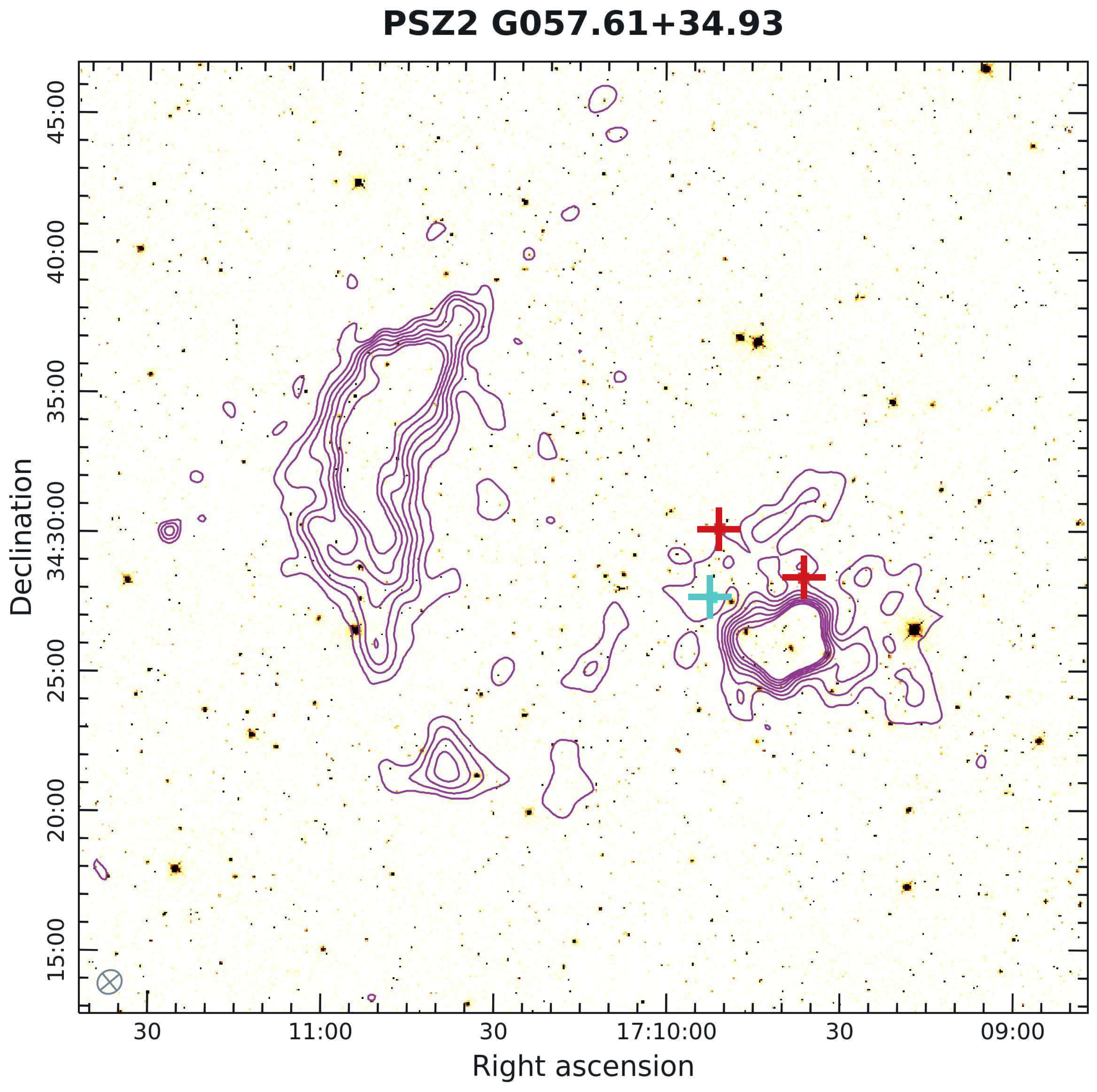}
                \caption{}
        \end{subfigure}
    	\begin{subfigure}{0.49\textwidth}
        	\centering
                \includegraphics[width=\linewidth]{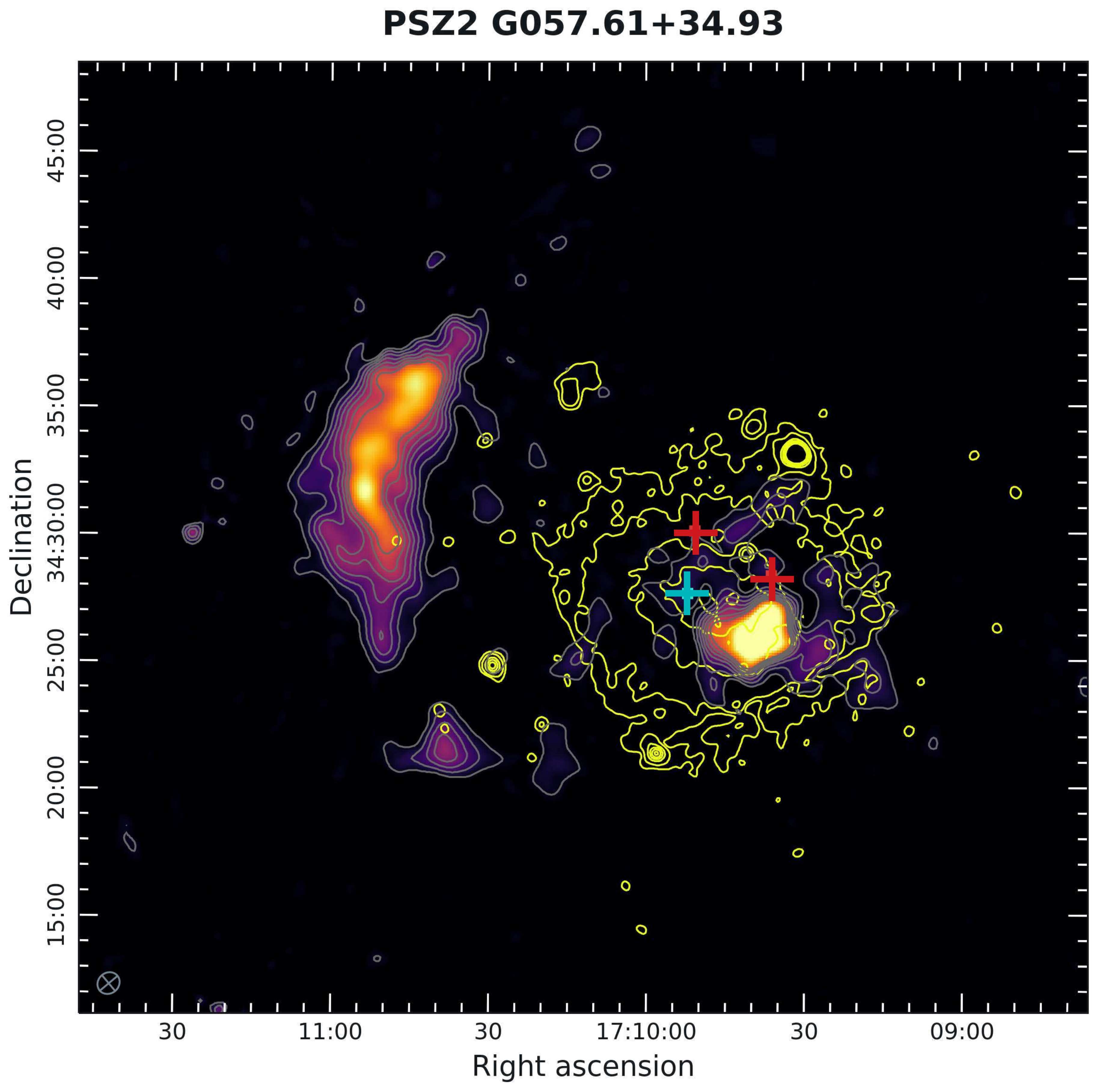}
                \caption{}
        \end{subfigure}
        \begin{subfigure}{0.49\textwidth}
        	\centering
                \includegraphics[width=\linewidth]{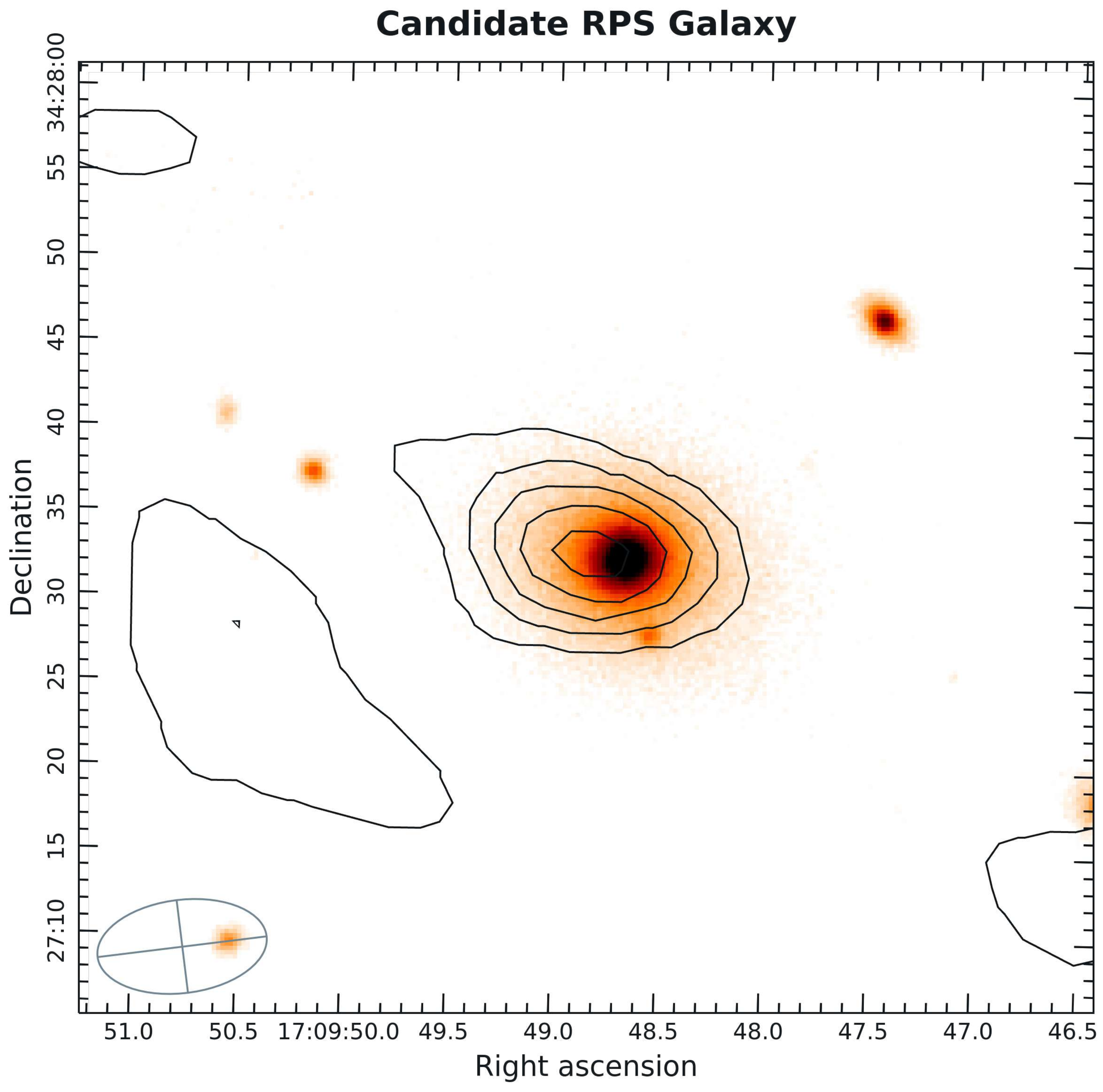}
                \caption{}
        \end{subfigure}

        \caption{\textbf{ Abell 2249 }. Panel~\textbf{(a)}: A plot of SDSS galaxies in the nearby field. Panel~\textbf{(b)}: LoTSS-2 low-resolution image contours at [3, 6, 12, 24, 48, 96] of $\sigma$ where $\sigma = 450 ~\rm{\mu Jy/beam}$ overlaid on DSS r band optical image. Panel~\textbf{(c)}: LoTSS - low-resolution image overlaid with X-ray contours from XMM- Newton 27 ks data. Panel~\textbf{(d)}:: A candidate Ram-pressure stripped galaxy; Pan-STARRS r band optical image overlaid with LoTSS-2 high-resolution image with contour levels of [3,6,9,12] at $\sigma = 80 ~\rm{\mu Jy/beam}$  }\label{fig:M72}
       
\end{figure*}

 This pair of clusters is located at a mean redshift of z = 0.084. Fig. \ref{fig:M72} (a) shows the plot of SDSS galaxies in the nearby field. The interacting system is seen to lie on the galaxy filament spanning roughly NE-SW and is present in a galaxy-rich neighbourhood. \citet{StrubleVelDisp} claim that the cluster has galactic velocity dispersion of $ 548 \rm{km s^{-1}}$ which is moderate, expecting a dynamical activity. \citet{PSZ2} report the cluster mass of $M_500$ $3.73 \times 10^{14}  M_\odot $. According to our analysis, the mass is around $M_500$ $5.2 \times 10^{14}  M_\odot $. The X-ray morphology appears to be clumpy and irregular and slightly extended along the filament which can be seen in panel (a).  The X-ray temperature of $5.837^{+0.376}_{-0.353}$ keV while the X-ray luminosity of  $L_x = 1.772\times 10^{44} \rm{erg}^{-1}$ has been reported by \citet{xraygclustsample} using the data from NASA Chandra observatory. This is suggestive of the fact that this cluster is moderately hot and relatively less luminous.

PSZ2 G057.61+34.93 is a system known for a characteristically large radio relic. \citet{A2249DiffShock} 
have identified this system as a diffusive shock acceleration candidate. The large relic may be from a merger event that happened along this filament. The presence of bright radio sources nearby the centres makes it difficult to confirm the presence of a halo, however, the point source subtracted image does show a diffuse radio emission. The angular extent of the relic is $\sim 834.938 \rm{kpc}$ with a diffuse radio flux density of $208.301 \pm 21.36$ mJy. Fig. \ref{fig:M72} (d) shows a candidate RPS galaxy; a possible indicator of the dynamical cluster environment causing the stripping of galactic gas.

\subsubsection{ PSZ2 G149.22+54.18 (M74)}
\begin{figure}
\centering
        
	 	\begin{subfigure}{\linewidth}
	 	\centering
                \includegraphics[width=0.85\linewidth]{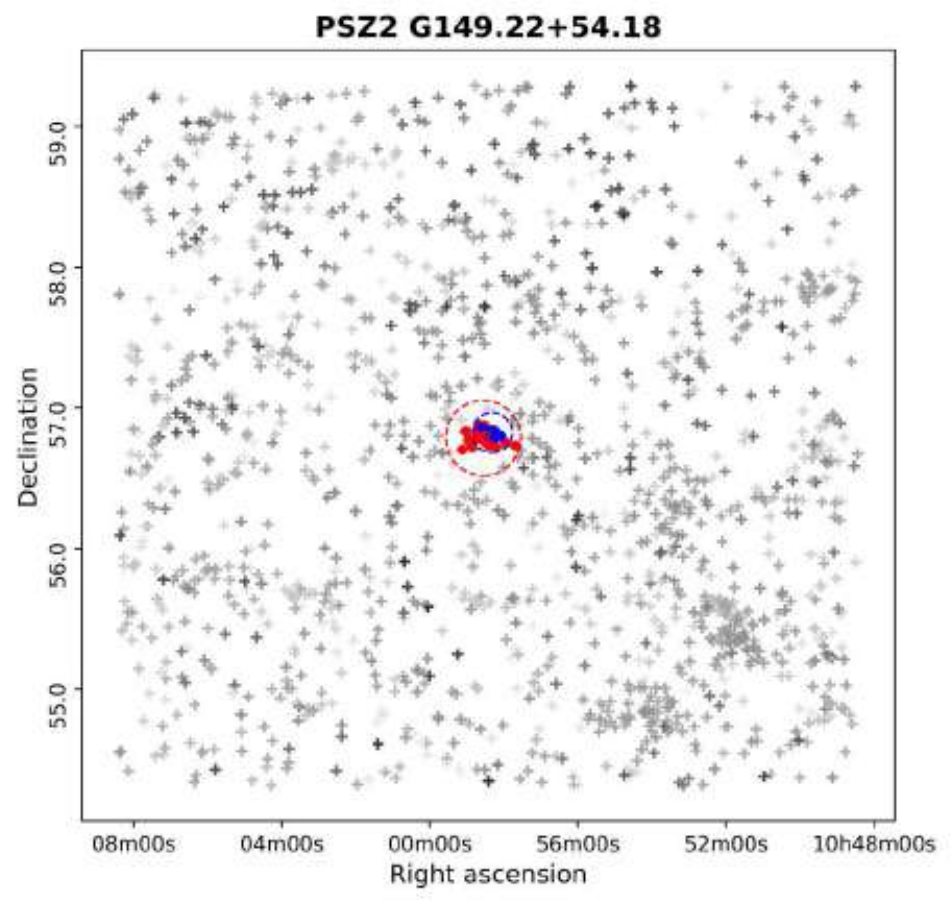}
                \caption{}
        \end{subfigure}
      \begin{subfigure}{\linewidth}
      \centering
                \includegraphics[width=0.75\linewidth]{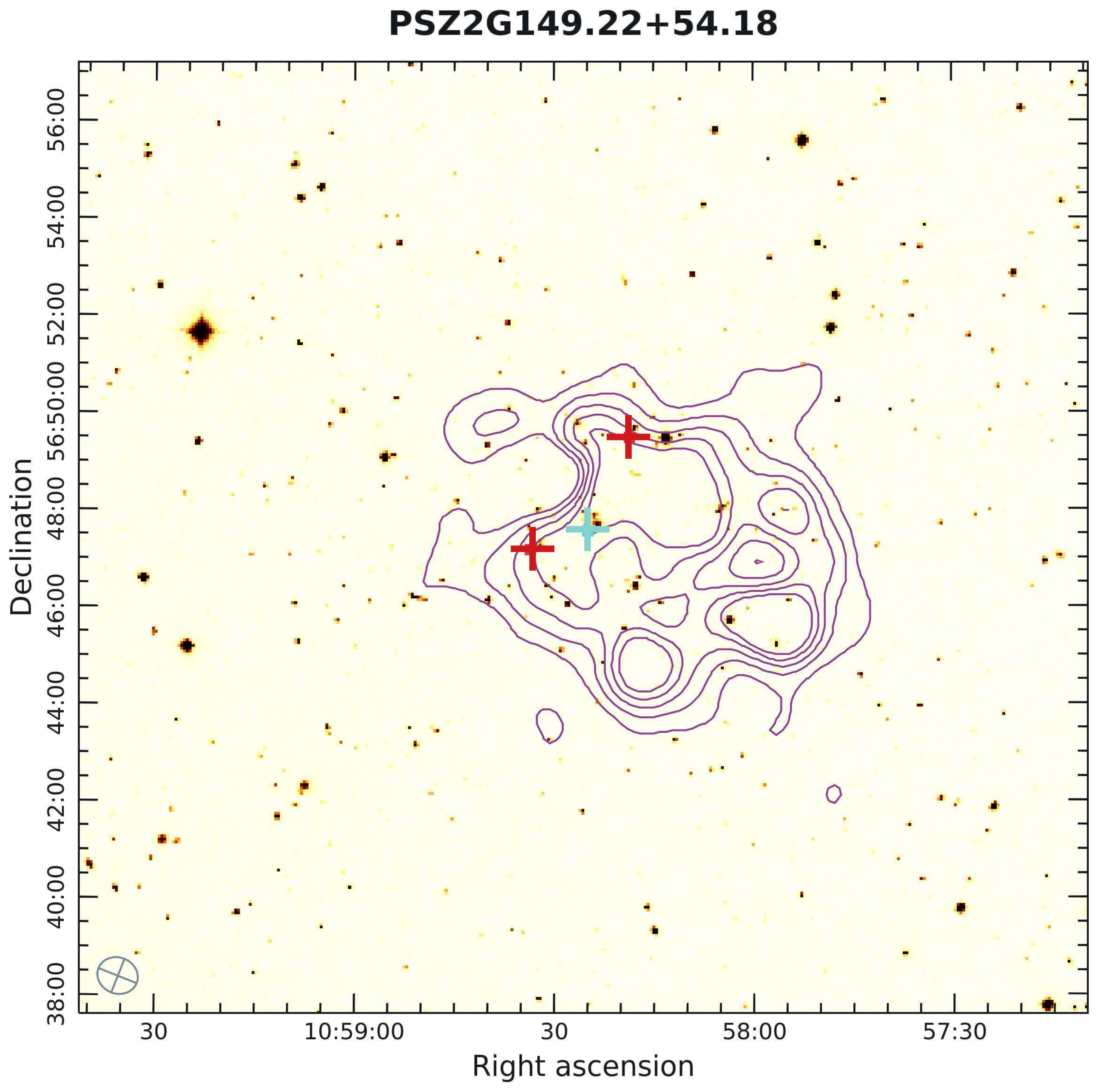}
                \caption{}
        \end{subfigure}
    	\begin{subfigure}{\linewidth}
        	\centering
                \includegraphics[width=0.75\linewidth]{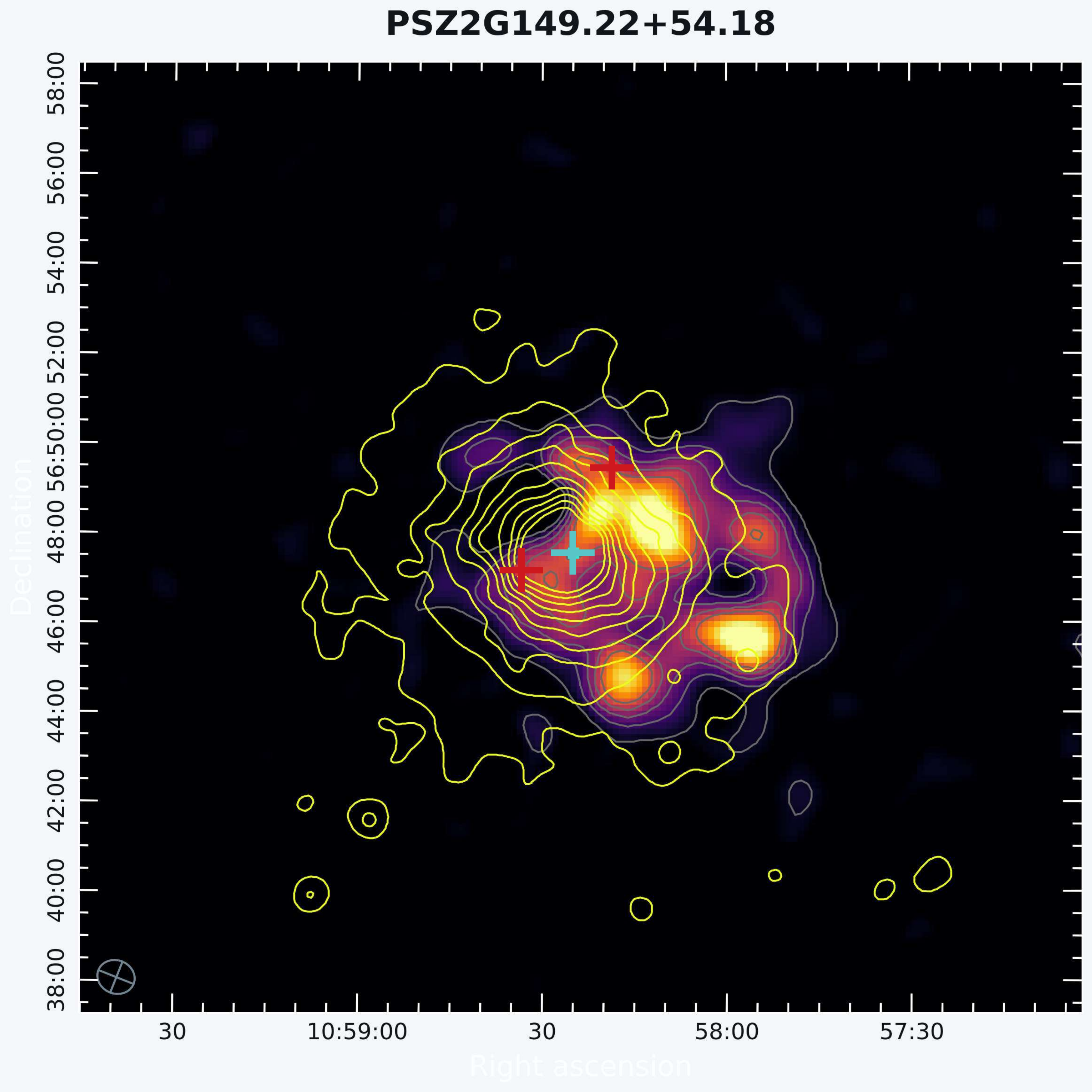}
                \caption{}
        \end{subfigure}

        \caption{\textbf{PSZ2 G149.22+54.18} Panel~\textbf{(a)}: A plot of SDSS galaxies. Panel~\textbf{(b)}: DSS r-band optical image overlaid with LoTSS-2 low resolution image contours at [3,6,9,12,15] of $\sigma$ where $\sigma_{\rm{rms}} = 500~\rm{\mu Jy/beam}$. Panel~\textbf{(c)}: Chandra X-ray contours (yellow) with exposure of 19~ks overlaid on LoTSS-2 point source subtracted image with same contour levels (grey)}\label{fig:M27}
       
\end{figure}

PSZ2 G149.22+54.18 is a couplet of interacting systems as flagged by our algorithm. The merger axis spans roughly along NW-SE. From Fig \ref{fig:M27}\textbf{a}, it can be seen that the system is present in the vicinity of other galaxy clusters that are present to the southwest. The cluster is relatively massive with mass of $5.9^{+0.2}_{-0.2} \times 10^{14}  M_\odot$ according to \citet{PSZ2}. We report the combined NFW  $M_{200}$ of this cluster as $3.866 \times 10^{14}  M_\odot $. \citep{xraygclustsample} have reported the X-ray luminosity of $L_x = 3.504 \times 10^{44} \rm{erg}^{-1}$ and a temperature of $9.937^{+1.282}_{-0.912}$ keV indicating this is a moderately luminous and very hot galaxy cluster. In panel \ref{fig:M27} \textbf{c} it can be seen that the X-ray morphology is roughly symmetrical and not clumpy, and overlaps the diffuse radio emission.\citet{YuanA1132} report an X-ray concentration parameter (c) of $0.1360 \pm 0.0034 $ which is a signature that the cluster is not relaxed (for relaxed clusters c $\sim$ 1). 

The cluster is known to host an ultra-steep spectrum radio halo, \citep{WilberA1132} with a spectral index of $-1.71 \pm 0.19 $. This can also be seen in LoTSS low-resolution images from Fig. \ref{fig:M27} \textbf{a} and \textbf{c}. The halo has a diffuse radio flux density of $ 305.6 \pm 58.7$ mJy. In the context of the connection between merging galaxy clusters and exhibiting radio halo, \citet{cucitimergerhalo} reports that this cluster is undergoing a merger event. A detailed study of radio emissions is performed by \citet{osingaA1132}. The authors also establish the connection with the radio-tailed galaxy towards the south of the cluster, which is also a possible indication of an active cluster medium undergoing interaction.

\subsubsection{PSZ2 G067.17+67.46 (M124)}
\begin{figure}
\centering
        
	 	\begin{subfigure}{\linewidth}
	 	\centering
                \includegraphics[width=0.85\linewidth]{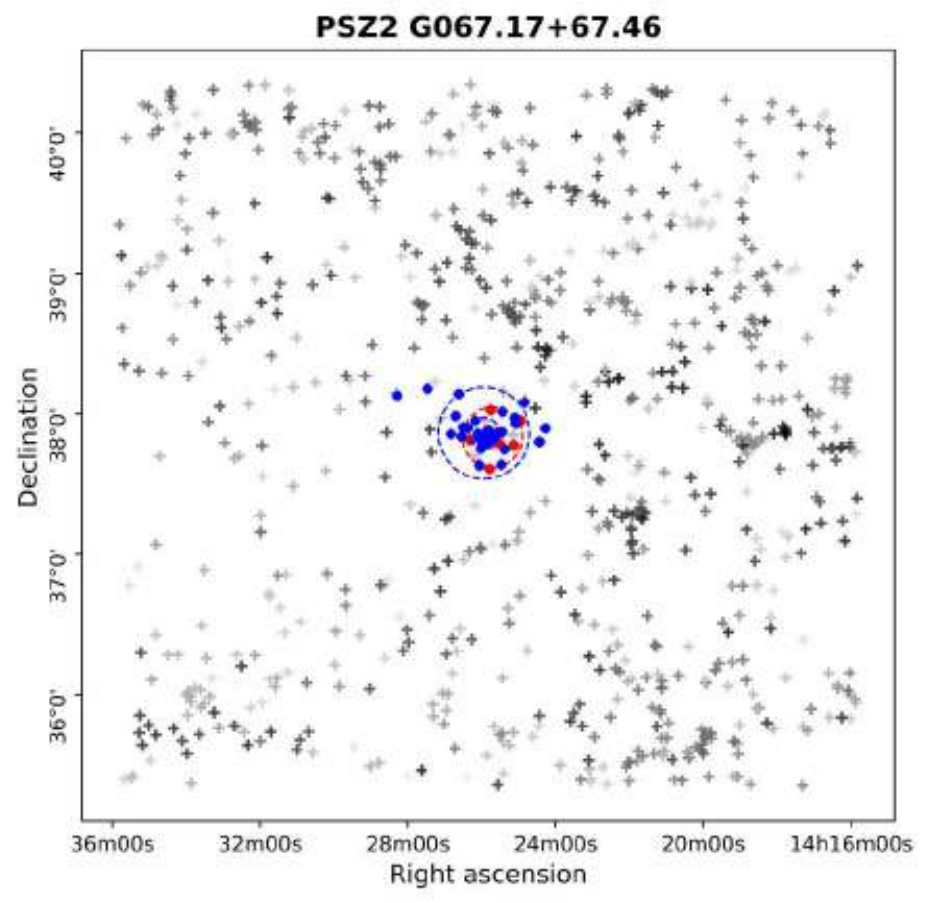}
                \caption{}
        \end{subfigure}
      \begin{subfigure}{\linewidth}
      \centering
                \includegraphics[width=0.75\linewidth]{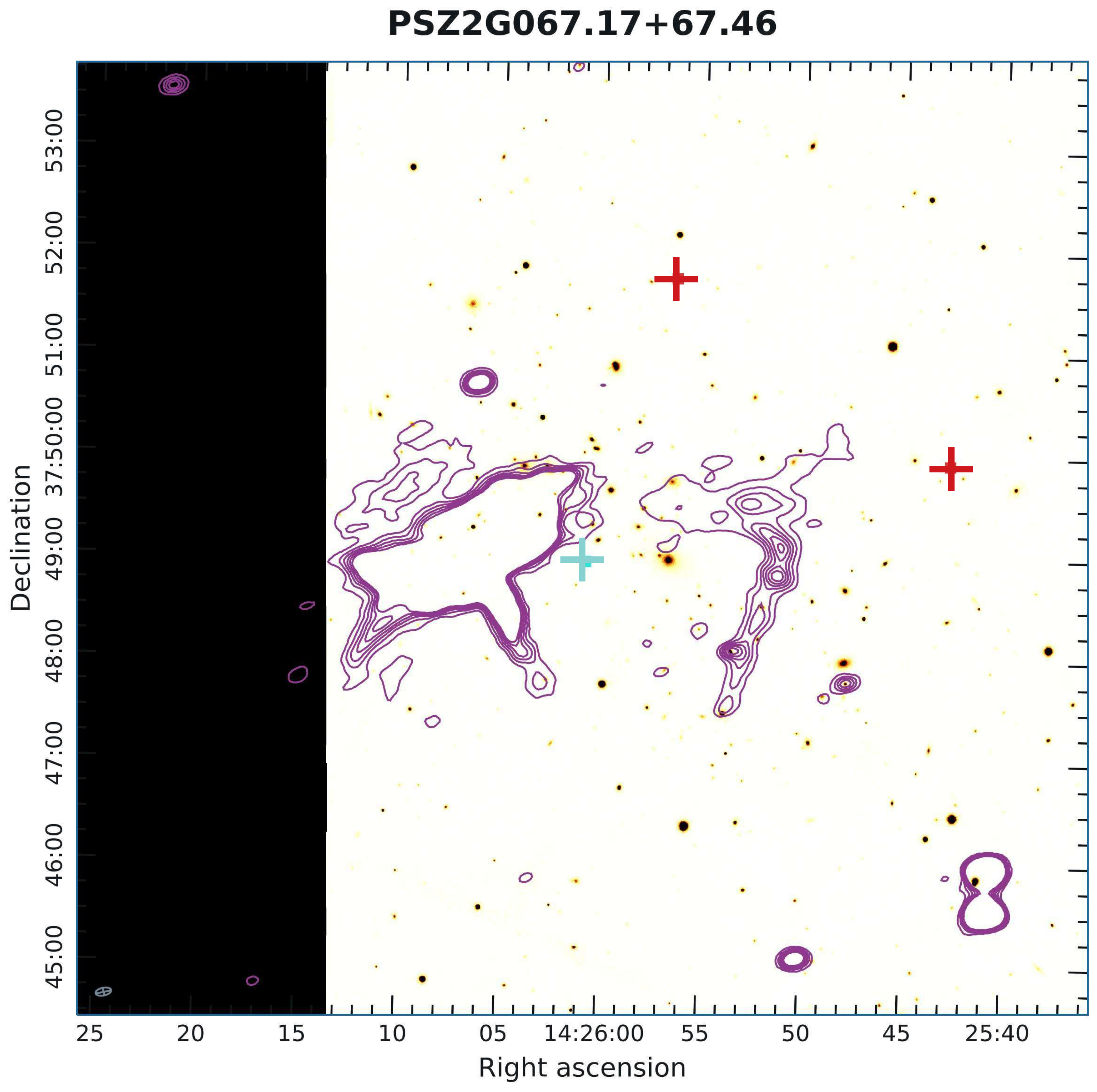}
                \caption{}
        \end{subfigure}
    	\begin{subfigure}{\linewidth}
        	\centering
                \includegraphics[width=0.75\linewidth]{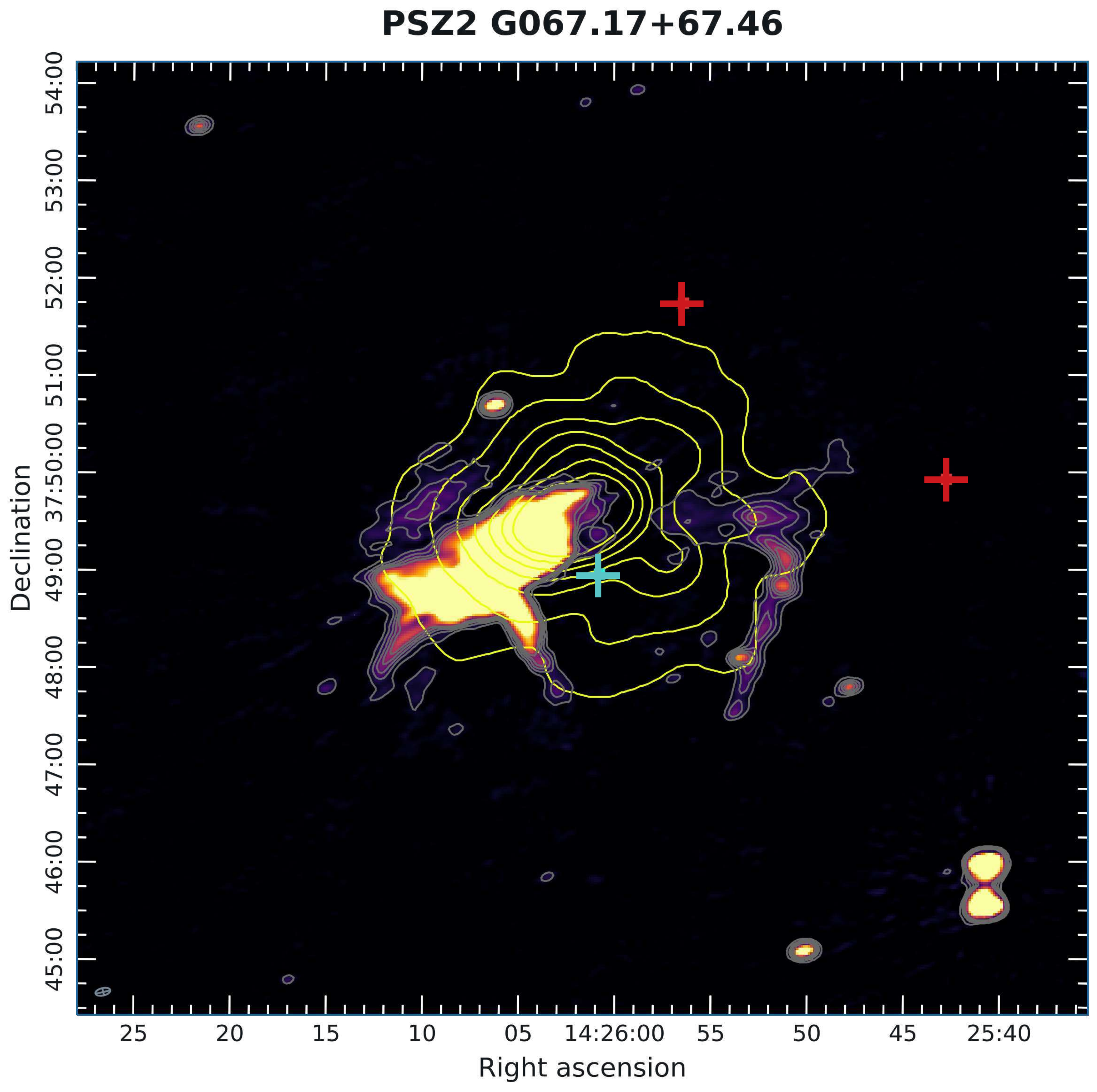}
                \caption{}
        \end{subfigure}

        \caption{\textbf{PSZ2 G067.17+67.46} Panel~\textbf{(a)}: A plot of SDSS galaxies. Panel~\textbf{(b)}: PAN-STARRS r-band optical image overlaid with LoTSS-2 low resolution image contours at [3,6,9,12,15] of $\sigma$ where $\sigma_{\rm{rms}} = 300~\rm{\mu Jy/beam}$. Panel~\textbf{(c)}: Chandra X-ray contours (yellow) with exposure of 28~ks overlaid on LoTSS-2 point source subtracted image with same contour levels (grey)}\label{fig:M56}
       
\end{figure}

 This merging system is located at the mean redshift of z = 0.167. A subtle filamentary structure can be seen just to the west of the identified system, roughly stretching from NE-SW (\ref{fig:M56}). PSZ2 G067.17+67.46 is a well-studied system and known to be dynamically active. \citet{barenaA1914} reported a velocity dispersion of $ \sigma = 1210^{+125}_{-110} \rm km/s$. The cluster is massive (M = $7.3 \pm 0.3 \times 10^{14}  M_\odot$, \citep{PSZ2}), hot ($T_x = 9.055^{+0.368}_{-0.287}$ keV), and highly luminous ($L_x = 9.214\times 10^{44} \rm{erg}^{-1}$) ,\citep{xraygclustsample}. We report the mass of the cluster to be around {$\sim 4 \times 10^{14}  M_\odot$} Although the overall X-ray morphology appears to be symmetric in Fig. \ref{fig:M56} \textbf{b}; there are recent reports of features indicating of dynamical activity in the ICM. \citet{RahmanA1914} report the presence of multiple shock fronts using the data from the {\it Chandra} X-ray observatory. The authors also suggested that there may be multiple merger axes corresponding to axial and equatorial shocks which have Mach numbers in the range of 1.13-1.64. In addition to this, with the help of morphological X-ray parameters, \citep{BuoteA1914} have shown that this cluster is not relaxed.

PSZ2 G067.17+67.46 exhibits complex radio features as seen in Fig. (\ref{fig:M56}) \textbf{c}. A detailed analysis of this diffuse emission is done by \citet{MandalA1914}. The spectrum of diffuse emission is ultra-steep in nature ($ \alpha = -2.17 \pm 0.11$). Earlier thought to be purely a halo, the authors conclude that the diffuse emission is arising from the combination of a radio phoenix candidate, a radio halo candidate and a tailed radio galaxy visible to the west. The eastern emission seen is bright with a flux density of $ 489 \pm 148$ mJy. The tailed galaxy is indicative of mildly relativistic electron injection, i.e. an evidence that this cluster is dynamically active.

\subsubsection{PSZ2 G113.29-29.69 (PPM8)}
\begin{figure}
\centering
        
	 	\begin{subfigure}{\linewidth}
	 	\centering
                \includegraphics[width=0.85\linewidth]{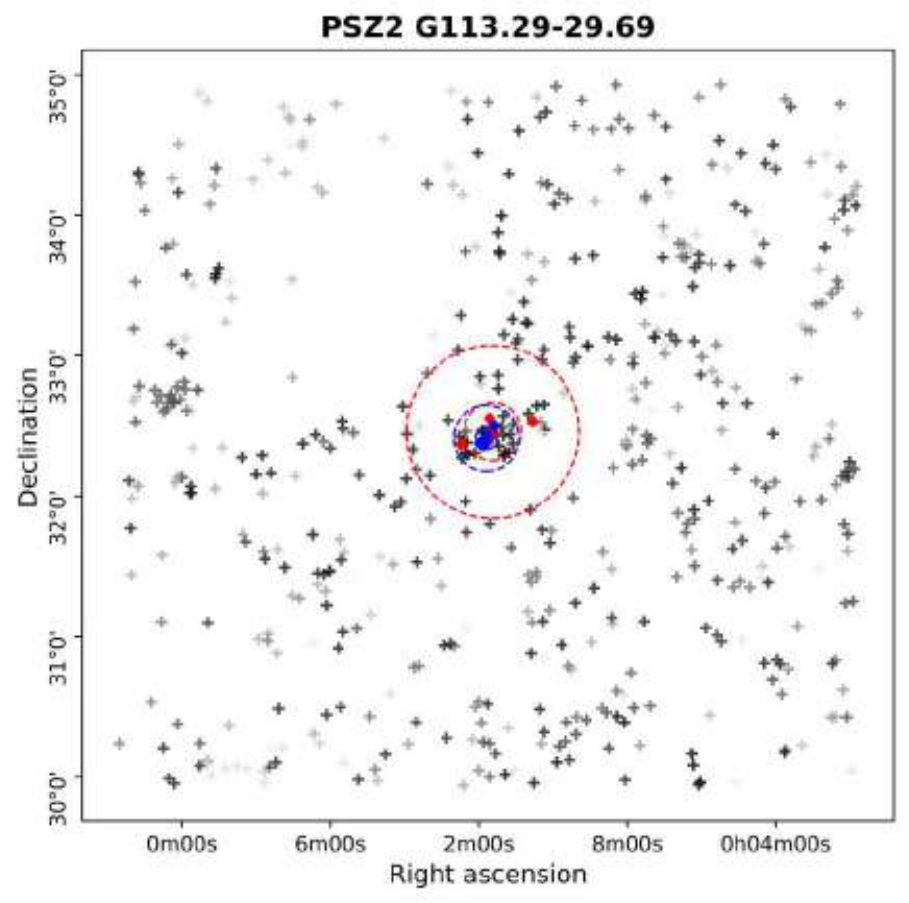}
                \caption{}
        \end{subfigure}
      \begin{subfigure}{\linewidth}
      \centering
                \includegraphics[width=0.75\linewidth]{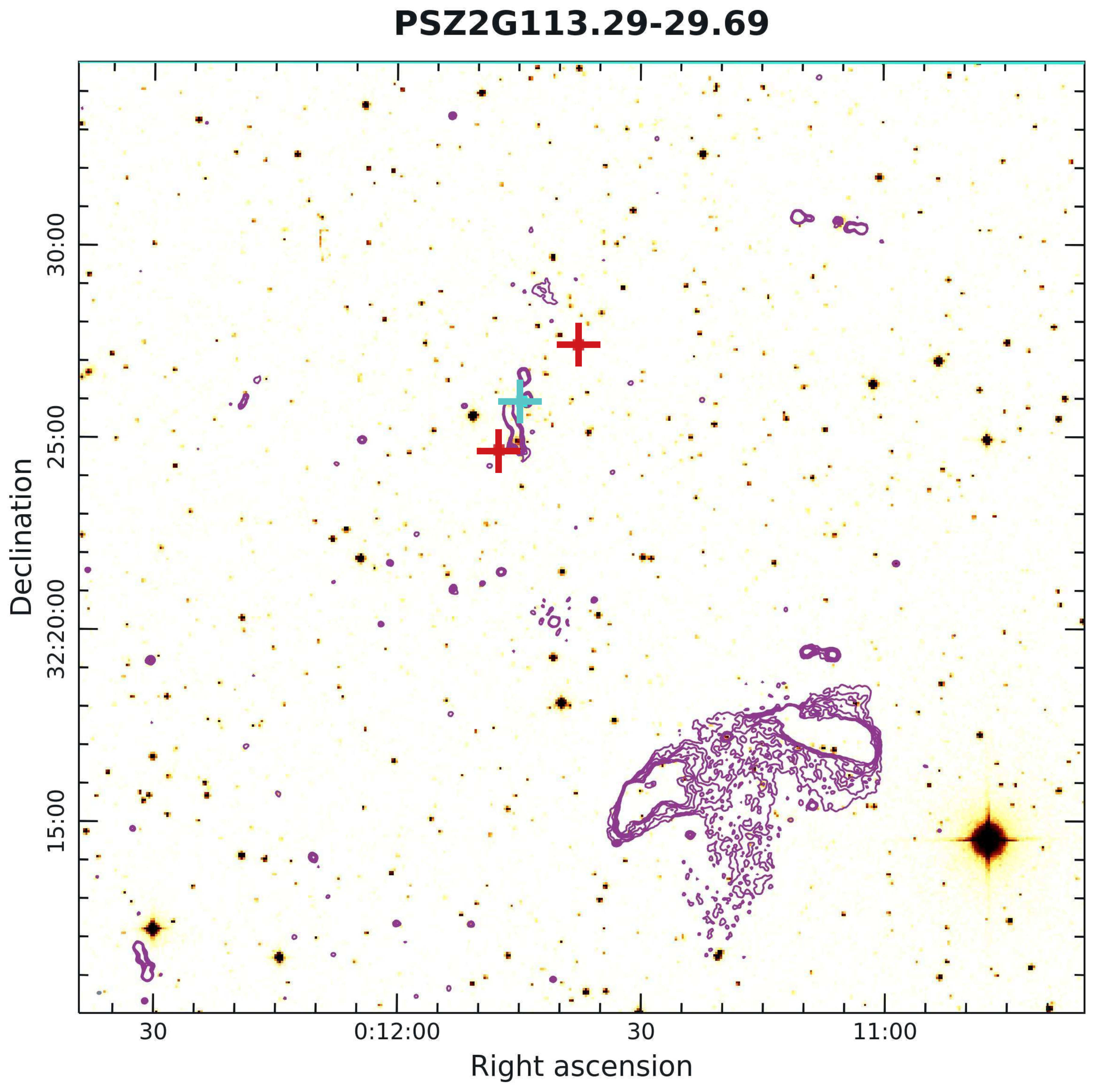}
                \caption{}
        \end{subfigure}
    	\begin{subfigure}{\linewidth}
        	\centering
                \includegraphics[width=0.75\linewidth]{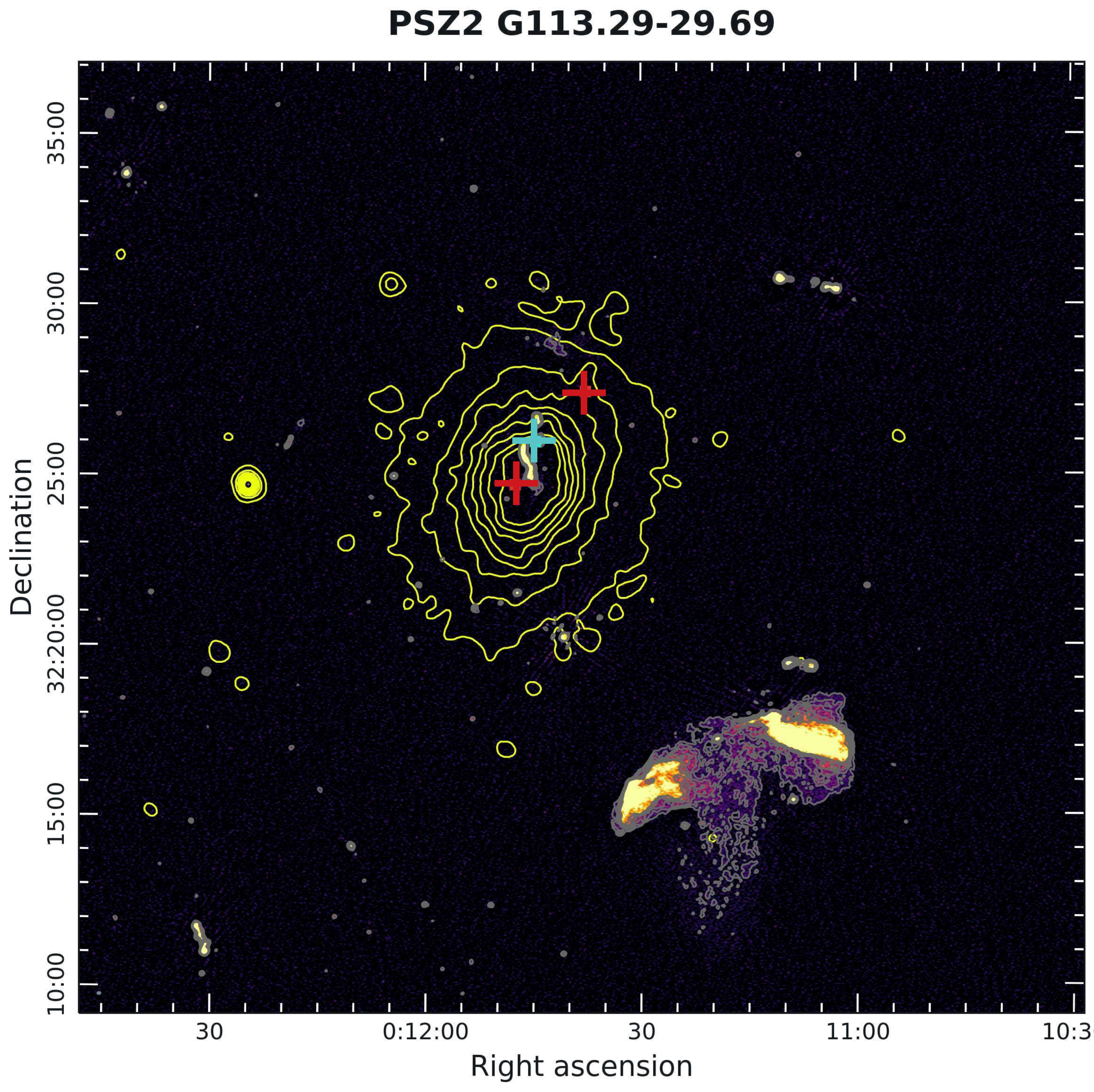}
                \caption{}
        \end{subfigure}

        \caption{\textbf{PSZ2 G113.29-29.69} Panel~\textbf{(a)}: A plot of SDSS galaxies. Panel~\textbf{(b)}: DSS r-band optical image overlaid with LoTSS-2 low resolution image contours at [3,6,9,12,15] of $\sigma$ where $\sigma_{\rm{rms}} = 100~\rm{\mu Jy/beam}$. Panel~\textbf{(c)}: XMM - Newton X-ray contours (yellow) with exposure of 36~ks overlaid on LoTSS-2 point source subtracted image with same contour levels (grey)}\label{fig:PPM6}
       
\end{figure}

Located in a galaxy-rich environment at the mean redshift of $z = 0.010 $, PSZ2 G113.29-29.69 is a merging system (see Fig.~\ref{fig:PPM6}). According to \citet{PSZ2}, the cluster has a mass of $3.71 \times 10^{14}  M_\odot$, has low-to-moderate X-ray luminosity of $L_x = 2.565 \times 10^{44} \rm{erg}^{-1}$; and is moderately hot with an X-ray temperature of $T_x = 4.674^{+0.389}_{-0.321} \rm{keV}$ \citep{xraygclustsample}. However, we estimate the upper limit on the cluster mass of $7.6 \times 10^{14}  M_\odot$.  The X-ray morphology appears to be symmetric, and mildly clumpy as seen in Fig. \ref{fig:PPM6} \textbf{c}. \citet{YuanDynamicalStates} report a relatively high centroid shift parameter ($ log_{10}w = -2.47 \pm 0.02 $), computed from Chandra X-ray observations, which possibly indicates that cluster may not be relaxed. Additionally, the authors also report an X-ray delta-parameter 
of $0.42 \pm 0.01 $, which, being positive, also points towards a non-relaxed dynamical state.

To the southwest of the flagged system, a disproportionately large diffuse radio emission can be seen (\ref{fig:PPM6} \textbf{c}). The flux density of this emission is high at $ 1741 \pm 178 $ mJy.  This emission appears generally bimodal, with an optical galaxy signature present at the centre, which indicates a radio AGN; however, the radio lobes also exhibit a disturbed structure, emanating to the south, in a roughly perpendicular direction to the AGN axis. This might be due to the dynamically active state of the cluster. Additionally, at the cluster centre, another diffuse source can be seen. \citet{PlanckLotssdr2} do not classify this into any category, however, from the morphology, it appears to be a tailed radio galaxy infalling towards the cluster centre.

\subsubsection{Abell 0084 (M7)}
\begin{figure}
\centering
        
	 	\begin{subfigure}{\linewidth}
	 	\centering
                \includegraphics[width=0.85\linewidth]{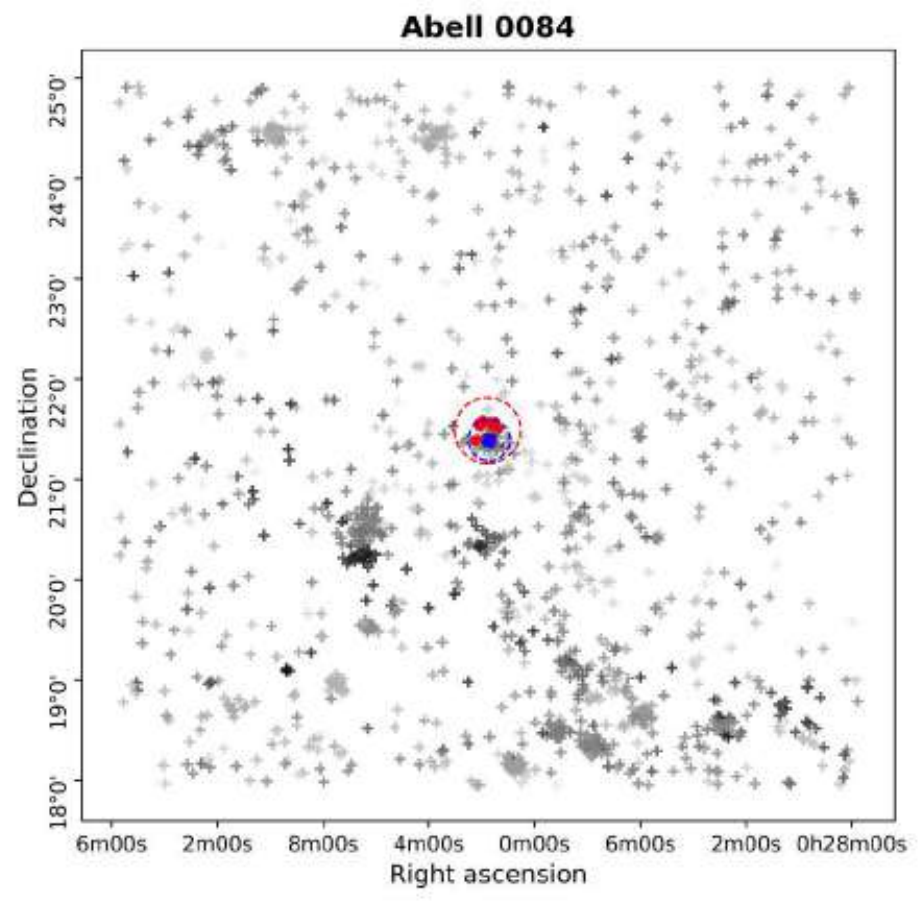}
                \caption{}
        \end{subfigure}
      \begin{subfigure}{\linewidth}
      \centering
                \includegraphics[width=0.75\linewidth]{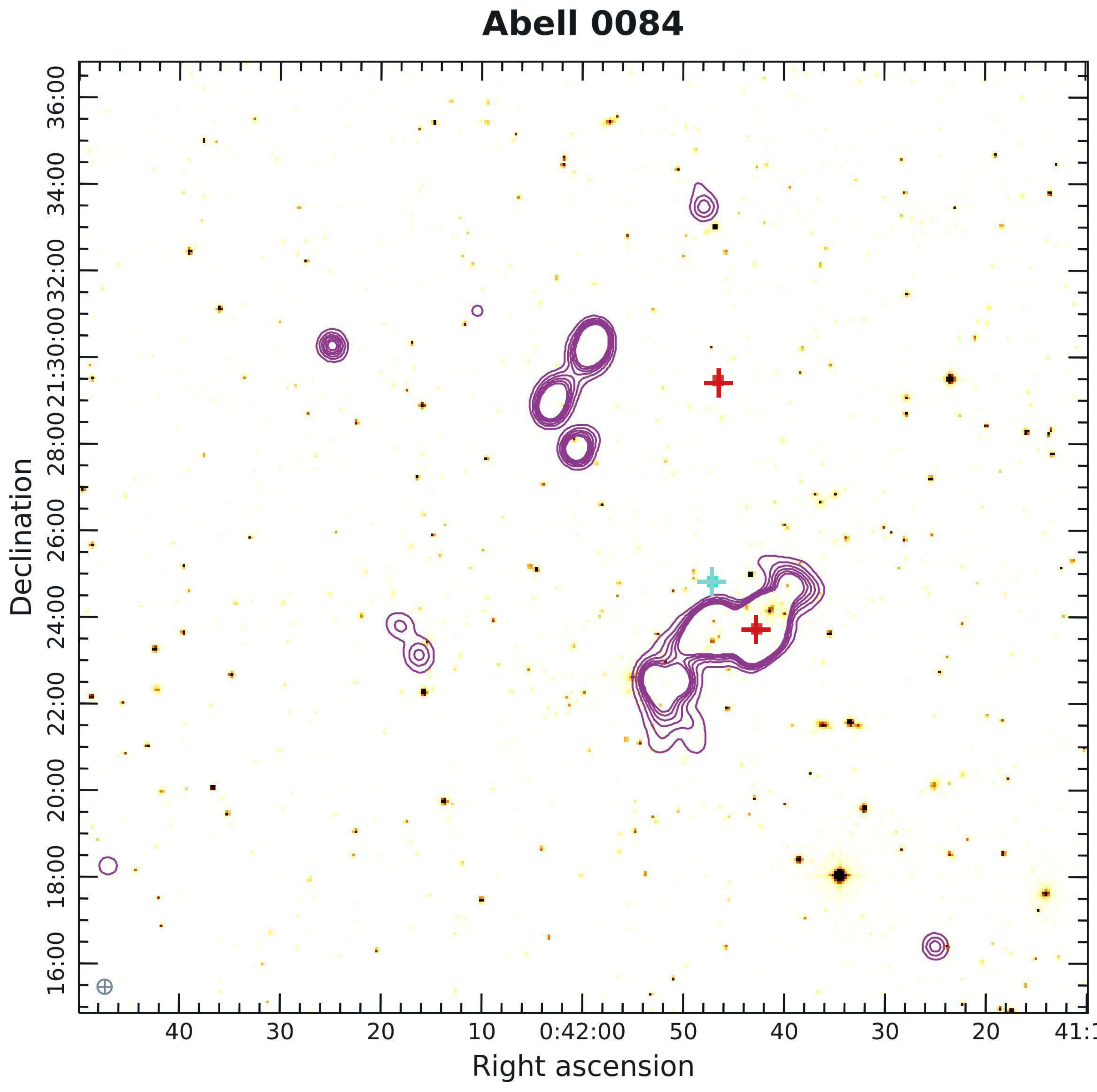}
                \caption{}
        \end{subfigure}
    	\begin{subfigure}{\linewidth}
        	\centering
                \includegraphics[width=0.75\linewidth]{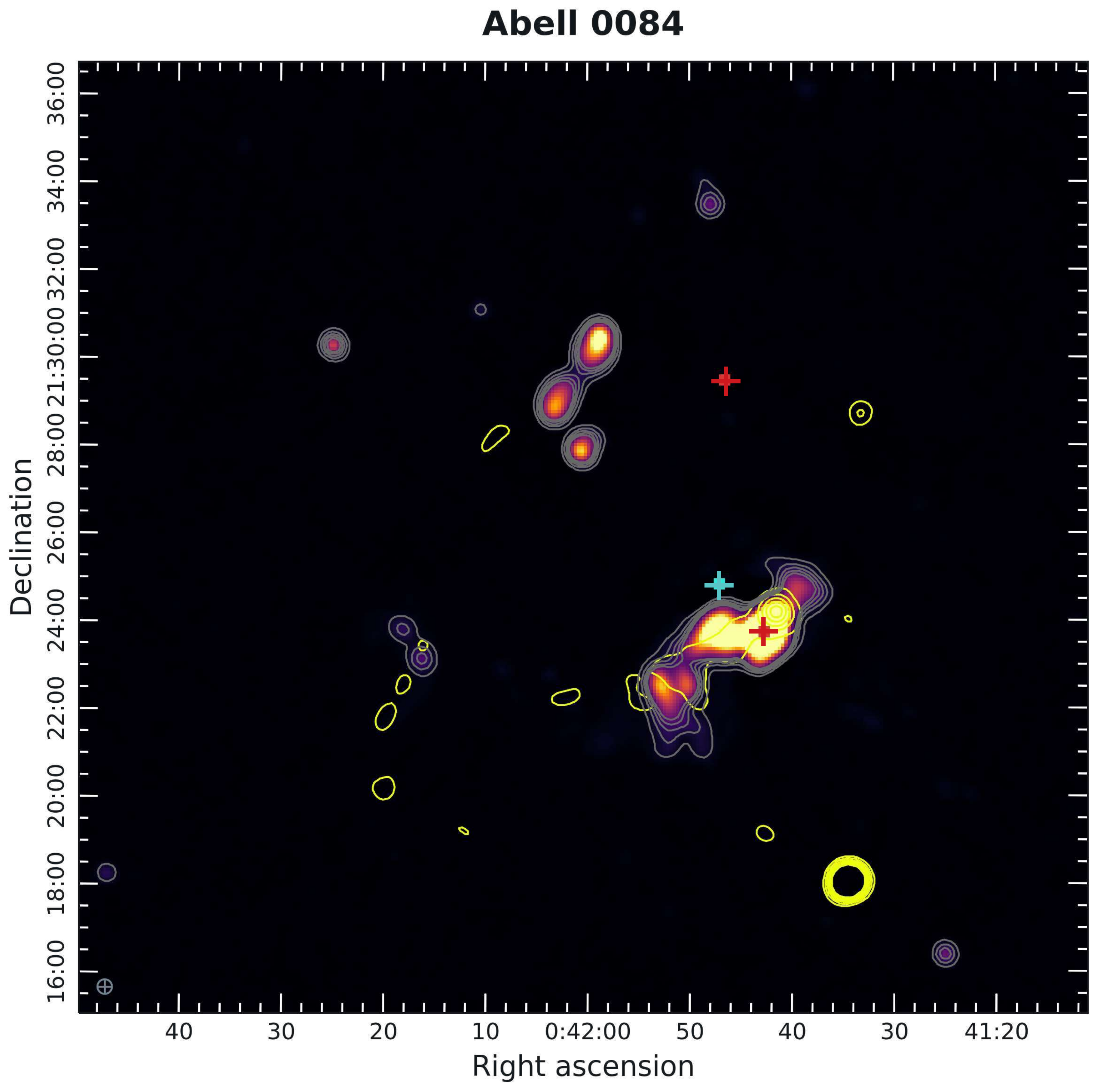}
                \caption{}
        \end{subfigure}

        \caption{\textbf{Abell 0084} Panel~\textbf{(a)}: A plot of SDSS galaxies. Panel~\textbf{(b)}: DSS r-band optical image overlaid with LoTSS-2 low resolution image contours at [3,6,9,12,15,18] of $\sigma$ where $\sigma_{\rm{rms}} = 150~\rm{\mu Jy/beam}$. Panel~\textbf{(c)}: XMM - Newton X-ray contours (yellow) with exposure of 8~ks overlaid on LoTSS-2 point source subtracted image with same contour levels (grey)}\label{fig:M3}
       
\end{figure}

 Abell 84 can be seen to lie in the vicinity of another cluster present to the southeast (\ref{fig:M3} \textbf{a}). This pair of interacting systems is located at the mean redshift of $ z = 0.103 $. We have computed the mass of this cluster as $1.99 \times 10^{14} M_\odot$ which suggests that this cluster is relatively small. \citet{StrazzulloA84} have reported the X-ray luminosity of $L_x = 1.82 \times 10^{44} \rm{erg}^{-1}$ in BAX 0.1-2.4 KeV band. This indicates that the cluster has low to moderate luminosity. The cluster does not show significant X-ray emission from ICM which may be due to its low mass. 

The radio features, as seen in \ref{fig:M3}\textbf{a}, appear generally bimodal, with diffuse emission present to the SW of the image in the LoTSS low-resolution images. The upper limit on the flux density of this emission is around $ 4339 \pm 86.80 $~mJy. \citet{NonPlanckLotss} reported this emission as a candidate radio halo. Deeper radio and X-ray observations are needed to fully confirm the origin and nature of this diffuse emission.

\subsubsection{Abell 1738 (M107)}
\begin{figure}
\centering
        
	 	\begin{subfigure}{\linewidth}
	 	\centering
                \includegraphics[width=0.85\linewidth]{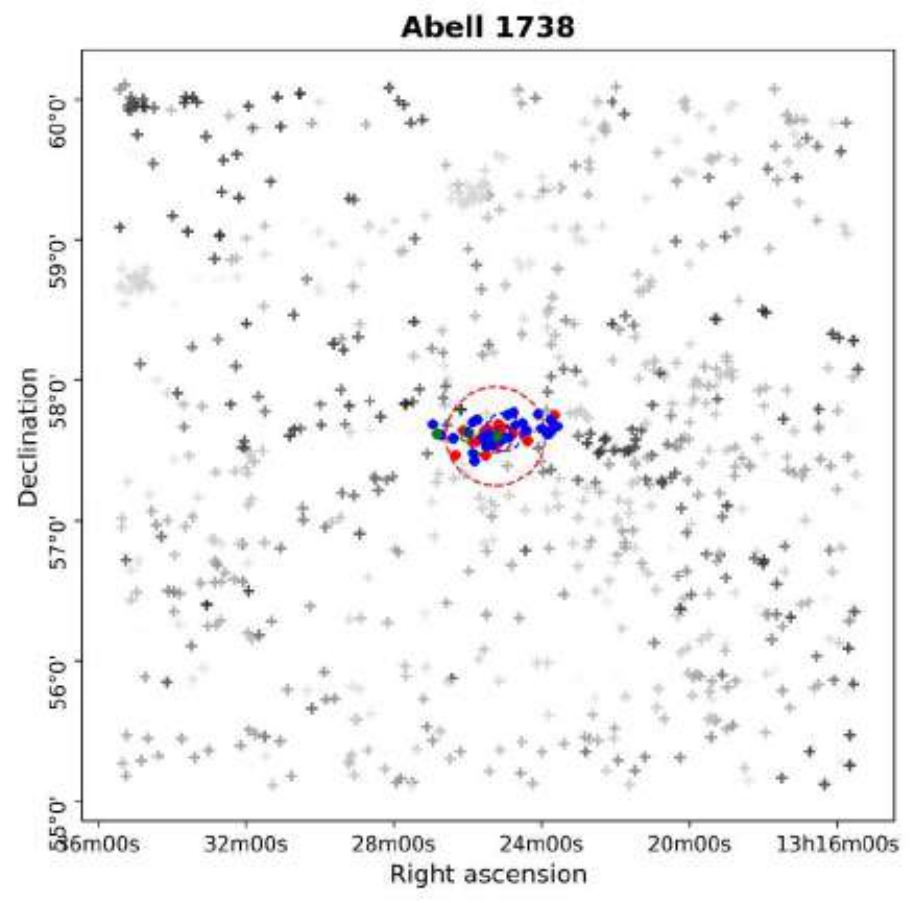}
                \caption{}
        \end{subfigure}
      \begin{subfigure}{\linewidth}
      \centering
                \includegraphics[width=0.75\linewidth]{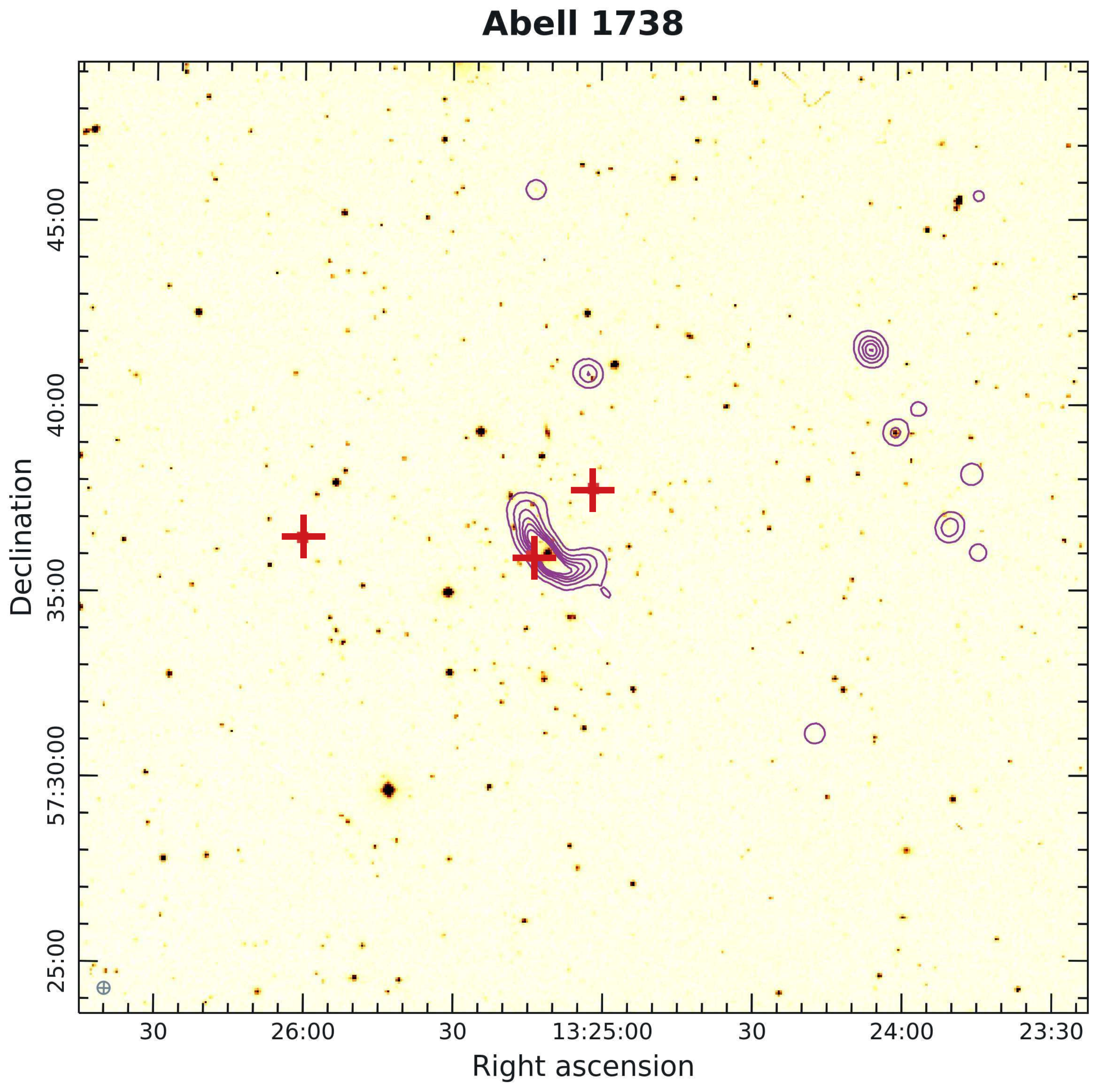}
                \caption{}
        \end{subfigure}
    	\begin{subfigure}{\linewidth}
        	\centering
                \includegraphics[width=0.75\linewidth]{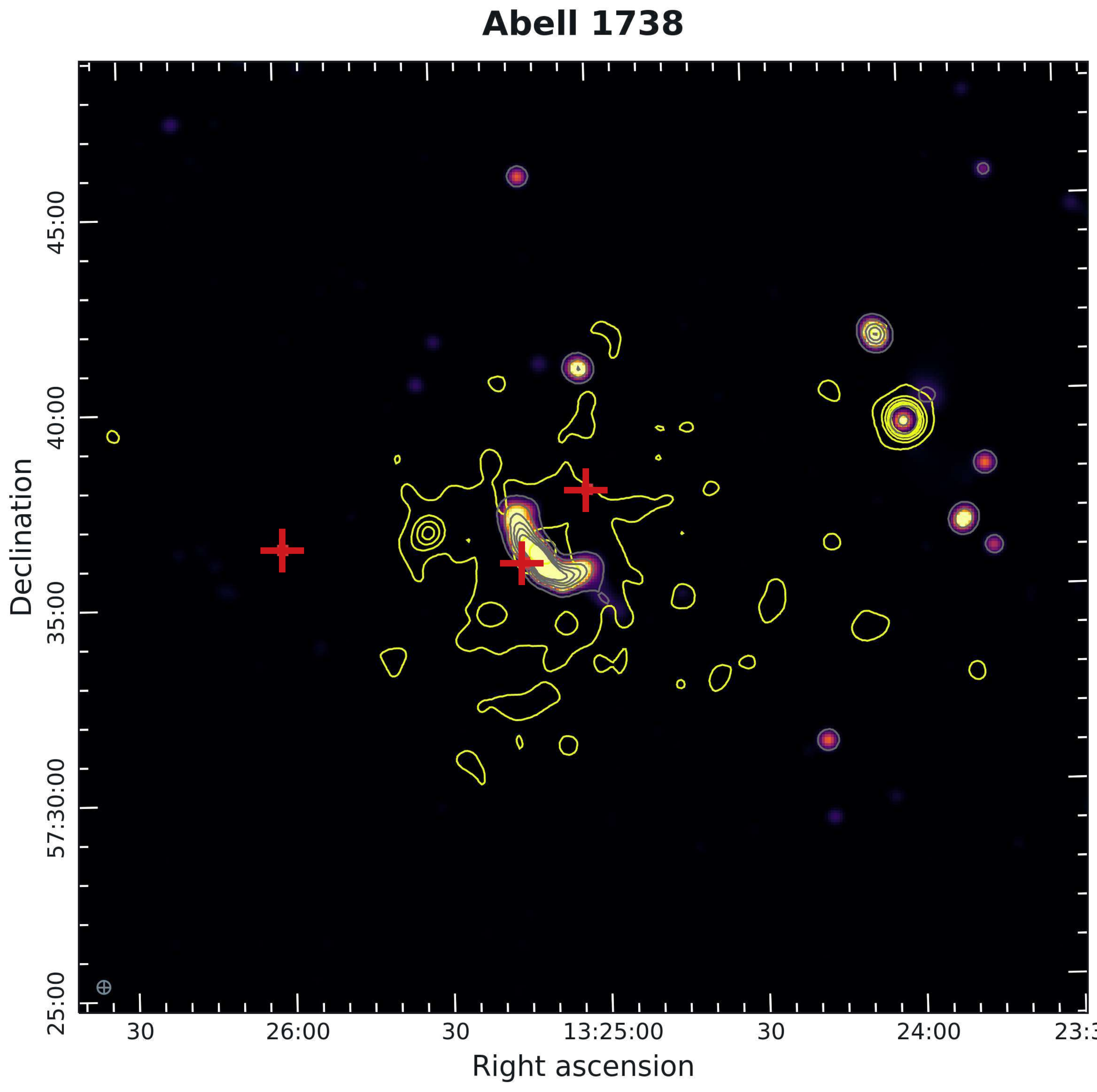}
                \caption{}
        \end{subfigure}

        \caption{\textbf{Abell 1738} Panel~\textbf{(a)}: A plot of SDSS galaxies. Panel~\textbf{(b)}: DSS r-band optical image overlaid with LoTSS-2 low resolution image contours at [3,6,9,12,15] of $\sigma$ where $\sigma_{\rm{rms}} = 6~\rm{mJy/beam}$. Panel~\textbf{(c)}: XMM - Newton X-ray contours (yellow) with exposure of 14~ks overlaid on LoTSS-2 point source subtracted image with same contour levels (grey)}\label{fig:M53}
       
\end{figure}

 A filament-like structure can be seen originating from the Northeast to the Southwest in Fig.~\ref{fig:M53}\textbf{a}. This is a triplet of merging system located at the mean redshift of 0.115. \citet{PSZ2} reported the cluster mass as $2.1 \times 10^{14} M_\odot$ which is same as our algorithm has estimated. The literature on this cluster is fairly recent and limited.  The X-ray morphology appears to be irregular and clumpy. Near the centre, the X-ray contours show a tail-like feature, which may be the hot matter originating from the galactic centre. 

The central diffuse radio emission has an upper limit on the flux density of $ 1712 \pm 34 $~mJy. From the morphology of the diffuse emission, it appears to be originating from a galaxy, which is supported by the optical signature in Fig. \ref{fig:M53}\textbf{b}. This may also be the case of Ram pressure stripping arising from the disturbed cluster medium. The direction of galactic material emanating due to this ram pressure is consistent with the centres identified in this work.  

\subsubsection{ WHLJ1007006.4+354041 (PPM29)} 
\begin{figure}
\centering
        
	 	\begin{subfigure}{\linewidth}
	 	\centering
                \includegraphics[width=0.85\linewidth]{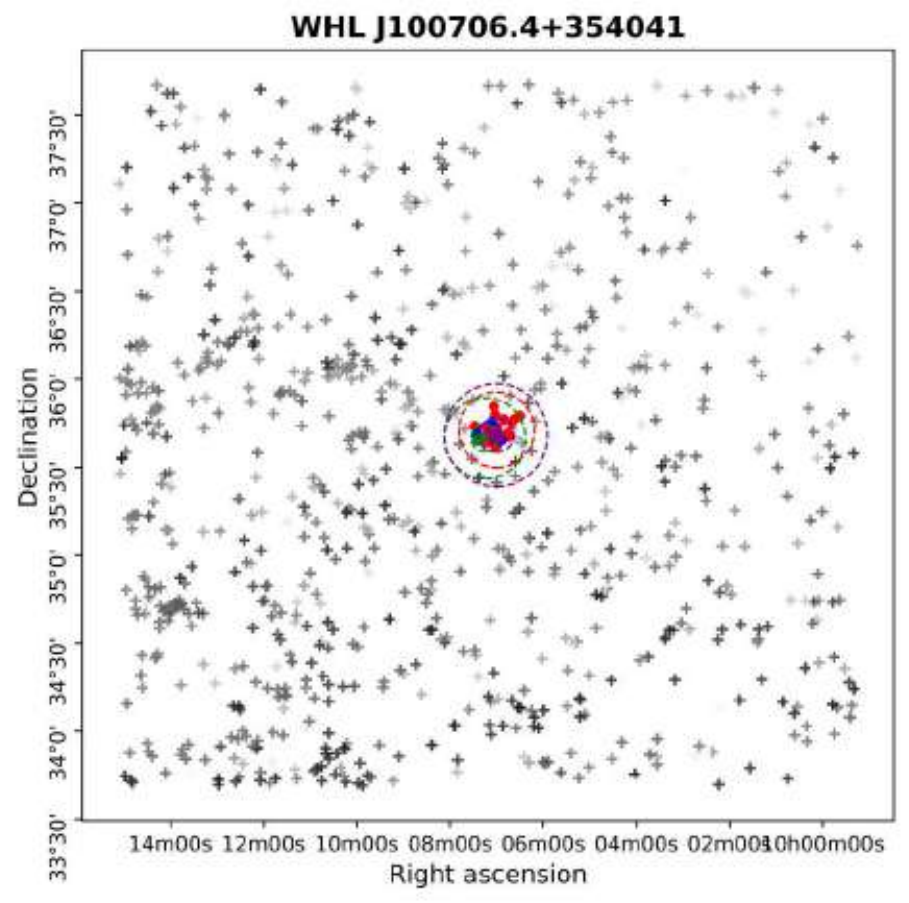}
                \caption{}
        \end{subfigure}
      \begin{subfigure}{\linewidth}
      \centering
                \includegraphics[width=0.75\linewidth]{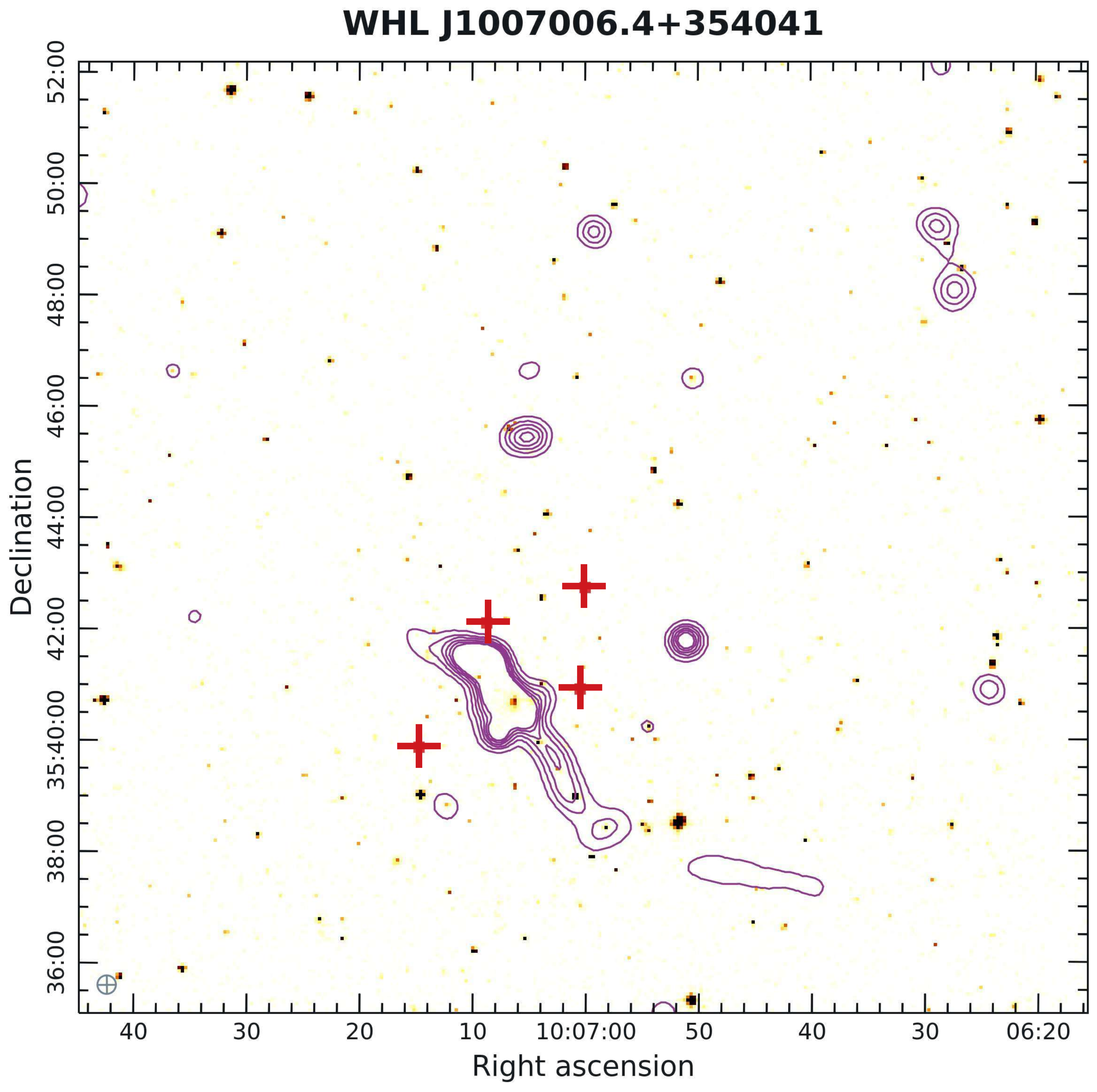}
                \caption{}
        \end{subfigure}
    	\begin{subfigure}{\linewidth}
        	\centering
                \includegraphics[width=0.75\linewidth]{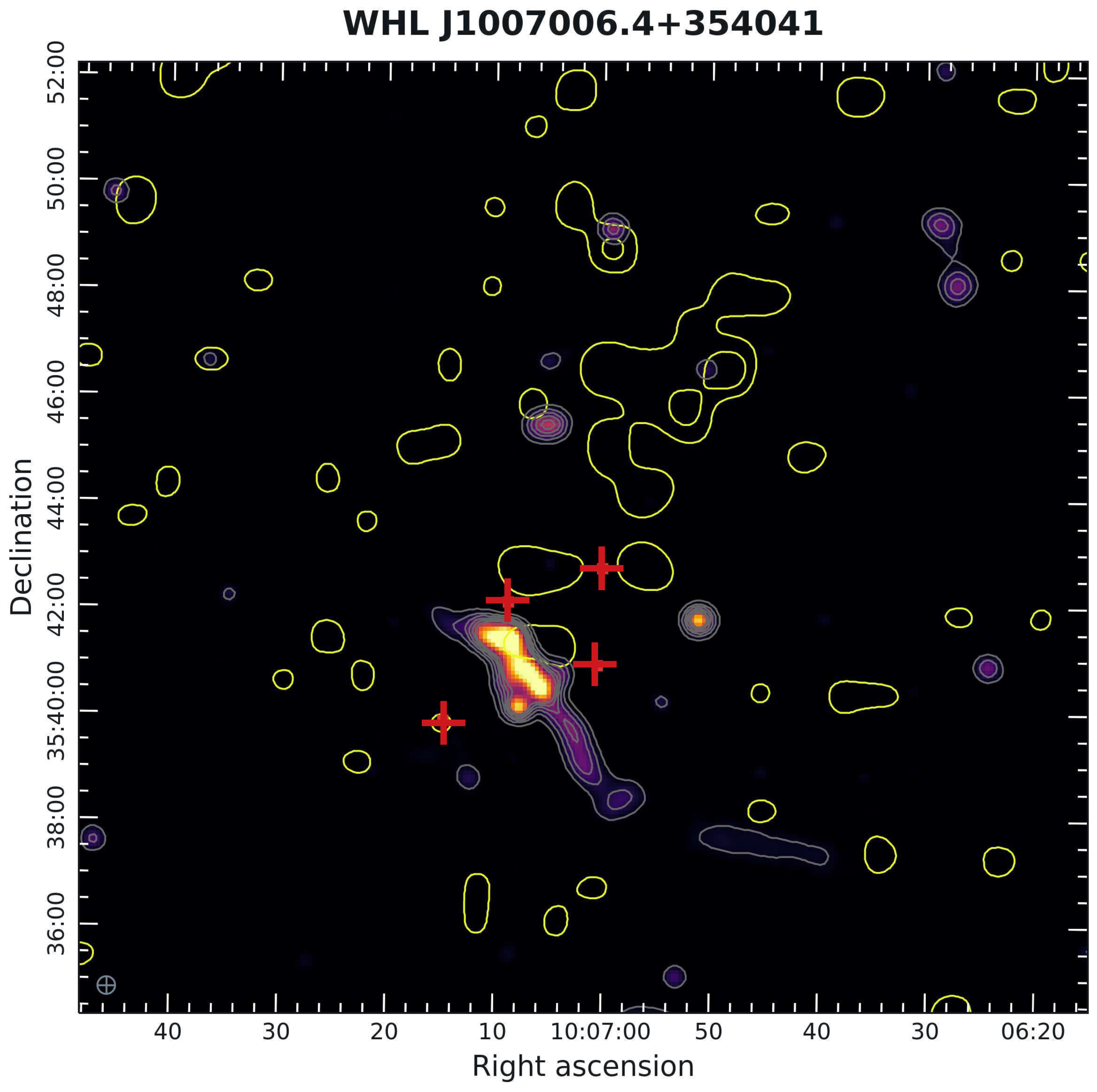}
                \caption{}
        \end{subfigure}

        \caption{\textbf{WHL1007006.4+354041} Panel~\textbf{(a)}: A plot of SDSS galaxies. Panel~\textbf{(b)}: DSS r-band optical image overlaid with LoTSS-2 low resolution image contours at [3,6,9,12,15,18] of $\sigma$ where $\sigma_{\rm{rms}} = 400~\rm{\mu Jy/beam}$. Panel~\textbf{(c)}: ROSAT X-ray contours (yellow) with exposure of 19~ks overlaid on LoTSS-2 point source subtracted image with same contour levels (grey)}\label{fig:PPM29}
       
\end{figure}

 A system of 4 interacting systems located at a mean redshift of $z = 0.14 $ has been flagged by our algorithm. From Fig.~\ref{fig:PPM29}\textbf{a}, it can be seen that the system is located in a galaxy-rich environment. We report the upper limit for the total mass of this system to be around $3.842 \times 10^{14} M_\odot$. ROSAT contours in Fig.~\ref{fig:PPM29}\textbf{a} show an elongated and clumpy morphology, however for accurate interpretation, detailed X-ray maps would be needed.

The system shows a diffuse radio source which does not overlap with X-ray contours. The larger elongation of this source appears to be perpendicular to the merger axis (NW-SE). All of these features primarily indicate that this could be a candidate relic. However, several optical signatures are present within the radio contours as seen in Fig.~\ref{fig:PPM29}\textbf{b}, suggesting that part of this emission could also be from galaxies. A dedicated multiwavelength study is needed to confirm the origin and nature of this source.    

\subsubsection{Abell 0021 -  PSZ2 G114.90-34.35 } (PPM10)

\begin{figure}
\centering
        
	 	\begin{subfigure}{0.8\linewidth}
	 	\centering
                \includegraphics[width=\linewidth]{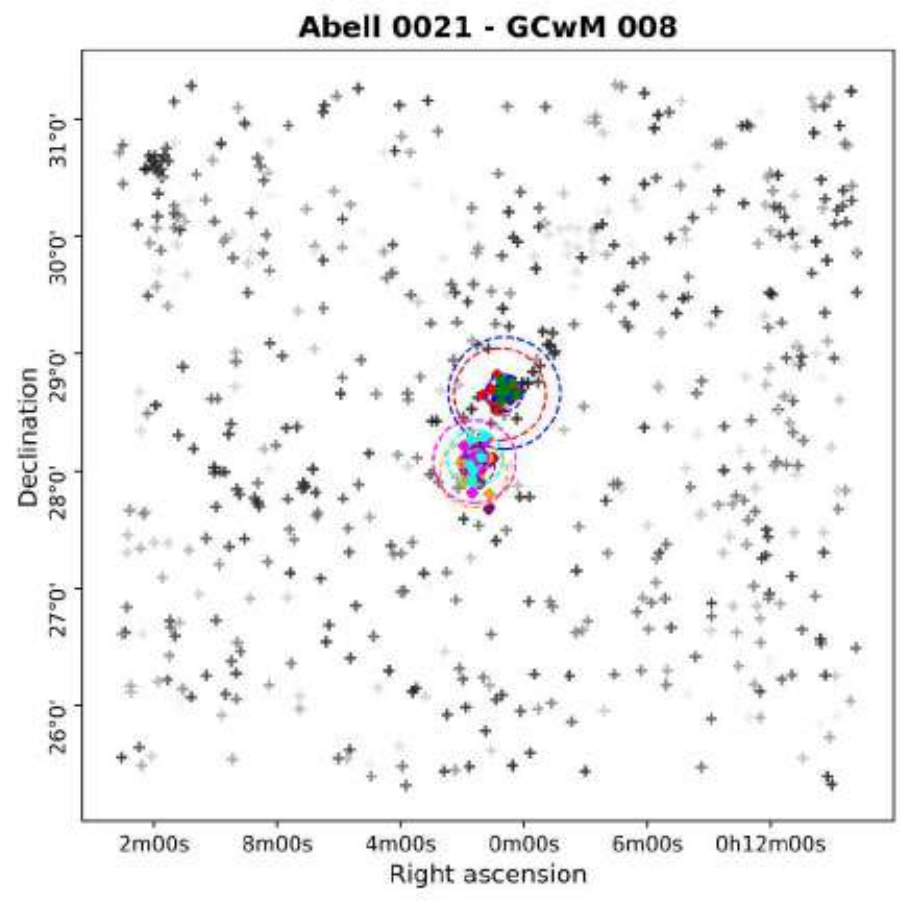}
                \caption{}
        \end{subfigure}
      \begin{subfigure}{0.8\linewidth}
      \centering
                \includegraphics[width=\linewidth]{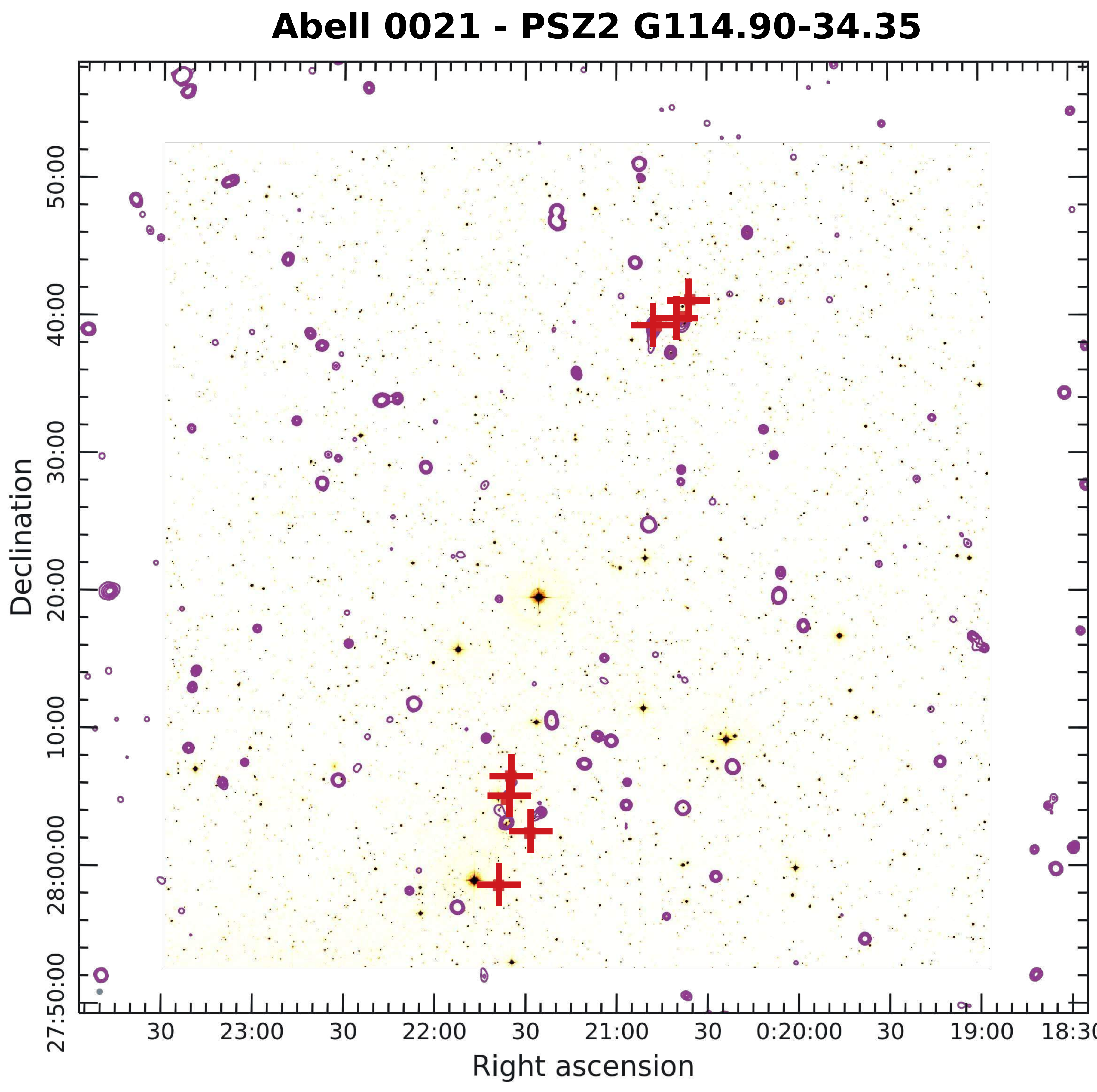}
                \caption{}
        \end{subfigure}
    	\begin{subfigure}{0.8\linewidth}
        	\centering
                \includegraphics[width=\linewidth]{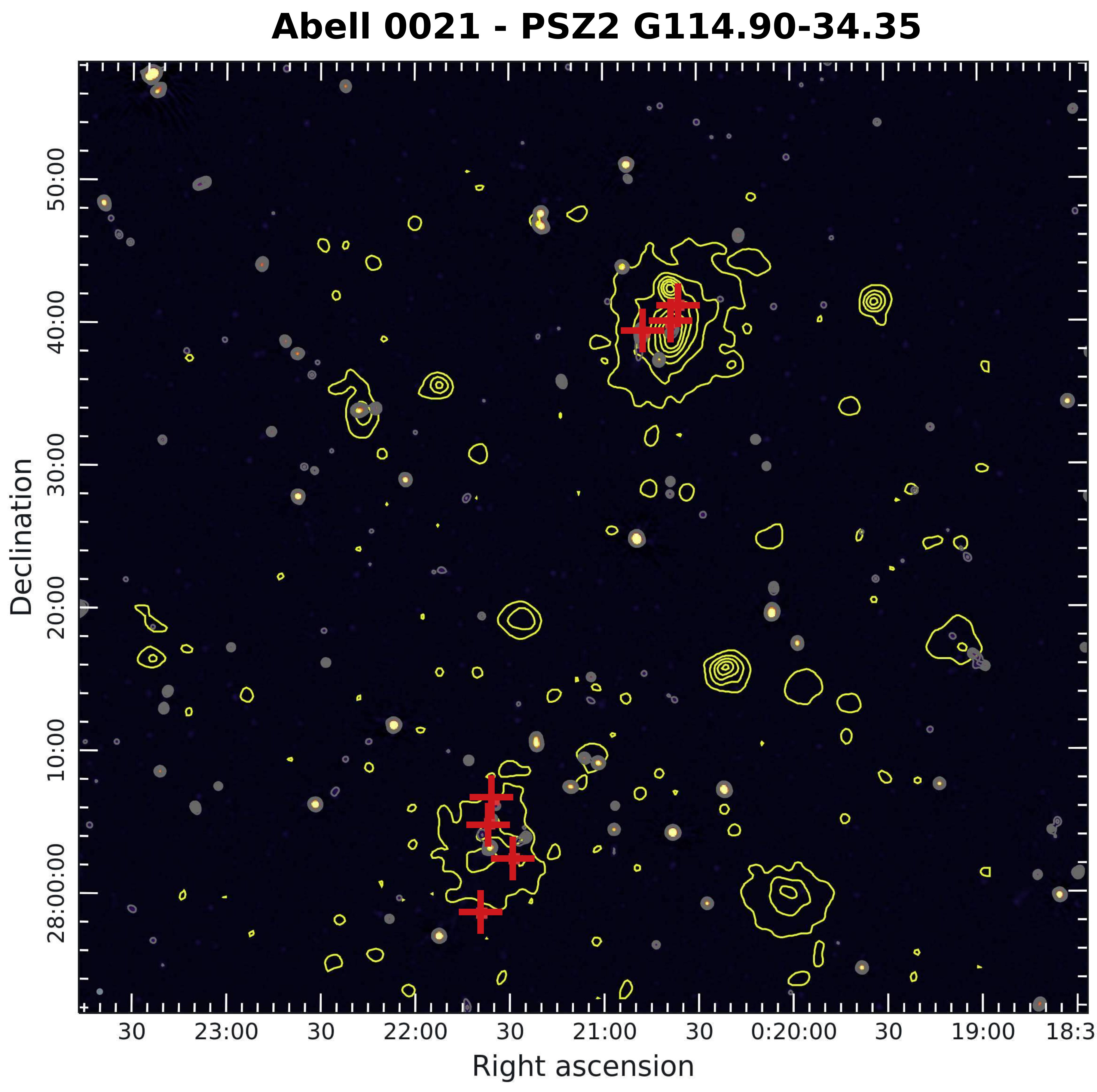}
                \caption{}
        \end{subfigure}

        \caption{\textbf{Abell 0021 -  PSZ2 G114.90-34.35} Panel~\textbf{(a)}: A plot of SDSS galaxies. Panel~\textbf{(b)}: DSS r-band optical image overlaid with LoTSS-2 low resolution image contours at [3,6,9,12,15,18] of $\sigma$ where $\sigma_{\rm{rms}} = 600~\rm{\mu Jy/beam}$. Panel~\textbf{(c)}: ROSAT X-ray contours (yellow) with exposure of 14~ks overlaid on LoTSS-2 point source subtracted image with same contour levels (grey)}\label{fig:PPM5}
       
\end{figure}

This system of two premerging/postmerging clusters is located at the mean redshift of $z = 0.093$. Several interacting substructures have been flagged by the algorithm in this work. A filament-like galaxy distribution can be seen to form in the central region of Fig.~\ref{fig:PPM5} \textbf{a}. Abell 0021 (Northern of the two) is a massive cluster with a weak lensing mass of $M_{500}$ of $6.1 \times 10^{14} M_\odot$ \citep{CCCPWLmass}. This is comparable with mass from our estimation of  $ \sim 4.4 \times 10^{14} M_\odot$. \citet{StrubleVelDisp} report a velocity dispersion of $ 621 km s^{-1}$ which is relatively low. The other member of this system is PSZ2 G114.90-34.35 with a mass of $M_{500}$ of $2.47 \times 10^{14} M_\odot$ \citep{PSZ2} is relatively less massive. The overall region shows several sources of diffuse X-ray emission as seen in Fig. \ref{fig:PPM5} \textbf{c}. Abell 0021 is reported be moderately hot ($T_x = 6.134^{+0.648}_{-0.434} \rm{keV}$) and of relatively low luminosity ( $L_x = 1.428 \times 10^{44} \rm{erg}^{-1}$ ) by \citet{xraygclustsample}. 
Although plenty of X-ray signatures are present, both the clusters do not show noticeable diffuse radio emission in LoTSS low-resolution images as seen in \ref{fig:PPM5} \textbf{c}. This is slightly surprising, as we estimate the combined interacting mass around $6.7 \times 10^{14} M_\odot$. Interestingly, \citet{BonjeanGasFilaments} have shown the presence of a gaseous filament connecting the two systems using an SZ map and galaxy field. This can also be seen in Fig.~\ref{fig:PPM5}\textbf{a}. Based on this report and our work, this site becomes an interesting candidate to observe WHIM filament and possibly a connecting radio bridge with upcoming SKA and other advanced interferometers.

\bsp	
\label{lastpage}
\end{document}